\documentclass[11pt]{article}
\usepackage{graphicx}
\usepackage{caption}
\usepackage{subcaption}
\usepackage{stfloats}
\usepackage{amsmath,amsthm,amssymb,amsfonts,mathtools,bbold,hyperref}
\usepackage{cite}
\usepackage{tikz}
\usepackage{color}
\usetikzlibrary{positioning}
\usetikzlibrary{decorations.pathreplacing}

\numberwithin{equation}{section}

\newcommand\numberthis{\addtocounter{equation}{1}\tag{\theequation}}

\DeclareMathOperator{\tr}{tr}

\newcommand{\hs}{\mathcal{H}}

\newcommand{\id}{\mathbb{1}}

\newcommand{\lp}{\left(}
\newcommand{\rp}{\right)}
\newcommand{\lb}{\left[}
\newcommand{\rb}{\right]}

\definecolor{brown}{HTML}{8B4513}
\definecolor{blue}{HTML}{007Aff}
\definecolor{green}{HTML}{94E700}
\definecolor{orange}{HTML}{FF8800}
\definecolor{red}{HTML}{FF0000}
\definecolor{pink}{HTML}{FF9797}
\definecolor{darkblue}{HTML}{03468F}
\definecolor{darkgreen}{HTML}{007355}

\begin{document}

\begin{titlepage}
{\ }
\vskip 1in

\begin{center}
{\LARGE{Bulk reconstruction and non-isometry in the \\
\vskip 0.05in backwards-forwards holographic black hole map}}
\vskip 0.5in {\Large Oliver DeWolfe\footnote{oliver.dewolfe@colorado.edu} and Kenneth Higginbotham\footnote{kenneth.higginbotham@colorado.edu}}
\vskip 0.2in {\it Department of Physics and \\
Center for Theory of Quantum Matter \\
390 UCB \\ University of Colorado \\ Boulder, CO 80309, USA}
\end{center}
\vskip 0.5in

\begin{abstract}\noindent
The backwards-forwards map, introduced as a generalization of the non-isometric holographic maps of the black hole interior of Akers, Engelhardt, Harlow, Penington, and Vardhan to include non-trivial dynamics in the effective description, has two possible formulations differing in when the post-selection is performed. While these two forms are equivalent on the set of dynamically generated states -- states formed from unitary time evolution acting on well-defined initial configurations of  infalling matter -- they differ on the generic set of states necessary to describe the apparent world of the infalling observer. We show that while both versions successfully reproduce the Page curve, the version involving post-selection as the final step, dubbed the backwards-forwards-post-selection (BFP) map, has the desirable properties of being non-isometric but isometric on average and providing state-dependent reconstruction of bulk operators, while the other version does not. Thus the BFP map is a suitable non-isometric code describing the black hole interior including interior interactions.
\end{abstract}

\end{titlepage}

\section{Introduction}

Recently, non-isometric codes have become a fascinating avenue for exploring how black hole interiors might be described by a theory of quantum gravity. The non-isometric holographic maps proposed by Akers, Engelhardt, Harlow, Penington, and Vardhan (PHEVA\footnote{We will continue to make use of this user-friendly permutation of the authors' initials.}) use post-selection to remove the extra degrees of freedom found in a semiclassical description of the interior \cite{akers_black_2022}. These codes have been successfully used to derive the quantum extremal surface (QES) formula \cite{ryu_holographic_2006,faulkner_quantum_2013,engelhardt_quantum_2015,penington_entanglement_2020,almheiri_entropy_2019}, obtain the Page curve \cite{page_information_1993}, realize a construction of the Python's lunch \cite{brown_pythons_2020,engelhardt_world_2021,engelhardt_finding_2022}, and provide a state-dependent reconstruction of interior operators \cite{akers_quantum_2022}, in a way consistent with expectations for the role of computational complexity \cite{harlow_quantum_2013}. The non-isometric property of these maps has also been observed in the context of two-dimensional conformal field theories and random tensor networks \cite{chandra_toward_2023}, further motivating their utility in understanding how semiclassical properties of black holes should be encoded in quantum gravity.

So far, realizations of non-isometric holographic maps -- which take a black hole state in the \textit{effective} (interior observer) description and map it to a \textit{fundamental} (exterior observer) description -- have been built on qudit toy models of black holes. Despite the inherent assumptions and simplifications in these qudit models, it is encouraging that non-isometric codes have been so successful even in these simplified contexts. The most detailed constructions involve building non-isometric maps out of the dynamics of the black hole itself. PHEVA constructed the first such dynamical holographic map out of the black hole dynamics in the fundamental description \cite{akers_black_2022}. In doing so, they assumed that the effective dynamics were trivial -- once inside the interior, nothing would happen to an infalling observer. Since an observer inside a large black hole should be able to do all sorts of physics experiments, it is natural to generalize this construction to include non-trivial  interactions in the effective description. 

The first step in this direction was taken by Kim and Preskill \cite{kim_complementarity_2023}. Concerned by issues surrounding post-selection that arose in the context of the final state proposal \cite{horowitz_black_2004,lloyd_unitarity_2014}, namely the possibility that the post-selection involved in the map could lead to violations of unitarity or to superpolynomial computational speedup violating the extended Church-Turing thesis \cite{aaronson_quantum_2005,deutsch_quantum_1997,susskind_horizons_2020}, they included interactions between an infalling robot and radiation modes (both inside and outside the horizon) in the effective description and applied PHEVA's dynamical non-isometric map. Kim and Preskill found that the new interactions combined with the post-selection in the non-isometric map to create small violations to the unitarity of black hole evaporation, but  that any computational speedups are limited by the size of the infaller's Hilbert space.

In a previous work \cite{dewolfe_non-isometric_2023}, we took a further step  by proposing that nontrivial interior dynamics should be reflected in the holographic map itself. We took the perspective that a holographic map from the effective to the fundamental description at a moment in time could be regarded as a composition of backwards time dynamics in the effective description until all infalling modes have been brought back outside the black hole, followed by forwards time dynamics in the fundamental description bringing them back in again. In the perspective of this ``backwards-forwards" map, post-selection arises because Hawking pairs annihilate and return to the vacuum during backwards effective time evolution. Because the map incorporates effective dynamics directly, it restores unitarity of evaporation and as well as avoiding computational speedup problems typically associated with post-selection.

Our previous work \cite{dewolfe_non-isometric_2023} focused on the action of the backwards-forwards map on \textit{dynamically generated states}: black hole states formed by the unitary evolution of initial configurations of non-singular infalling matter. These states form a subspace of the full effective Hilbert space. On this subspace, we showed that the backwards-forwards map can be transformed to a ``post-selection version'' which is equivalent to PHEVA's dynamical map when effective interactions are removed. All of the results in \cite{dewolfe_non-isometric_2023} are appropriate for this dynamically generated subspace.

However, PHEVA's non-isometric maps were applied to more than just the dynamically generated subspace. In particular, their calculations \cite{akers_black_2022} showing that non-isometric holographic maps act isometrically on subexponentially complex states, and that their maps provide state-dependent reconstruction of bulk operators, required the maps' actions on generic states -- general linear combinations of the dynamically generated states and the null states annihilated by the holographic map. These states cannot be reached by any dynamical perturbation of a black hole state formed from unitary collapse -- no observer or robot could fall into the black hole and perform a unitary operation to take the black hole out of the dynamically generated subspace. However, they are states in the effective description Hilbert space, and a well-defined holographic map should have an action on all such states. Moreover, they can be perceived by an interior observer's measurement, considered as involving a unitary operation that entangles an observer and their apparatus with the system they are measuring. This entangling unitary is dynamical and cannot take the black hole state out of the dynamically generated subspace alone;  however, the observer could (purposefully or naively) choose a basis for their apparatus that involves null states outside of the dynamically generated subspace. Thus when they observe their apparatus, the observer could see the wavefunction collapse to a state inaccessible by dynamics alone.

Thus to demonstrate that the backwards-forwards holographic map is an appropriate interacting generalization of the maps of \cite{akers_black_2022}, and can successfully reproduce properties like isometry on subexponential states and bulk reconstruction, we must also investigate its action on generic states. This action must be treated carefully; for a generic state of the effective description, backwards time evolution will not return Hawking modes to the expected maximally entangled states before they annihilate. Such a mismatch could lead to singularities forming in the past.

We will see that when acting on generic states, the two presentations of the backwards-forwards map that coincided on dynamically generated states are no longer equivalent. The original formulation of the map involves post-selecting on the Hawking states as soon as they return to the vacuum, while the other version saves post-selection for the end, more closely analogous to the PHEVA map. The aim of this work is to determine which (if either) of the two versions of the backwards-forwards map can work well on generic states. This will involve making sure both maps are defined so as to avoid past singularities, and verifying whether they satisfy two criteria established by PHEVA in \cite{akers_black_2022}: that the holographic map should act isometrically on generic states for Haar-typical fundamental dynamics, and that it should provide a reconstruction of interior unitary operators in the effective description up to exponentially small errors, in a fashion consistent with entanglement wedge reconstruction. We will find that the first formulation of the backwards-forwards map -- which we refer to as the BF map -- does not satisfy these criteria, while the alternate version -- referred to here as the backwards-forwards-post-selection (BFP) map -- does. Reproducing a number of calculations of \cite{akers_black_2022}, we see that while the derivation of the Page curve succeeds for both maps,  isometry on average and bulk reconstruction follow only for the BFP map, in a way analogous to the original PHEVA calculations with generally small corrections.

This paper will be organized as follows. Sec.~\ref{sec:maps} will review the definitions of the holographic maps proposed in both \cite{akers_black_2022} and \cite{dewolfe_non-isometric_2023} and establish the notational conventions taken here. We will also discuss how the backwards-forwards and BFP maps can avoid the creation of a past singularity during the backwards evolution. For completeness, we will demonstrate here that our maps can be used to obtain the same QES formula and Page curve as PHEVA's dynamical maps. Sec.~\ref{sec:dyn_gen} will review our results from \cite{dewolfe_non-isometric_2023} for the action of our maps on the dynamically generated subspace and provide new clarifying details. Sec.~\ref{sec:generic} will check the two criteria above for both the backwards-forwards and BFP maps. Finally, we will conclude in Sec.~\ref{sec:conc} where we will discuss the assumptions made in this work and opportunities for future studies.

\section{The backwards-forwards holographic maps} \label{sec:maps}

We begin by defining the black hole model, its effective description and fundamental description time dynamics, and the backwards-forwards holographic maps that will be used throughout the paper. 

\subsection{Effective and fundamental descriptions and dynamics}
The two descriptions of the black hole system are taken to be described by systems of qudits \cite{akers_black_2022} and thus are represented as finite-dimensional Hilbert spaces. The black hole degrees of freedom in the effective description are divided into left-moving (radially ingoing) modes $\ell$ and right-moving (radially outgoing) modes $r$. In the fundamental description, internal black hole degrees of freedom are labeled by $B$. Both descriptions share the same exterior: $R_\text{in}$ describes infalling modes, and $R_\text{out}$ contains the Hawking radiation. Thus we get the following factorizations of the effective and fundamental Hilbert spaces,
\begin{equation}
    \hs_\text{eff} = \hs_\ell \otimes \hs_r \otimes \hs_{R_\text{in}} \otimes \hs_{R_\text{out}}, \qquad \hs_\text{fun} = \hs_B \otimes \hs_{R_\text{in}} \otimes \hs_{R_\text{out}}.
\end{equation}
When needed, we will include a reference system $\hs_L$ in the effective description that keeps track of (and purifies) the $\ell$ modes that fell into the black hole from $R_\text{in}$. When referring to different  Hilbert space factors we will often drop the $\hs$ and just refer to them as $\ell$, $r$ and so forth.

\begin{figure}
    \centering
    \subfloat{

\scalebox{1.15}{
\begin{tikzpicture}[line width=1.1pt]

\draw[red] (0,0) -- (0,0.5);
\draw[red] (0.5,0) -- (0.5,0.5);
\draw[red] (1,0) -- (1,0.5);
\draw[red] (1.5,0) -- (1.5,0.5);
\draw (2,0) -- (2,0.5);
\draw (2.5,0) -- (2.5,0.5);

\node at (0.75,-0.35) {$R_\text{in}$};
\node at (2.25,-0.35) {$|\psi\rangle_f$};

\draw[red] (0,0.5) -- (0,1.5);
\draw[red] (0.5,0.5) -- (0.5,1.5);
\draw (1,1) -- (1,1.5);
\draw (1.5,1) -- (1.5,1.5);
\draw (2,1) -- (2,1.5);
\draw (2.5,1) -- (2.5,1.5);

\draw[fill=gray!30] (0.75,0.5) rectangle (2.75,1);
\node at (1.75,0.75) {$U_0$};
\node at (-1,0.75) {$t=0$};

\draw[red] (0,1.5) -- (0,2.5);
\draw (0.5,2) -- (0.5,2.5);
\draw (1,2) -- (1,2.5);
\draw (1.5,2) -- (1.5,2.5);
\draw[blue] (2,2) -- (2,2.5);
\draw[blue] (2.5,2) -- (2.5,2.5);

\draw[fill=gray!30] (0.25,1.5) rectangle (2.75,2);
\node at (1.5,1.75) {$U_1$};
\node at (-1,1.75) {$t=1$};

\draw (0,3) -- (0,3.5);
\draw (0.5,3) -- (0.5,3.5);
\draw[blue] (1,3) -- (1,3.5);
\draw[blue] (1.5,3) -- (1.5,3.5);
\draw[blue] (2,2.5) -- (2,3.5);
\draw[blue] (2.5,2.5) -- (2.5,3.5);

\draw[fill=gray!30] (-0.25,2.5) rectangle (1.75,3);
\node at (0.75,2.75) {$U_2$};
\node at (-1,2.75) {$t=2$};

\node at (0.25,3.85) {$B$};
\node at (1.75,3.85) {$R_\text{out}$};

\end{tikzpicture}
}}
    \hfill
    \subfloat{

\scalebox{1.2}{
\begin{tikzpicture}[line width=1.1pt]

\draw[red] (0,0) -- (0,0.5);
\draw[red] (0.5,0) -- (0.5,0.5);
\draw[red] (1,0) -- (1,0.5);
\draw[red] (1.5,0) -- (1.5,0.5);

\node at (0.75,-0.35) {$R_\text{in}$};

\draw[red] (0,0.5) -- (0,1.5);
\draw[red] (0.5,0.5) -- (0.5,1.5);
\draw[orange] (1,1) -- (1,1.5);
\draw[orange] (1.5,1) -- (1.5,1.5);

\draw[brown,fill=brown!20] (0.75,0.5) rectangle (1.75,1);
\node[brown] at (1.25,0.75) {$\hat{U}_0$};
\node at (-1,0.75) {$t=0$};

\draw[red] (0,1.5) -- (0,2.5);
\draw[orange] (0.5,2) -- (0.5,2.5);
\draw[orange] (1,2) -- (1,2.5);
\draw[orange] (1.5,2) -- (1.5,2.5);
\draw[green] (3.75,1.5) -- (2,2) -- (2,2.5);
\draw[green] (3.75,1.643) -- (2.5,2) -- (2.5,2.5);
\draw[blue] (3.75,1.5) -- (5.5,2) -- (5.5,2.5);
\draw[blue] (3.75,1.643) -- (5,2) -- (5,2.5);

\node at (3.75,1.15) {$|\text{MAX}\rangle_{r,R_\text{out}}$};
\draw[brown, fill=brown!20] (0.25,1.5) rectangle (1.75,2);
\node[brown] at (1,1.75) {$\hat{U}_1$};
\node at (-1,1.75) {$t=1$};

\draw[orange] (0,3) -- (0,3.5);
\draw[orange] (0.5,3) -- (0.5,3.5);
\draw[orange] (1,3) -- (1,3.5);
\draw[orange] (1.5,3) -- (1.5,3.5);
\draw[green] (2,3) -- (2,3.5);
\draw[green] (2.5,3) -- (2.5,3.5);
\draw[green] (3.75,2.5) -- (3,3) -- (3,3.5);
\draw[green] (3.75,2.833) -- (3.5,3) -- (3.5,3.5);
\draw[blue] (3.75,2.5) -- (4.5,3) -- (4.5,3.5);
\draw[blue] (3.75, 2.833) -- (4,3) -- (4,3.5);
\draw[blue] (5,2.5) -- (5,3.5);
\draw[blue] (5.5,2.5) -- (5.5,3.5);

\draw[brown, fill=brown!20] (-0.25,2.5) rectangle (2.75,3);
\node[brown] at (1.25,2.75) {$\hat{U}_2$};
\node at (-1,2.75) {$t=2$};

\node at (0.75,3.85) {$\ell$};
\node at (2.75,3.85) {$r$};
\node at (4.75,3.85) {$R_\text{out}$};

\end{tikzpicture}
}}
    \caption{An example of fundamental dynamics $U_t$ (left) and \textcolor{brown}{effective dynamics} $\hat{U}_t$ (right) until $t=2$ for $n_0=4$ and $m_0=2$. Lines are colored by qudit type: \textcolor{red}{red} for $R_\text{in}$, \textcolor{blue}{blue} for $R_\text{out}$, black for $B$ and $f$, \textcolor{orange}{orange} for $\ell$, and \textcolor{green}{green} for $r$.}
    \label{fig:fun_eff_dyn}
\end{figure}
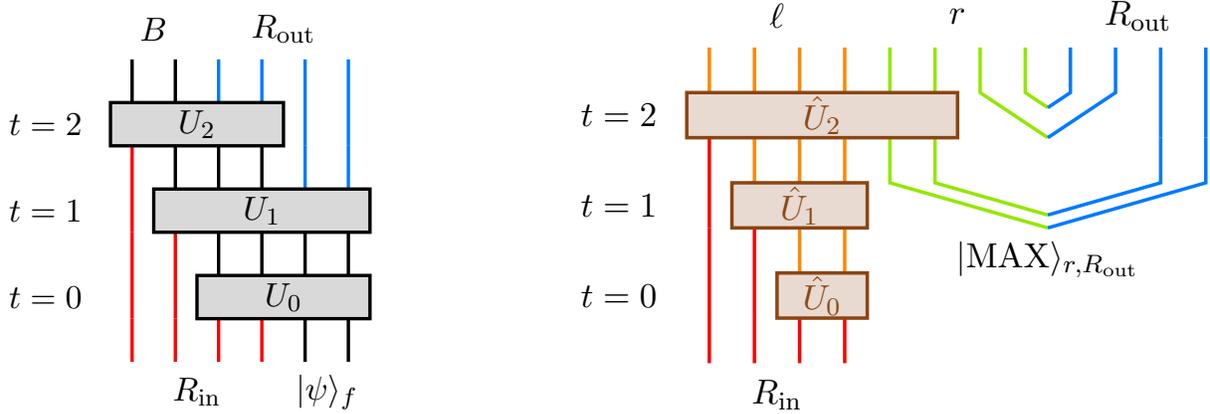

The fundamental description characterizes the point of view of the external observer, who perceives the black hole as an object that accepts infalling degrees of freedom, scrambles them, and re-radiates them as Hawking radiation. This is pictured on the left-hand-side of Fig.~\ref{fig:fun_eff_dyn}; at time zero a set of modes falling in from $R_\text{in}$ form the black hole, and are joined by a set of fixed degrees of freedom $|\psi\rangle_f$ coming down from high energies during the initial collapse. These are together processed by a unitary transformation $U_0$ into black hole modes $B$. At each succeeding time step, another unitary transformation $U_t$ processes the modes in $B$ and one new infalling degree of freedom from $R_\text{in}$,  outputting two Hawking modes into $R_\text{out}$ while the rest remain in $B$. If $m_0$ infallers from $R_\text{in}$ are joined by $n_0-m_0$ fixed degrees of freedom $f$, the unitaries describing the fundamental dynamics are maps,
\begin{equation} \label{eq:fun_dyn}
    U_t\,:\,
        \begin{cases}
            R_\text{in}^{(m_0)} f^{(n_0-m_0)} \rightarrow B^{(n_0)}, & t=0 \\[0.2cm]
            R_\text{in}^{(1)} B^{(n_0 + 1 - t)} \rightarrow B^{(n_0 - t)} R_\text{out}^{(2)}, & t>0
        \end{cases}
\end{equation}
where parenthetical superscripts have been used to denote the number of qudits from each factor.
In \cite{akers_black_2022}, it was assumed that each $U_t$ was drawn from the Haar measure. Recent work by Kim and Preskill \cite{kim_complementarity_2023} suggests that pseudorandom unitaries could offer a more realistic description of each $U_t$.

The effective description describes the semiclassical gravity picture seen by an observer behind the black hole horizon. In this description, modes fall into the black hole and are joined at each time step by the creation of Hawking pairs, one particle in the pair remaining in $r$ inside the horizon and the other escaping into $R_\text{out}$ as Hawking radiation. These Hawking pairs are created in the maximally entangled state,
\begin{equation}
    |\text{MAX}\rangle_{r,R_\text{out}} = \frac{1}{\sqrt{q}} \sum_{j=1}^{q} |j\rangle_r |j\rangle_{R_\text{out}},
\end{equation}
where $q$ is the dimension of the qudits, 
reflecting the assumed smoothness of spacetime at the horizon. Dealing with this pair creation backwards will be a subtlety for backwards time evolution.
In addition,
our previous work \cite{dewolfe_non-isometric_2023} introduced new non-trivial dynamics $\hat{U} \equiv \hat{U}_t \hat{U}_{t-1} \ldots \hat{U}_0$ for the interior of the black hole in the effective description. Each $\hat{U}_t$ acts on effective degrees of freedom as
\begin{equation} \label{eq:eff_dyn}
    \hat{U}_t \,:\, 
        \begin{cases}
            R_\text{in}^{(m_0)} \rightarrow \ell^{(m_0)}, & t=0 \\[0.2cm]
            R_\text{in}^{(1)} \ell^{(m_0 - 1 + t)} r^{(2(t-1))} \rightarrow \ell^{(m_0 + t)} r^{(2(t-1))}, & t>0 \,.
        \end{cases}
\end{equation}
An example of these dynamics is shown in the right panel of Fig.~\ref{fig:fun_eff_dyn}. The $\hat{U}_t$ represent the semiclassical gravity interactions perceived by the infalling observer, but for this toy model we have so far made no restrictions on their precise form; later on we will see that some assumptions about their average properties will be useful.

\subsection{Non-isometric codes for the black hole interior}
At any moment in time $t$, we expect there to be a holographic map $V_t$:
\begin{equation}
    V_t: \hs_\text{eff} \to \hs_\text{fun}
\end{equation}
carrying a configuration in the effective description to a configuration in the fundamental description, and acting as the identity on $R_\text{in} \otimes R_\text{out}$.  The generic non-isometric map $V\,:\, \ell r \rightarrow B$ defined in \cite{akers_black_2022} takes the form 
\begin{equation} \label{eq:PHEVA_gen}
    V = \sqrt{|P|} \langle\phi|_P U_\text{H} |\psi\rangle_f \,,
\end{equation}
where $U_\text{H}$ is a typical block unitary drawn from the Haar measure with no additional structure, $|\psi\rangle_f$ is the insertion of fixed degrees of freedom, and $\langle\phi|_P$ represents post-selecting a subsystem $P$ (the complement of $B$ in the output of $U_\text{H}$) on some specified state. The prefactor $\sqrt{|P|}$ is included to ensure correct normalization of the state after post-selection.

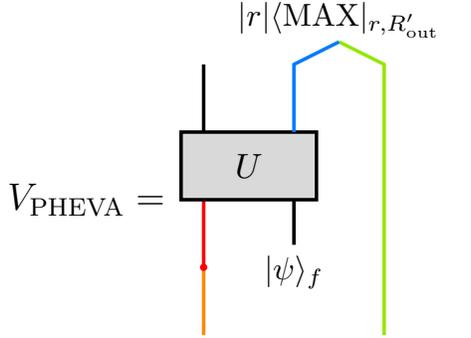
\begin{figure}
     \centering

\scalebox{1.2}{
\begin{tikzpicture}[line width=1.1pt]

\node[font=\large] at (-1.3,2.75) {$V_\text{PHEVA} =$};

\draw[blue] (3,1.25) -- (3,4.25);

\draw[orange] (0,1.25) -- (0,2);
\draw[red] (0,2) -- (0,2.75);
\filldraw[red] (0,2) circle (0.6pt);
\draw[green] (2,1.25) -- (2,4.25) -- (1.5,4.5);
\draw (1,2.25) -- (1,2.75);

\node[font=\small] at (1,1.95) {$|\psi\rangle_f$};

\draw[fill=gray!30] (-0.25,2.75) rectangle (1.25,3.5);
\node at (0.5,3.125) {$U$};

\draw (0,3.5) -- (0,4.25);
\draw[blue] (1,3.5) -- (1,4.25) -- (1.5,4.5);
\node[font=\small] at (1.5,4.75) {$|r|\langle\text{MAX}|_{r,R'_\text{out}}$};

\end{tikzpicture}
}
     \caption{A schematic representation of the PHEVA dynamical holographic map for the black hole interior \cite{akers_black_2022}. Unlike in Fig.~\ref{fig:fun_eff_dyn}, each line here represents all qudits in that factor of the Hilbert space.}
     \label{fig:V_PHEVA}
 \end{figure}

The authors of \cite{akers_black_2022} further provided a realization of a map of the form (\ref{eq:PHEVA_gen}) constructed out of the fundamental description dynamics: modes in $\ell$ are fed into fundamental time evolution, while modes in $r$ are post-selected together with the $R_\text{out}$ modes coming out of the fundamental dynamics, giving a map of the form
\begin{equation}
\label{eq:V_PHEVA}
    V_\text{PHEVA} = |r| \langle \text{MAX} |_{r,R'_\text{out}} U |\psi\rangle_f \,.
\end{equation}
Here there are two copies of the Hawking radiation $R_\text{out}$, and we label the one that is output of the fundamental dynamics as $R'_\text{out}$. When the input $rR_\text{out}$ modes are in the maximally entangled state $| \text{MAX} \rangle_{r,R'_\text{out}}$, the map functions as a teleportation protocol, transferring the state of the degrees of freedom in $R'_\text{out}$ to those in $R_\text{out}$. Fig.~\ref{fig:V_PHEVA} gives a circuit representation of PHEVA's dynamical holographic map.

The PHEVA map was possible for a case with no effective description dynamics, where each mode in $\ell$ can be identified uniquely with a mode in $R_\text{in}$; but in general, it is unnatural to use the state of the $\ell$ modes existing well beyond the horizon as an input to fundamental dynamics, which expects modes just crossing the horizon. The ``backwards-forwards" map of \cite{dewolfe_non-isometric_2023} was designed to generalize $V_\text{PHEVA}$ and resolve this issue, functioning as a holographic map taking into account effective description dynamics.

\subsection{The backwards-forwards (BF) map}
The philosophy of the backwards-forwards (BF)  map is that the holographic map is a composition of backwards time dynamics in the effective description, followed by forward time dynamics in the fundamental description.  Beginning with a state on $\ell rR_\text{out}$ at some time step $t$, the map first evolves these effective degrees of freedom backwards in the effective description using $\hat{U}^\dagger$. At each time step in the backwards evolution, one qudit from $\ell$ leaves the black hole and rejoins $R_\text{in}$. Furthermore, at each time-step two Hawking pairs in the maximally entangled state disappear. Once the backwards evolution is complete and there are no more black hole degrees of freedom, the map performs forwards fundamental time evolution $U$ on $R_\text{in}$ to bring the black hole into the fundamental description. Thus the map is aware of both kinds of dynamics.

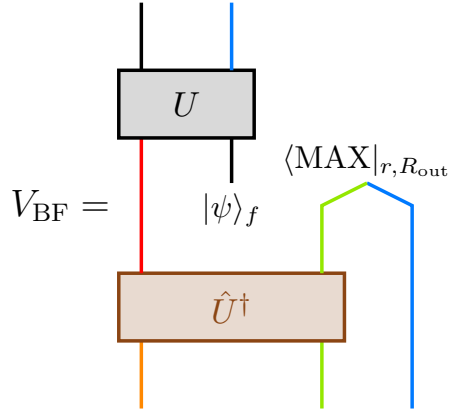
\begin{figure}
     \centering

\scalebox{1.2}{
\begin{tikzpicture}[line width=1.1pt]

\node[font=\large] at (-0.9,2) {$V_\text{BF} =$};

\draw[orange] (0,-0.25) -- (0,0.5);
\draw[green] (2,-0.25) -- (2,0.5);
\draw[blue] (3,-0.25) -- (3,2) -- (2.5,2.25);

\draw[brown,fill=brown!20] (-0.25,0.5) rectangle (2.25,1.25);
\node[brown] at (1,0.875) {$\hat{U}^\dagger$};

\draw[red] (0,1.25) -- (0,2.75);
\draw[green] (2,1.25) -- (2,2) -- (2.5,2.25);
\node[font=\small] at (2.5,2.5) {$\langle\text{MAX}|_{r,R_\text{out}}$};
\draw (1,2.25) -- (1,2.75);
\node[font=\small] at (1,1.95) {$|\psi\rangle_f$};

\draw[fill=gray!30] (-0.25,2.75) rectangle (1.25,3.5);
\node at (0.5,3.125) {$U$};

\draw (0,3.5) -- (0,4.25);
\draw[blue] (1,3.5) -- (1,4.25);

\end{tikzpicture}
}
     \caption{A schematic representation of the backwards-forwards (BF) map defined in equation (\ref{eq:V_BF}). Unlike in Fig.~\ref{fig:fun_eff_dyn}, each line here represents all qudits in that factor of the Hilbert space.}
     \label{fig:V_BF}
 \end{figure}

 Since the forward time evolution of the effective description is $\hat{U} | \text{MAX} \rangle_{r,R_\text{out}}$ where we have concatenated the $\hat{U}_t$ at each time step to a single $\hat{U}$, and the forward time evolution of the fundamental description is $U |\psi\rangle_f$ with the $U_t$ concatenated into $U$, we obtain the backwards-forwards (BF) map \cite{dewolfe_non-isometric_2023},
\begin{equation} \label{eq:V_BF}
    V_\text{BF} = U |\psi\rangle_f \langle \text{MAX} |_{r,R_\text{out}} \hat{U}^\dagger \,.
\end{equation}
Inverting the effective dynamics in this way has produced a post-selection $\langle \text{MAX} |_{r,R_\text{out}}$ that removes the Hawking pairs as we evolve back to their moment of creation. Equivalently, we can imagine decoupling each pair from the dynamics as we get back to its time of creation, and then post-selecting on all of them at $t=0$. The backwards-forwards map is presented in Fig.~\ref{fig:V_BF}.

\subsection{The backwards-forwards-post-selection (BFP) map}
\begin{figure}
    \centering

\scalebox{1.2}{
\begin{tikzpicture}[line width=1.1pt]

\node[font=\large] at (-0.9,2) {$V_\text{BFP} =$};

\draw[orange] (0,-0.25) -- (0,0.5);
\draw[green] (2,-0.25) -- (2,0.5);
\draw[blue] (3,-0.25) -- (3,4.25);

\draw[brown,fill=brown!20] (-0.25,0.5) rectangle (2.25,1.25);
\node[brown] at (1,0.875) {$\hat{U}^\dagger$};

\draw[red] (0,1.25) -- (0,2.75);
\draw[green] (2,1.25) -- (2,4.25) -- (1.5,4.5);
\draw (1,2.25) -- (1,2.75);
\node[font=\small] at (1,1.95) {$|\psi\rangle_f$};

\draw[fill=gray!30] (-0.25,2.75) rectangle (1.25,3.5);
\node at (0.5,3.125) {$U$};

\draw (0,3.5) -- (0,4.25);
\draw[blue] (1,3.5) -- (1,4.25) -- (1.5,4.5);
\node[font=\small] at (1.5,4.75) {$|r|\langle\text{MAX}|_{r,R'_\text{out}}$};

\end{tikzpicture}
}
    \caption{A schematic representation of the backwards-forwards-post-selection (BFP) map defined in equation (\ref{eq:V_BFP}). Unlike in Fig.~\ref{fig:fun_eff_dyn}, each line here represents all qudits in that factor of the Hilbert space.}
    \label{fig:V_BFP}
\end{figure}
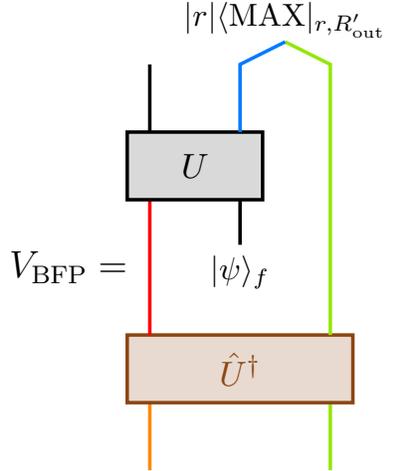

The backwards-forwards map presented in  equation~(\ref{eq:V_BF}) and Fig.~\ref{fig:V_BF} is not obviously a generalization of the PHEVA map $V_\text{PHEVA}$ (\ref{eq:V_PHEVA}), which involves post-selection on the output of the fundamental dynamics.
However, when the state acted on by the BF map (\ref{eq:V_BF}) is a {\em dynamically generated state} -- a state that evolved from non-singular initial conditions of matter infalling into the black hole -- the BF map can be recast in another form which is of the PHEVA type. For such a dynamically generated state, the backwards time evolution simply follows the state's original generation in reverse, and each Hawking pair is guaranteed to be in the maximally entangled state at the time-step it disappears. Thus if we think of decoupling and freezing each Hawking pair at the appropriate time-step, when we get to $t=0$ the $r R_\text{out}$ factor of the Hilbert space will indeed be in the state $| \text{MAX} \rangle_{r,R_\text{out}}$. Rather than post-selecting on it, we can keep it hanging around (we can imagine storing it in a ``quantum memory") while the forwards time evolution of the fundamental description takes place. Since the fundamental dynamics also has output Hawking radiation, we will have two factors of $R_\text{out}$, and we label the one coming from the fundamental dynamics as $R'_\text{out}$. Finally after the fundamental dynamics is complete, we post-select on $r$ which has remained in our quantum memory with $R'_\text{out}$, rather than with $R_\text{out}$. As with the PHEVA map (\ref{eq:V_PHEVA}), this functions as a teleportation protocol, transferring the state of the degrees of freedom in $R'_\text{out}$ to those in $R_\text{out}$. This defines the backwards-forwards-post-selection (BFP) version of the map, 
\begin{equation} \label{eq:V_BFP}
    V_\text{BFP} = |r| \langle \text{MAX} |_{r,R'_\text{out}} U |\psi\rangle_f \hat{U}^\dagger,
\end{equation}
where again the factor of $|r|$ is included for normalization,
as illustrated in Fig.~\ref{fig:V_BFP}. It is of the general PHEVA form (\ref{eq:PHEVA_gen}), and when the effective dynamics is trivial $\hat{U} = \mathbb{1}$, it reduces to the dynamical holographic map $V_\text{PHEVA}$ (\ref{eq:V_PHEVA}).

\subsection{Action of the maps on non-dynamically generated states}
When we are imagining acting the maps $V_\text{BF}$ and $V_\text{BFP}$ on states that were created by forward time evolution from non-singular initial configurations of matter, the backwards time evolution acts in a straightforward way that just reverses this evolution, and the two maps are equivalent. However, it is well-known that the Hilbert space of the effective description after the Page time will obtain many more degrees of freedom than exist in the fundamental description. Many of these states are null states, expected to be annihilated by the holographic map. A generic state will be a linear combination of dynamically generated states and null states. We need to be somewhat careful about the definition of the backwards evolution on one of these generic states, due to the Hawking modes coming into and out of the spectrum.

One way to think about the generation of the Hawking pairs is the following. The interior of a black hole grows as it evolves in time, and as it does, spacetime near the horizon expands. Similar to adiabatically increasing the width of a square well, the energy levels of quantum fields on this expanding spacetime will be red-shifted, bringing field modes down from high energies. These new modes may be occupied, which we interpret as the generation of Hawking radiation.\footnote{In principle, a well-defined theory of semiclassical gravity requires an ultraviolet cutoff to keep fluctuations of the stress-energy tensor (and therefore the metric) finite when including backreaction \cite{kiem_black_1995}. We will not need an explicit ultraviolet cutoff here, but having such a cutoff in mind may be conceptually useful when considering these Hawking modes descending from high energies.}

During the backwards evolution of a black hole state in the effective description, radiation modes will be blue-shifted back to high energy scales. We assume that if these radiation modes are occupied in the particular way that they are generated (i.e.\ maximally entangled) they will disappear nicely at high energies, exiting smoothly from the effective theory and no longer participating in interactions with modes remaining at lower energies. However, if we begin with a generic configuration in the effective description, radiation modes may approach high energy scales in the wrong state during the backwards evolution. One may anticipate this could cause some catastrophe in the UV, creating a singularity (or white hole) in the past \cite{akers_black_2022,kiem_black_1995,t_hooft_quantum_1985}. 

Let us consider how to properly implement the two maps to avoid such singularities. The BF map is more straightforward: it involves post-selection on each Hawking pair being in the correct maximally entangled state that it would be generated in. Thus if a particular set of Hawking modes is in the ``wrong" state as they are lifted back up out of the spectrum, post-selection simply annihilates the state.

For the BFP map, however, we do not immediately post-select on the Hawking modes as they leave the spectrum; we wish to keep them around to post-select after the fundamental dynamics. Thus, we define the BFP map such that as each set of Hawking modes is lifted out of the spectrum, it is frozen (stored in our ``quantum memory") and undergoes no further evolution. We may think of this memory as a record of the occupation of the various modes as they leave the effective theory. Since the modes are frozen at this point, they do not evolve to arbitrary high energies and no singularity will form. We keep these degrees of freedom until they are needed for the post-selection at the end of the BFP map. Since we found in \cite{dewolfe_non-isometric_2023} that exterior interactions drop out of the backwards-forwards map, we can also freeze out modes that leave the black hole when they are converted from $\ell$ back to $R_\text{in}$ during the backwards evolution. 

The backwards effective evolution portion of the BFP map is completed once all modes have been frozen as either infalling matter $R_\text{in}$ or radiation $rR_\text{out}$. At this point, we feed $R_\text{in}$ into the fundamental dynamics at the appropriate time-steps. The fundamental dynamics should be performed for a length of time given by the size of radiation modes frozen out during the backwards evolution, $t = (1/2) \log_q |R_\text{out}|$. The fundamental dynamics will produce black hole degrees of freedom $B$ and radiation $R'_\text{out}$.  The backwards-forwards map then performs post-selection on the stored $r$ qudits and the radiation output of the fundamental dynamics $R'_\text{out}$.

Once we are considering the action of the BF and BFP maps on generic states, it is not obvious that they are equivalent. Indeed, we will find that they are not. In the next section, we consider the action of the maps on dynamically generated states in more detail, and show that they both can be used to derive the Page curve. In the section following the next, we consider their action on generic states and show that only the BFP map successfully reproduces isometry on average and state-dependent reconstruction of bulk states.

Let us make a few remarks before turning to the next section. We note that this implementation of freezing out the Hawking modes during the holographic map is suitable for this kind of toy model of the black hole, but a deeper description would involve understanding the effective description degrees of freedom as modes in semiclassical gravity and understanding the role of backreaction on the geometry. Additionally, we emphasize that these holographic maps are just maps between two different descriptions of the black hole at a moment in time. While they are constructed using time evolution unitaries, they do not represent the dynamics or experiences of any observer. In particular, post-selection is not part of the effective dynamics of the black hole.

\begin{figure}
    \centering
    \scalebox{1.2}{
\begin{tikzpicture}[line width = 1.1pt]

\draw (0,3) -- (4,3);
\node (A) at (5,3) {$B$, $f$};
\draw[blue] (0,2.5) -- (4,2.5);
\node[below=0.5cm of A.west,anchor=west] (B) {$R_\text{out}$};
\draw[green] (0,2) -- (4,2);
\node[below=0.5cm of B.west,anchor=west] (C) {$r$};
\draw[orange] (0,1.5) -- (4,1.5);
\node[below=0.5cm of C.west,anchor=west] (D) {$\ell$};
\draw[red] (0,1) -- (4,1);
\node[below=0.5cm of D.west,anchor=west] (E) {$R_\text{in}$};
\draw[brown] (0,0.5) -- (4,0.5);
\node[below=0.5cm of E.west,anchor=west] (F) {$L$};
\draw[black, fill=gray!30] (1,-0.5) rectangle (3,0);
\node at (2,-0.25) {$U_t$};
\node[below=0.75cm of F.west,anchor=west] (G) {Fundamental dynamics};
\draw[brown, fill=brown!20] (1,-1.5) rectangle (3,-1);
\node[brown,font=\small] at (2,-1.25) {$\hat{U}_t$};
\node[below=1cm of G.west,anchor=west] (F) {Effective dynamics};

\end{tikzpicture}
}
    \caption{The color scheme used to identify qudits and dynamics in circuit diagrams.}
    \label{fig:colors}
\end{figure}
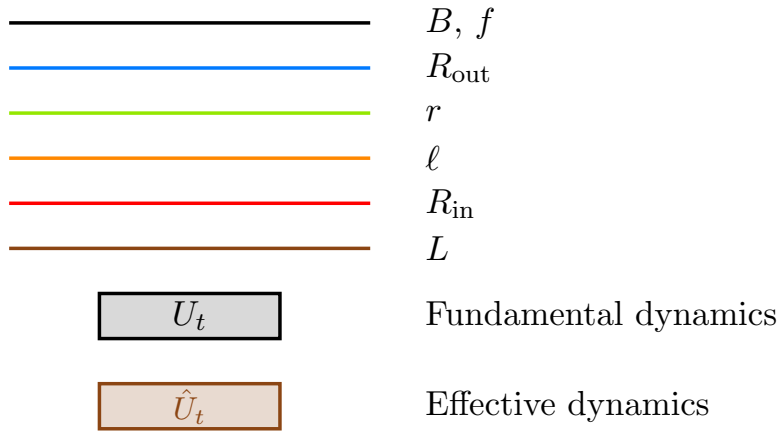

The color scheme for labeling different factors of the Hilbert space in our circuit diagrams is summarized in Fig.~\ref{fig:colors}. In addition, we will denote density matrices and pure states in the fundamental description with an uppercase $\Psi_i, \, |\Psi_i\rangle$, with the index $i$ labeling an orthonormal basis $\{ |\Psi_i\rangle \}$ of $\hs_\text{fun}$,
\begin{equation}
    \langle\Psi_j|\Psi_i\rangle = \delta_{ij}.
\end{equation}
States in the effective description will be distinguished with a hat: $\hat{\Psi}_i, \, |\hat{\Psi}_i\rangle$. Again, $i$ labels an orthonormal basis $\{ |\hat{\Psi}_i\rangle \}$ of $\hs_\text{eff}$. We will leave hats off of states of $R_\text{in}$ since they are shared by both descriptions.

\section{Action on the dynamically generated subspace} \label{sec:dyn_gen}

Now that we have defined our holographic maps, we begin with the action of $V_\text{BF}$ and $V_\text{BFP}$ on dynamically generated states. Here we will review essential results from our previous work in \cite{dewolfe_non-isometric_2023} and present new results to help clarify other properties of these maps. 

We will show that both the $V_\text{BF}$ and $V_\text{BFP}$ act isometrically on dynamically generated states. This fact has important consequences. Work by Kim and Preskill \cite{kim_complementarity_2023} brought forward concerns that the post-selection present in the dynamical holographic map proposed by \cite{akers_black_2022} might lead to superpolynomial computational complexity, violating the quantum extended Church-Turing thesis \cite{aaronson_quantum_2005,deutsch_quantum_1997,susskind_horizons_2020}. Post-selection has also been shown to lead to undesirable artifacts such as superluminal signaling and speedups to Grover's algorithm \cite{bao_grover_2016}. By including non-trivial effective dynamics in the backwards-forwards map directly, we have shown that these maps do act isometrically on dynamically generated states despite the presence of post-selection in the map. Therefore, both backwards-forwards maps avoid these issues typically associated with post-selection. 

While both maps  $V_\text{BF}$ and $V_\text{BFP}$ act isometrically on dynamically generated states of the effective description, only $V_\text{BF}$ acts unitarily. We show that $V_\text{BFP}^\dagger$ in general maps dynamically generated states of the fundamental description to generic states of the effective description, and thus is not itself an isometry. Projecting the result of $V_\text{BFP}^\dagger$ on the dynamically generated subspace restores unitarity, however. We also show that both $V_\text{BF}$ and $V_\text{BFP}$ are equivariant, commuting with time evolution for the two descriptions. We conclude the section with a demonstration that both maps successfully reproduce the Page curve.

In this section we will always assume that black hole states are generated from unitary dynamics and are a part of the dynamically generated subspace. In the effective description, we define a dynamically generated state using the effective dynamics,
\begin{equation} \label{eq:dyn_gen_eff}
    |\hat{\Psi}_i(t)\rangle_{\ell rR_\text{out}} \equiv \hat{U} |\Psi_i\rangle_{R_\text{in}} |\text{MAX}\rangle_{r,R_\text{out}},
\end{equation}
where we've included $(t)$ in the ket to indicate this state is an element of the dynamically generated subspace and to emphasize the time dependent nature of these states. Dynamically generated states in the effective description are characterized by the initialization of the Hawking pairs $rR_\text{out}$ in the maximally entangled state. Thus $|\hat{\Psi}(t)\rangle$ lives in a subspace of dimension $|R_\text{in}|$.

We also define a dynamically generated subspace of the fundamental description. We imagine that the fixed states $f$ are completely determined by the conditions of the collapse that formed the black hole; therefore, they carry no additional information than what is already in $R_\text{in}$. We define a dynamically generated state of the fundamental description using the fundamental dynamics,
\begin{equation} \label{eq:dyn_gen_fun}
    |\Psi_i(t)\rangle_{BR_\text{out}} \equiv U |\Psi_i\rangle_{R_\text{in}} |\psi\rangle_f.
\end{equation}
The dynamically generated subspace of the fundamental description defined by these states has dimension $|R_\text{in}|$, the same size as the dynamically generated subspace of the effective description.

It will be useful to define projectors onto these dynamically generated subspaces. For the effective description, we define
\begin{equation} \label{eq:proj_dyn_eff}
    \hat{P}_t \equiv \sum_i |\hat{\Psi}_i(t)\rangle\langle\hat{\Psi}_i(t)|,
\end{equation}
with the sum ranging over an orthonormal basis $\{ |\hat{\Psi}_i(t)\rangle \}$ of the dynamically generated subspace. Similarly, we define a projector onto the dynamically generated subspace of the fundamental description,
\begin{equation} \label{eq:proj_dyn_fun}
    P_t \equiv \sum_i |\Psi_i(t)\rangle\langle\Psi_i(t)|,
\end{equation}
with the sum ranging over a similar orthonormal basis $\{ |\Psi_i(t)\rangle \}$. We note that both of these projectors are time dependent since the dynamically generated states they are constructed from depend on the unitary dynamics $U_t$ and $\hat{U}_t$.

\subsection{The BF map acts unitarily}
We begin with the action of the backwards-forwards map $V_\text{BF}$ (\ref{eq:V_BF}) on the dynamically generated states (\ref{eq:dyn_gen_eff}) of the effective description. Recall that the backwards-forwards map begins by performing backwards effective evolution. This backwards evolution includes the annihilation of Hawking pairs to the vacuum, which we interpret as post-selection on $\langle \text{MAX}|_{r,R_\text{out}}$. Thus acting on (\ref{eq:dyn_gen_eff}) with $V_\text{BF}$, we find
\begin{align*}
    V_\text{BF} |\hat{\Psi}_i(t)\rangle_{\ell rR_\text{out}} &= \lp U |\psi\rangle_f \langle\text{MAX}|_{r,R_\text{out}} \hat{U}^\dagger \rp \hat{U} |\Psi_i\rangle_{R_\text{in}} |\text{MAX}\rangle_{r,R_\text{out}} \\
        &= U |\Psi_i\rangle_{R_\text{in}} |\psi\rangle_f \\
        &= |\Psi_i(t)\rangle_{BR_\text{out}}. \numberthis
\end{align*}
$V_\text{BF}$ maps a dynamically generated state of the effective description to a dynamically generated state of the fundamental description. Similarly, we can ask how the adjoint of the backwards-forwards map, 
\begin{equation}
    V_\text{BF}^\dagger = \hat{U} |\text{MAX}\rangle_{r,R_\text{out}} \langle\psi|_f U^\dagger
\end{equation}
acts on dynamically generated states of the fundamental description. $V_\text{BF}^\dagger$ acts on states in the fundamental description, first evolving them backwards using fundamental dynamics. Post-selection is performed on $f$ being in the fixed state $\langle\psi|_f$ as they are blue-shifted to high energies in the backwards evolution. Maximally entangled radiation is then inserted on $rR_\text{out}$ before forwards effective dynamics are applied. Acting on (\ref{eq:dyn_gen_fun}), the adjoint gives
\begin{align*}
    V_\text{BF}^\dagger |\Psi_i(t)\rangle_{BR_\text{out}} &= \lp \hat{U} |\text{MAX}\rangle_{r,R_\text{out}} \langle\psi|_f U^\dagger \rp U |\Psi_i\rangle_{R_\text{in}} |\psi\rangle_f \\
        &=  \hat{U} |\Psi_i\rangle_{R_\text{in}} |\text{MAX}\rangle_{r,R_\text{out}} \\
        &= |\hat{\Psi}_i(t)\rangle_{\ell rR_\text{out}}. \numberthis
\end{align*}
$V_\text{BF}^\dagger$ then maps a dynamically generated state of the fundamental description to a dynamically generated state of the effective description. We see that the backwards-forwards map and its adjoint take states back and forth between the dynamically generated subspaces of both descriptions; it never maps a dynamically generated state to a generic state of the other Hilbert space. Furthermore, inner products between states in each dynamically generated subspace are preserved, and $V_\text{BF}$ acts unitarily on the dynamically generated subspace of both descriptions. We express this fact as operator equations using the projectors (\ref{eq:proj_dyn_eff}) and (\ref{eq:proj_dyn_fun}),
\begin{gather}
    \hat{P}_t V_\text{BF}^\dagger V_\text{BF} \hat{P}_t = \hat{P}_t \\
    P_t V_\text{BF} V_\text{BF}^\dagger P_t = P_t
\end{gather}
It is good that $V_\text{BF}$ acts unitarily, since the backwards-forwards map is constructed directly from backwards and forwards unitary evolution. Furthermore, the above results are exact -- no averaging over $U$ or $\hat{U}$ was needed to find them.

We also note that the backwards-forwards map is equivariant: effective description time evolution followed by the holographic map is equivalent to the holographic map followed by fundamental description time evolution,
\begin{equation} \label{eq:equivariance}
    V_{t+1} \hat{U}_{t+1} = U_{t+1} V_t.
\end{equation}
This was an important property of the dynamical non-isometric code first proposed by PHEVA in \cite{akers_black_2022}. $V_\text{BF}$ is manifestly equivariant, as can be seen by substituting (\ref{eq:V_BF}) into (\ref{eq:equivariance}).

\subsection{The BFP map acts isometrically}
We now turn to the action of the backwards-forwards-post-selection (BFP) map (\ref{eq:V_BFP}) on the dynamically generated subspace. The fact that the post-selection acts at the end  of the map on $r R'_\text{out}$, instead of earlier on $r R_\text{out}$, 
will have interesting consequences for the action of the BFP map on dynamically generated states. 

First, we check the action of $V_\text{BFP}$ on dynamically generated states of the effective description (\ref{eq:dyn_gen_eff}),
\begin{align*}
    V_\text{BFP} |\hat{\Psi}_i(t)\rangle_{\ell rR_\text{out}} &= |r| \lp \langle\text{MAX}|_{r,R'_\text{out}} U |\psi\rangle_f \hat{U}^\dagger \rp \hat{U} |\Psi_i\rangle_{R_\text{in}} |\text{MAX}\rangle_{r,R_\text{out}} \\
        &= |r| \langle\text{MAX}|_{r,R'_\text{out}} |\Psi_i(t)\rangle_{BR'_\text{out}} |\text{MAX}\rangle_{r,R_\text{out}}. \numberthis
\end{align*}
Using a circuit representation, 
\begin{equation} \label{eq:BFP_dyn_gen_eff}
    V_\text{BFP} |\hat{\Psi}_i(t)\rangle_{\ell rR_\text{out}} = 
    \vcenter{
    \hbox{
    \scalebox{1}{
\begin{tikzpicture}[line width=1.1pt]

\draw[red] (0,2) -- (0,2.75);
\draw[green] (2.5,1.75) -- (2,2) -- (2,4.25) -- (1.5,4.5);
\draw (1,2) -- (1,2.75);
\draw[blue] (2.5,1.75) -- (3,2) -- (3,4.25);
\node[font=\small] at (0,1.7) {$|\Psi\rangle_{R_\text{in}}$};
\node[font=\small] at (1,1.7) {$|\psi\rangle_f$};
\node[font=\small] at (2.5,1.4) {$|\text{MAX}\rangle_{r,R_\text{out}}$};

\draw[fill=gray!30] (-0.25,2.75) rectangle (1.25,3.5);
\node at (0.5,3.125) {$U$};

\draw (0,3.5) -- (0,4.25);
\draw[blue] (1,3.5) -- (1,4.25) -- (1.5,4.5);
\node[font=\small] at (1.5,4.75) {$|r|\langle\text{MAX}|_{r,R'_\text{out}}$};

\end{tikzpicture}
}
    }
    },
\end{equation}
we see that $V_\text{BFP}$ maps a dynamically generated state of the effective description to a dynamically generated state of the fundamental description followed by a teleportation protocol acting on the radiation degrees of freedom. We can remove the teleportation by unbending the radiation lines, finding
\begin{equation}
    V_\text{BFP} |\hat{\Psi}_i(t)\rangle_{\ell rR_\text{out}} = |\Psi_i(t)\rangle_{BR_\text{out}}.
\end{equation}
$V_\text{BFP}$ indeed maps states in the dynamically generated subspace of the effective description to dynamically generated states of the fundamental description. Furthermore, inner products of these states are preserved. Taking an inner product of (\ref{eq:BFP_dyn_gen_eff}) with itself gives
\begin{equation} \label{eq:V_daggerV}
    \langle\hat{\Psi}_j(t)| V_\text{BFP}^\dagger V_\text{BFP} |\hat{\Psi}_i(t)\rangle = |r|^2 
    \vcenter{
    \hbox{
    \scalebox{0.85}{
\begin{tikzpicture}[line width=1.1]

\draw[red] (0,0) -- (0,0.75);
\node[font=\small] at (0,-0.3) {$|\Psi_i\rangle_{R_\text{in}} $};
\draw (1,0) -- (1,0.75);
\node[font=\small] at (1,-0.3) {$ |\psi\rangle_f $};

\draw[fill=gray!30] (-0.25,0.75) rectangle (1.25,1.5);
\node at (0.5,1.125) {$ U $};

\draw (0,1.5) -- (0,4.5);
\draw[blue] (1,1.5) -- (1,2.25) -- (1.5,2.5);
\draw[green] (1.5,2.5) -- (2,2.25) -- (2,0) -- (2.5,-0.25);
\draw[blue] (2.5,-0.25) -- (3,0) -- (3,6) -- (2.5,6.25);
\draw[green] (2.5,6.25) -- (2,6) -- (2,3.75) -- (1.5,3.5);
\draw[blue] (1.5,3.5) -- (1,3.75) -- (1,4.5);

\node[font=\small] at (1.5,2.75) {$\langle\text{MAX}|_{r,R'_\text{out}}$};
\node[font=\small] at (1.5,3.25) {$|\text{MAX}\rangle_{r,R'_\text{out}}$};
\node[font=\small] at (2.5,-0.55) {$|\text{MAX}\rangle_{r,R_\text{out}}$};
\node[font=\small] at (2.5,6.55) {$\langle\text{MAX}|_{r,R_\text{out}}$};

\draw[fill=gray!30] (-0.25,4.5) rectangle (1.25,5.25);
\node at (0.5,4.875) {$ U^\dagger $};
\draw[red] (0,5.25) -- (0,6);
\node[font=\small] at (0,6.3) {$ \langle\Psi_j|_{R_\text{in}} $};
\draw (1,5.25) -- (1,6);
\node[font=\small] at (1,6.3) {$ \langle\psi|_f $};

\end{tikzpicture}
}
    }
    }
\end{equation}
Straightening the green and blue lines adds a factor of $1/|r|^2$, which cancels with the factor of $|r|^2$ inserted to maintain normalization after post-selection. Thus, (\ref{eq:V_daggerV}) reduces to
\begin{equation}
    \big( \langle \Psi_j|_{R_\text{in}} \langle \psi|_f \big) U^\dagger U \big( |\Psi_i\rangle_{R_\text{in}} |\psi\rangle_f \big) = \delta_{ij}.
\end{equation}
Indeed, the BFP map acts isometrically on dynamically generated states in the effective description, as would be expected from a generic teleportation protocol. In fact, this result did not require any averaging over $U$ or $\hat{U}$ -- inner products of states given by (\ref{eq:dyn_gen_eff}) are preserved exactly.

\begin{figure} 
    \centering

\scalebox{0.8}{
\begin{tikzpicture}[line width=1.1pt]


\node[scale=1.5] at (-1.8,1.25) {$V_\text{BFP} \hat{U} = $};

\draw[red] (0,-1.75) -- (0,-1);
\draw[green] (2.5,-2) -- (2,-1.75) -- (2,-1);
\draw[blue] (2.5,-2) -- (3,-1.75) -- (3,4.25);
\node[font=\small] at (2.5,-2.35) {$|\text{MAX}\rangle_{r,R_\text{out}}$};

\draw[brown,fill=brown!20] (-0.25, -1) rectangle (2.25,-0.25);
\node[brown] at (1,-0.625) {$\hat{U}$};

\draw[orange] (0,-0.25) -- (0,0.5);
\draw[green] (2,-0.25) -- (2,0.5);

\draw[brown,fill=brown!20] (-0.25,0.5) rectangle (2.25,1.25);
\node[brown] at (1,0.875) {$\hat{U}^\dagger$};

\draw[red] (0,1.25) -- (0,2.75);
\draw[green] (2,1.25) -- (2,4.25) -- (1.5,4.5);
\draw (1,2.25) -- (1,2.75);
\node[font=\small] at (1,1.95) {$|\psi\rangle_f$};

\draw[fill=gray!30] (-0.25,2.75) rectangle (1.25,3.5);
\node at (0.5,3.125) {$U$};

\draw (0,3.5) -- (0,4.25);
\draw[blue] (1,3.5) -- (1,4.25) -- (1.5,4.5);
\node[font=\small] at (1.5,4.75) {$|r|\langle\text{MAX}|_{r,R'_\text{out}}$};


\draw[->, line width=1.3mm] (4.7,1.25) -- (5.8,1.25);

\begin{scope}[shift={(7.5,-1.9)}]
\draw[red] (0,2) -- (0,2.75);
\draw[green] (2.5,1.75) -- (2,2) -- (2,4.25) -- (1.5,4.5);
\draw (1,2) -- (1,2.75);
\draw[blue] (2.5,1.75) -- (3,2) -- (3,4.25);
\node[font=\small] at (1,1.7) {$|\psi\rangle_f$};
\node[font=\small] at (2.5,1.4) {$|\text{MAX}\rangle_{r,R_\text{out}}$};

\draw[fill=gray!30] (-0.25,2.75) rectangle (1.25,3.5);
\node at (0.5,3.125) {$U$};

\draw (0,3.5) -- (0,4.25);
\draw[blue] (1,3.5) -- (1,4.25) -- (1.5,4.5);
\node[font=\small] at (1.5,4.75) {$|r|\langle\text{MAX}|_{r,R'_\text{out}}$};
\end{scope}]


\draw[->, line width=1.2mm] (12.2,1.25) -- (13.3,1.25);

\begin{scope}[shift={(15,-1.9)}]
\draw[red] (0,2) -- (0,2.75);
\draw (1,2) -- (1,2.75);
\node[font=\small] at (1,1.7) {$|\psi\rangle_f$};

\draw[fill=gray!30] (-0.25,2.75) rectangle (1.25,3.5);
\node at (0.5,3.125) {$U$};

\draw (0,3.5) -- (0,4.25);
\draw[blue] (1,3.5) -- (1,4.25);
\end{scope}]

\end{tikzpicture}
}
    \caption{Demonstrating equivariance is satisfied by $V_\text{BFP}$. The left figure shows time evolution in the effective description composed with the BFP map.}
    \label{fig:equivariance}
\end{figure}
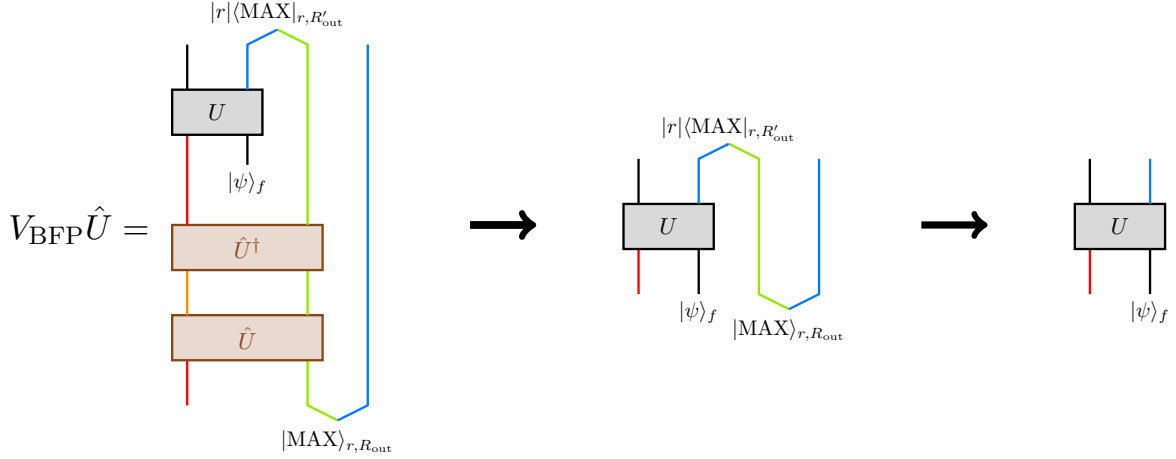

We now turn to the action of the adjoint of the BFP map,
\begin{equation}
    V_\text{BFP}^\dagger = |r| \hat{U} \langle\psi|_f U^\dagger |\text{MAX}\rangle_{r,R'_\text{out}} 
\end{equation}
on dynamically generated states of the fundamental description (\ref{eq:dyn_gen_fun}). The adjoint of the BFP map differs from the adjoint of the backwards-forwards map in an important way: instead of using radiation $R_\text{out}$ from the dynamically generated fundamental state in the backwards evolution of the fundamental description, $V_\text{BFP}^\dagger$ applies $U^\dagger$ to the maximally entangled radiation $R'_\text{out}$ inserted at the beginning of the map. The remainder of the adjoint map works similarly, with post-selection on the backwards evolved fixed states $\langle\psi|_f$. Acting on (\ref{eq:dyn_gen_fun}), the adjoint gives
\begin{equation} \label{eq:BFP_dagger_dyn_gen_fun}
    V_\text{BFP}^\dagger |\Psi_i(t)\rangle_{BR_\text{out}} = \lp |r| \hat{U} \langle\psi|_f U^\dagger |\text{MAX}\rangle_{r,R'_\text{out}} \rp U |\Psi_i\rangle_{R_\text{in}} |\psi\rangle_f = |r| 
        \vcenter{
        \hbox{
        \scalebox{0.85}{
\begin{tikzpicture}[line width=1.1pt]

\begin{scope}[yscale=-1]

\draw[orange] (0,-0.25) -- (0,0.5);
\draw[green] (2,-0.25) -- (2,0.5);
\draw[blue] (3,-0.25) -- (3,5);

\draw[brown,fill=brown!20] (-0.25,0.5) rectangle (2.25,1.25);
\node[brown] at (1,0.875) {$\hat{U}$};

\draw[red] (0,1.25) -- (0,2.75);
\draw[green] (2,1.25) -- (2,4.25) -- (1.5,4.5);
\draw (1,2.25) -- (1,2.75);
\node[font=\small] at (1,1.95) {$\langle\psi|_f$};

\draw[fill=gray!30] (-0.25,2.75) rectangle (1.25,3.5);
\node at (0.5,3.125) {$U^\dagger$};

\draw (0,3.5) -- (0,5);
\draw[blue] (1,3.5) -- (1,4.25) -- (1.5,4.5);
\node[scale=0.7] at (1.5,4.75) {$|\text{MAX}\rangle_{r,R'_\text{out}}$};

\draw[fill=gray!30] (-0.25,5) rectangle (3.25,5.75);
\node at (1.5,5.375) {$U$};

\draw[red] (0,5.75) -- (0,6.5);
\draw (3,5.75) -- (3,6.5);

\node[font=\small] at (0,6.85) {$|\Psi_i\rangle_{R_\text{in}}$};
\node[font=\small] at (3,6.85) {$|\psi\rangle_f$};
    
\end{scope}

\end{tikzpicture}
}
        }
        } 
\end{equation}
We see that $r$ will not be in the maximally entangled state with $R_\text{out}$ before the application of $\hat{U}$ (as is required of a dynamically generated state) because it was inserted in the maximally entangled state with $R'_\text{out}$ instead. Therefore, $V_\text{BFP}^\dagger$ does not produce a state with support only in the dynamically generated subspace,
\begin{equation}
    V_\text{BFP}^\dagger |\Psi_i(t)\rangle_{BR_\text{out}} \neq |\hat{\Psi}_i(t)\rangle_{\ell rR_\text{out}}.
\end{equation}
Instead, $V_\text{BFP}^\dagger$ maps dynamically generated states of the fundamental description to generic states of the effective Hilbert space. Furthermore, $V_\text{BFP}$ does not even preserve the inner product of dynamically generated fundamental states. Taking the inner product of (\ref{eq:BFP_dagger_dyn_gen_fun}) with itself and averaging over $U$ using (\ref{eq:dU2}), we find
\begin{equation}
    \int dU\, \langle\Psi_j(t)| V_\text{BFP} V_\text{BFP}^\dagger |\Psi_i(t)\rangle = \frac{|r|^2}{|B|^2|r|^2-1} \lp |B|\cdot|\ell|\cdot|r| + |B|^2 - 1 - \frac{|B|\cdot|\ell|}{|r|} \rp \delta_{ij},
\end{equation}
which generically deviates from unit norm. $V_\text{BFP}^\dagger$ does not act isometrically on dynamically generated states of the fundamental description, and therefore $V_\text{BFP}$ cannot act as a unitary on the dynamically generated subspaces of both descriptions. This emphasizes the fact that the BFP map does not represent time dynamics in either description. As before, we can summarize these results as operator equations:
\begin{align}
    & \hat{P}_t V_\text{BFP}^\dagger V_\text{BFP} \hat{P}_t = \hat{P}_t \\
    \int dU \, & P_t V_\text{BFP} V_\text{BFP}^\dagger P_t = \frac{|r|^2}{|B|^2|r|^2-1} \lp |B|\cdot|\ell|\cdot|r| + |B|^2 - 1 - \frac{|B|\cdot|\ell|}{|r|} \rp P_t
\end{align}
While the result of $V_\text{BFP}^\dagger$ is not a dynamically generated state, it does have non-zero overlap with the dynamically generated subspace. We could force the result of $V_\text{BFP}^\dagger$ into the dynamically generated subspace of the effective description by applying the projector (\ref{eq:proj_dyn_eff}),
\begin{equation}
    \hat{P}_t V_\text{BFP}^\dagger |\Psi_i(t)\rangle_{BR_\text{out}} = |\hat{\Psi}_i(t)\rangle_{\ell rR_\text{out}}.
\end{equation}
In fact, adding the projector restores preservation of the inner product without the need for averaging,
\begin{equation}
    \langle\Psi_j(t)|V_\text{BFP} \hat{P}_t V_\text{BFP}^\dagger |\Psi_i(t)\rangle = \delta_{ij},
\end{equation}
where we used $\hat{P}_t^\dagger = \hat{P}_t^2 = \hat{P}_t$. Therefore $\hat{P}_t V_\text{BFP}^\dagger$ acts isometrically, whereas $V_\text{BFP}^\dagger$ alone does not,
\begin{equation}
    P_t V_\text{BFP} \hat{P}_t V_\text{BFP}^\dagger P_t = P_t.
\end{equation}
The BFP map is also equivariant, although demonstrating that it satisfies (\ref{eq:equivariance}) is not as straightforward as it was for the backwards-forwards map. Fig.~\ref{fig:equivariance} demonstrates that composing effective dynamics with $V_\text{BFP}$ indeed gives fundamental dynamics, proving equivariance.\footnote{This result is essentially the same as the result of (\ref{eq:BFP_dyn_gen_eff}), but we repeat it here for clarity.} Our work in \cite{dewolfe_non-isometric_2023} went further to show that equivariance still holds for $V_\text{BFP}$ in the presence of exterior interactions between $R_\text{in}$ and $R_\text{out}$, even though these exterior interactions drop out of the holographic map.

\subsection{Relationship between BF and BFP} \label{sec:relationship}
Before turning to the derivation of the Page curve, we compare the two maps $V_\text{BF}$ and $V_\text{BFP}$ on dynamically generated states. One of the most important results in our previous work \cite{dewolfe_non-isometric_2023} demonstrated the equivalence of $V_\text{BF}$ and $V_\text{BFP}$ on dynamically generated states of the effective description. The transformations relating the two are reproduced in Fig.~\ref{fig:transformations} below, and we can represent the result as an operator equation:
\begin{equation} \label{eq:relate_Vs}
    V_\text{BF} \hat{P}_t = V_\text{BFP} \hat{P}_t.
\end{equation}
We note that the second transformation in Fig.~\ref{fig:transformations} -- replacing the projector $|\text{MAX}\rangle\langle\text{MAX}|_{r,R_\text{out}}$ with the identity $\id$ -- is only possible because we are restricting to the dynamically generated subspace. This restriction guarantees that the Hawking pairs $rR_\text{out}$ will be returned to the maximally entangled state by the backwards effective dynamics $\hat{U}^\dagger$, on which the projector acts trivially.

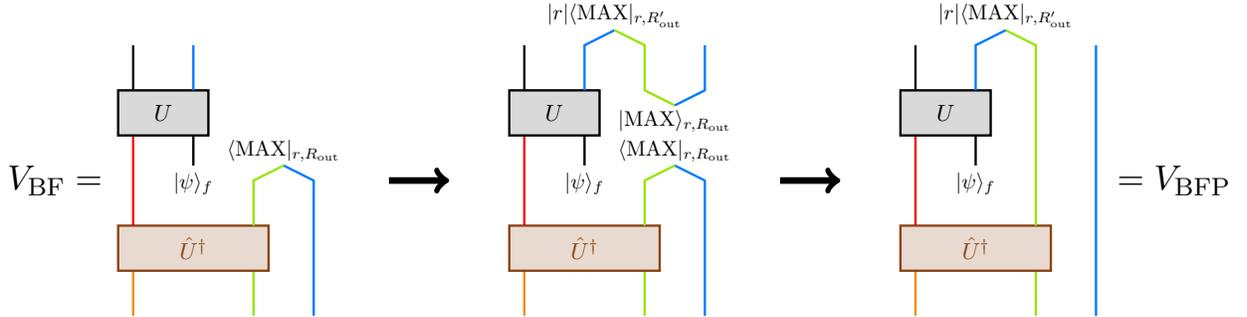
\begin{figure}[!ht]
    \centering

\scalebox{0.8}{
\begin{tikzpicture}[line width=1.1pt]

\node[scale=1.5] at (-1.3,2) {$V_\text{BF} =$};

\draw[orange] (0,-0.25) -- (0,0.5);
\draw[green] (2,-0.25) -- (2,0.5);
\draw[blue] (3,-0.25) -- (3,2) -- (2.5,2.25);

\draw[brown,fill=brown!20] (-0.25,0.5) rectangle (2.25,1.25);
\node[brown] at (1,0.875) {$\hat{U}^\dagger$};

\draw[red] (0,1.25) -- (0,2.75);
\draw[green] (2,1.25) -- (2,2) -- (2.5,2.25);
\node[font=\small] at (2.5,2.5) {$\langle\text{MAX}|_{r,R_\text{out}}$};
\draw (1,2.25) -- (1,2.75);
\node[font=\small] at (1,1.95) {$|\psi\rangle_f$};

\draw[fill=gray!30] (-0.25,2.75) rectangle (1.25,3.5);
\node at (0.5,3.125) {$U$};

\draw (0,3.5) -- (0,4.25);
\draw[blue] (1,3.5) -- (1,4.25);

\draw[->, line width=1.1mm] (4.25,2) -- (5.25,2);


\begin{scope}[shift={(6.5,0)}]

\draw[orange] (0,-0.25) -- (0,0.5);
\draw[green] (2,-0.25) -- (2,0.5);
\draw[blue] (3,-0.25) -- (3,2) -- (2.5,2.25);

\draw[brown,fill=brown!20] (-0.25,0.5) rectangle (2.25,1.25);
\node[brown] at (1,0.875) {$\hat{U}^\dagger$};

\draw[red] (0,1.25) -- (0,2.75);
\draw[green] (2,1.25) -- (2,2) -- (2.5,2.25);
\node[font=\small] at (2.5,2.5) {$\langle\text{MAX}|_{r,R_\text{out}}$};
\draw (1,2.25) -- (1,2.75);
\node[font=\small] at (1,1.95) {$|\psi\rangle_f$};

\draw[fill=gray!30] (-0.25,2.75) rectangle (1.25,3.5);
\node at (0.5,3.125) {$U$};

\draw (0,3.5) -- (0,4.25);
\draw[blue] (1,3.5) -- (1,4.25) -- (1.5,4.5);
\draw[green] (1.5,4.5) -- (2,4.25) -- (2,3.5) -- (2.5,3.25);
\draw[blue] (2.5,3.25) -- (3,3.5) -- (3,4.25);

\node[font=\small] at (1.5,4.75) {$|r|\langle\text{MAX}|_{r,R'_\text{out}}$};
\node[font=\small] at (2.5,3) {$|\text{MAX}\rangle_{r,R_\text{out}}$};

\end{scope}

\draw[->, line width=1.1mm] (10.75,2) -- (11.75,2);


\begin{scope}[shift={(13,0)}]

\draw[orange] (0,-0.25) -- (0,0.5);
\draw[green] (2,-0.25) -- (2,0.5);
\draw[blue] (3,-0.25) -- (3,4.25);

\draw[brown,fill=brown!20] (-0.25,0.5) rectangle (2.25,1.25);
\node[brown] at (1,0.875) {$\hat{U}^\dagger$};

\draw[red] (0,1.25) -- (0,2.75);
\draw[green] (2,1.25) -- (2,4.25) -- (1.5,4.5);
\draw (1,2.25) -- (1,2.75);
\node[font=\small] at (1,1.95) {$|\psi\rangle_f$};

\draw[fill=gray!30] (-0.25,2.75) rectangle (1.25,3.5);
\node at (0.5,3.125) {$U$};

\draw (0,3.5) -- (0,4.25);
\draw[blue] (1,3.5) -- (1,4.25) -- (1.5,4.5);
\node[font=\small] at (1.5,4.75) {$|r|\langle\text{MAX}|_{r,R'_\text{out}}$};

\node[scale=1.5] at (4.3,2) {$= V_\text{BFP}$};
    
\end{scope}

\end{tikzpicture}
}
    \caption{The transformations used in \cite{dewolfe_non-isometric_2023} to equate $V_\text{BF}$ and $V_\text{BFP}$ on the dynamically generated subspace of the effective description.}
    \label{fig:transformations}
\end{figure}

We interpret the inverse transformations (going from right to left in Fig.~\ref{fig:transformations}) as the analog of the ``straightening procedure'' of \cite{akers_black_2022} that relates PHEVA's dynamical holographic map to the fundamental dynamics. When interior interactions in the effective description were included by \cite{kim_complementarity_2023} without adjusting the map, this straightening procedure led to partially transposed unitaries in the fundamental dynamics that spoiled the unitarity of black hole evaporation. By including interior interactions in our maps directly, we have demonstrated that the isometric BFP map can be related to unitary backwards-forwards map in a way that does not lead to partially transposed unitaries. The unitarity of black hole evaporation is preserved. 

\begin{figure} 
    \centering

\scalebox{0.8}{

\begin{tikzpicture}[line width=1.1pt]

\begin{scope}[yscale=-1]

\node[scale=1.5] at (-1.3,2) {$V_\text{BF}^\dagger =$};

\draw[orange] (0,-0.25) -- (0,0.5);
\draw[green] (2,-0.25) -- (2,0.5);
\draw[blue] (3,-0.25) -- (3,2) -- (2.5,2.25);

\draw[brown,fill=brown!20] (-0.25,0.5) rectangle (2.25,1.25);
\node[brown] at (1,0.875) {$\hat{U}$};

\draw[red] (0,1.25) -- (0,2.75);
\draw[green] (2,1.25) -- (2,2) -- (2.5,2.25);
\node[font=\small] at (2.5,2.5) {$|\text{MAX}\rangle_{r,R_\text{out}}$};
\draw (1,2.25) -- (1,2.75);
\node[font=\small] at (1,1.95) {$\langle\psi|_f$};

\draw[fill=gray!30] (-0.25,2.75) rectangle (1.25,3.5);
\node at (0.5,3.125) {$U^\dagger$};

\draw (0,3.5) -- (0,4.25);
\draw[blue] (1,3.5) -- (1,4.25);

\draw[->, line width=1.1mm] (4.25,2) -- (5.25,2);


\begin{scope}[shift={(6.5,0)}]

\draw[orange] (0,-0.25) -- (0,0.5);
\draw[green] (2,-0.25) -- (2,0.5);
\draw[blue] (3,-0.25) -- (3,2) -- (2.5,2.25);

\draw[brown,fill=brown!20] (-0.25,0.5) rectangle (2.25,1.25);
\node[brown] at (1,0.875) {$\hat{U}$};

\draw[red] (0,1.25) -- (0,2.75);
\draw[green] (2,1.25) -- (2,2) -- (2.5,2.25);
\node[font=\small] at (2.5,2.5) {$|\text{MAX}\rangle_{r,R_\text{out}}$};
\draw (1,2.25) -- (1,2.75);
\node[font=\small] at (1,1.95) {$\langle\psi|_f$};

\draw[fill=gray!30] (-0.25,2.75) rectangle (1.25,3.5);
\node at (0.5,3.125) {$U^\dagger$};

\draw (0,3.5) -- (0,4.25);
\draw[blue] (1,3.5) -- (1,4.25) -- (1.5,4.5);
\draw[green] (1.5,4.5) -- (2,4.25) -- (2,3.5) -- (2.5,3.25);
\draw[blue] (2.5,3.25) -- (3,3.5) -- (3,4.25);

\node[font=\small] at (1.5,4.75) {$|r||\text{MAX}\rangle_{r,R'_\text{out}}$};
\node[font=\small] at (2.5,3) {$\langle\text{MAX}|_{r,R_\text{out}}$};

\end{scope}

\draw[->, line width=1.1mm] (10.75,2) -- (11.75,2);
\draw[red, line width=3] (10.8,2.5) -- (11.7,1.5);
\draw[red, line width=3] (10.8,1.5) -- (11.7,2.5);


\begin{scope}[shift={(13,0)}]

\draw[orange] (0,-0.25) -- (0,0.5);
\draw[green] (2,-0.25) -- (2,0.5);
\draw[blue] (3,-0.25) -- (3,4.25);

\draw[brown,fill=brown!20] (-0.25,0.5) rectangle (2.25,1.25);
\node[brown] at (1,0.875) {$\hat{U}$};

\draw[red] (0,1.25) -- (0,2.75);
\draw[green] (2,1.25) -- (2,4.25) -- (1.5,4.5);
\draw (1,2.25) -- (1,2.75);
\node[font=\small] at (1,1.95) {$\langle\psi|_f$};

\draw[fill=gray!30] (-0.25,2.75) rectangle (1.25,3.5);
\node at (0.5,3.125) {$U^\dagger$};

\draw (0,3.5) -- (0,4.25);
\draw[blue] (1,3.5) -- (1,4.25) -- (1.5,4.5);
\node[font=\small] at (1.5,4.75) {$|r||\text{MAX}\rangle_{r,R'_\text{out}}$};

\node[scale=1.5] at (4.3,2) {$= V_\text{BFP}^\dagger$};
    
\end{scope}

\end{scope}

\end{tikzpicture}

}
    \caption{The transformations of Fig.~\ref{fig:transformations} cannot relate the adjoints of the backwards-forwards map (left) and the BFP map (right) since fundamental dynamics do not guarantee that the Hawking pairs $rR_\text{out}$ will be in the maximally entangled state.}
    \label{fig:adjoint_transform}
\end{figure}
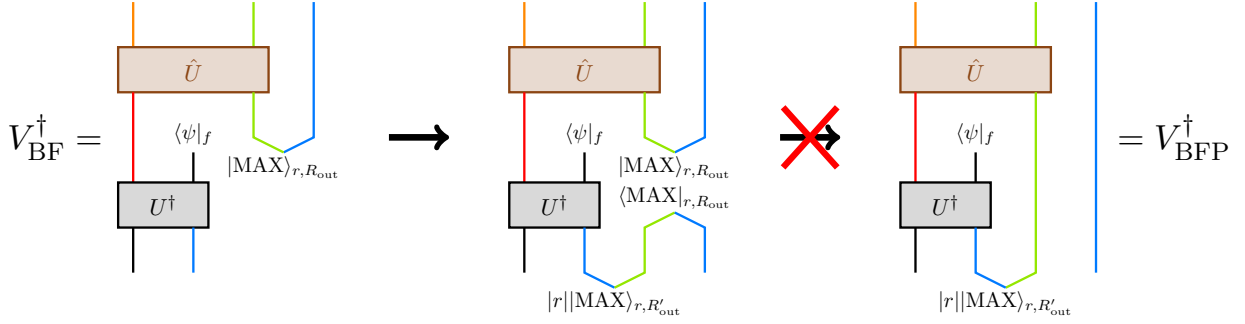

However, it is not possible to relate the adjoints $V_\text{BF}^\dagger$ and $V_\text{BFP}^\dagger$ with similar transformations,
\begin{equation}
    V_\text{BF}^\dagger P_t \neq V_\text{BFP}^\dagger P_t,
\end{equation}
since (\ref{eq:BFP_dagger_dyn_gen_fun}) demonstrated that $V_\text{BFP}^\dagger$ maps fundamental dynamically generated states outside of the dynamically generated subspace of the effective description. This is further demonstrated by Fig.~\ref{fig:adjoint_transform}; the projector onto the maximally entangled state cannot be replaced with the identity since fundamental dynamics do not guarantee that the Hawking pairs $rR_\text{out}$ will be in the maximally entangled state. 

Instead, we found that it is $\hat{P}_t V_\text{BFP}^\dagger$ that acts isometrically on the dynamically generated subspace of the fundamental description. Therefore we expect
\begin{equation}
    \hat{P}_t V_\text{BF}^\dagger P_t = \hat{P}_t V_\text{BFP}^\dagger P_t
\end{equation}
to be true. Indeed, projecting (\ref{eq:relate_Vs}) onto the dynamically generated subspace of the fundamental description and taking the overall adjoint gives exactly this relation between the adjoints of the two maps.

\subsection{Page curve} \label{sec:page}
The work of \cite{akers_black_2022} goes further to derive the QES formula for the entropy of the radiation with their dynamical map, obtaining the Page curve. We show here that both the BF and BFP maps also reproduce the same QES formula and Page curve.

First, we note that the effective dynamics we presented in \cite{dewolfe_non-isometric_2023} don't involve a mechanism that allows information to escape the black hole. Because the horizon causally separates the interior and exterior of the black hole, we don't include any unitaries $\hat{U}_t$ in the effective description that connects $\ell r$ modes with $R_\text{out}$. Thus each $\hat{U}_t$ only keeps information behind the horizon. A naive calculation of the radiation's entropy shows that it can only increase, giving the Hawking curve.

Holographic maps provide a proxy for these missing dynamics to calculate the correct entropy of the radiation. Applying either holographic map to the effective dynamics, we map to the fundamental description where the calculation of the radiation's entropy does give the Page curve. Calculating the fine-grained entropy $S_2(\hat{\Psi}_{R_\text{out}})$ of the radiation after application of the $V_\text{BF}$ or $V_\text{BFP}$ requires an average over $U$. We will perform this integration using the spin network method described in Appendix~\ref{app:int}. Fig.~\ref{fig:Page_curve} shows the leading two contributions to the average over $U$ of $e^{-S_2(\hat{\Psi}_{R_\text{out}})}$ using $V_\text{BFP}$ for $n_0=4$ and $m_0=2$ up until $t=2$. The result for $V_\text{BF}$ is the same and can be found by straightening the bent radiation lines. Generalizing to arbitrary $n_0$, $m_0$, and $t$, these evaluate to
\begin{equation}
    q^{-2t} \qquad \text{and} \qquad q^{-(n_0+t)} e^{-S_2(\hat{\Psi}_{R_\text{in}})} 
\end{equation}
respectively, where we've used the large $q$ approximation. These terms will dominate at different times; taking a logarithm, we find the same Page curve as in  \cite{akers_black_2022},
\begin{equation} \label{eq:QES}
    S_2(\hat{\Psi}_{R_\text{out}}) = \min \Big[ 2t\log q, \, (n_0 - t) \log q + S_2(\Psi_{R_\text{in}}) \Big].
\end{equation}
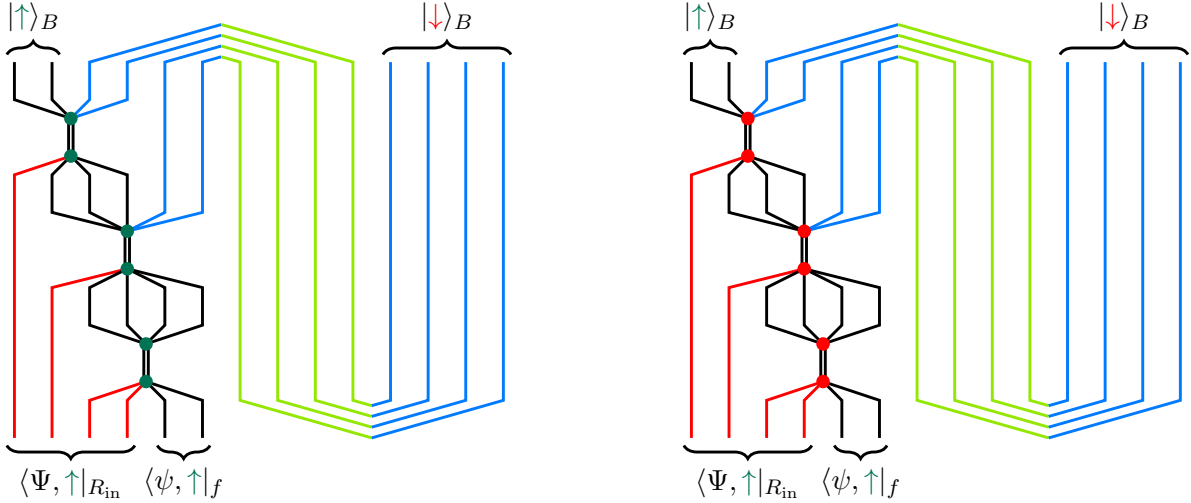
\begin{figure} 
    \centering
    \scalebox{1}{
\begin{tikzpicture}[line width=1.1pt]

\draw[decorate,decoration={brace,amplitude=5pt,mirror}] (-0.1,-0.1) -- (1.6,-0.1);
\node at (0.75,-0.6) {$\langle\Psi,\textcolor{darkgreen}{\uparrow}|_{R_\text{in}}$};

\draw[decorate,decoration={brace,amplitude=5pt,mirror}] (1.9,-0.1) -- (2.6,-0.1);
\node at (2.25,-0.6) {$\langle\psi,\textcolor{darkgreen}{\uparrow}|_f$};

\draw[red] (0,0) -- (0,3.5) -- (0.75,3.75);
\draw[red] (0.5,0) -- (0.5,2) -- (1.5,2.25);
\draw[red] (1,0) -- (1,0.5) -- (1.75,0.75);
\draw[red] (1.5,0) -- (1.5,0.5) -- (1.75,0.75);
\draw (2,0) -- (2,0.5) -- (1.75,0.75);
\draw (2.5,0) -- (2.5,0.5) -- (1.75,0.75);

\draw[double] (1.75,0.75) -- (1.75,1.25);

\draw (1.75,1.25) -- (1,1.5) -- (1,2) -- (1.5,2.25);
\draw (1.75,1.25) -- (1.5,1.5) -- (1.5,2.25);
\draw (1.75,1.25) -- (2,1.5) -- (2,2) -- (1.5,2.25);
\draw (1.75,1.25) -- (2.5,1.5) -- (2.5,2) -- (1.5,2.25);

\draw[double] (1.5,2.25) -- (1.5,2.75);

\draw (1.5,2.75) -- (0.5,3) -- (0.5,3.5) -- (0.75,3.75);
\draw (1.5,2.75) -- (1,3) -- (1,3.5) -- (0.75,3.75);
\draw (1.5,2.75) -- (1.5,3.5) -- (0.75,3.75);
\draw[blue] (1.5,2.75) -- (2,3) -- (2,5) -- (2.75,5.214);
\draw[blue] (1.5,2.75) -- (2.5,3) -- (2.5,5) -- (2.75,5.071);

\draw[double] (0.75,3.75) -- (0.75,4.25);

\draw (0.75,4.25) -- (0,4.5) -- (0,5);
\draw (0.75,4.25) -- (0.5,4.5) -- (0.5,5);
\draw[blue] (0.75,4.25) -- (1,4.5) -- (1,5) -- (2.75,5.5);
\draw[blue] (0.75,4.25) -- (1.5,4.5) -- (1.5,5) -- (2.75,5.357);

\draw[green] (2.75,5.071) -- (3,5) -- (3,0.5) -- (4.75,0);
\draw[green] (2.75,5.214) -- (3.5,5) -- (3.5,0.5) -- (4.75,0.143);
\draw[green] (2.75,5.357) -- (4,5) -- (4,0.5) -- (4.75,0.286);
\draw[green] (2.75,5.5) -- (4.5,5) -- (4.5,0.5) -- (4.75,0.429);

\draw[blue] (4.75,0.429) -- (5,0.5) -- (5,5);
\draw[blue] (4.75,0.286) -- (5.5,0.5) -- (5.5,5);
\draw[blue] (4.75,0.143) -- (6,0.5) -- (6,5);
\draw[blue] (4.75,0) -- (6.5,0.5) -- (6.5,5);

\draw[decorate,decoration={brace,amplitude=5pt}] (-0.1,5.1) -- (0.6,5.1);
\node at (0.25,5.6) {$|\textcolor{darkgreen}{\uparrow}\rangle_B$};

\draw[decorate,decoration={brace,amplitude=5pt}] (4.9,5.1) -- (6.6,5.1);
\node at (5.75,5.6) {$|\textcolor{red}{\downarrow}\rangle_B$};


\filldraw[darkgreen] (1.75,0.75) circle (2pt);
\filldraw[darkgreen] (1.75,1.25) circle (2pt);

\filldraw[darkgreen] (1.5,2.25) circle (2pt);
\filldraw[darkgreen] (1.5,2.75) circle (2pt);

\filldraw[darkgreen] (0.75,3.75) circle (2pt);
\filldraw[darkgreen] (0.75,4.25) circle (2pt);


\begin{scope}[shift={(9,0)}]

\draw[decorate,decoration={brace,amplitude=5pt,mirror}] (-0.1,-0.1) -- (1.6,-0.1);
\node at (0.75,-0.6) {$\langle\Psi,\textcolor{darkgreen}{\uparrow}|_{R_\text{in}}$};

\draw[decorate,decoration={brace,amplitude=5pt,mirror}] (1.9,-0.1) -- (2.6,-0.1);
\node at (2.25,-0.6) {$\langle\psi,\textcolor{darkgreen}{\uparrow}|_f$};

\draw[red] (0,0) -- (0,3.5) -- (0.75,3.75);
\draw[red] (0.5,0) -- (0.5,2) -- (1.5,2.25);
\draw[red] (1,0) -- (1,0.5) -- (1.75,0.75);
\draw[red] (1.5,0) -- (1.5,0.5) -- (1.75,0.75);
\draw (2,0) -- (2,0.5) -- (1.75,0.75);
\draw (2.5,0) -- (2.5,0.5) -- (1.75,0.75);

\draw[double] (1.75,0.75) -- (1.75,1.25);

\draw (1.75,1.25) -- (1,1.5) -- (1,2) -- (1.5,2.25);
\draw (1.75,1.25) -- (1.5,1.5) -- (1.5,2.25);
\draw (1.75,1.25) -- (2,1.5) -- (2,2) -- (1.5,2.25);
\draw (1.75,1.25) -- (2.5,1.5) -- (2.5,2) -- (1.5,2.25);

\draw[double] (1.5,2.25) -- (1.5,2.75);

\draw (1.5,2.75) -- (0.5,3) -- (0.5,3.5) -- (0.75,3.75);
\draw (1.5,2.75) -- (1,3) -- (1,3.5) -- (0.75,3.75);
\draw (1.5,2.75) -- (1.5,3.5) -- (0.75,3.75);
\draw[blue] (1.5,2.75) -- (2,3) -- (2,5) -- (2.75,5.214);
\draw[blue] (1.5,2.75) -- (2.5,3) -- (2.5,5) -- (2.75,5.071);

\draw[double] (0.75,3.75) -- (0.75,4.25);

\draw (0.75,4.25) -- (0,4.5) -- (0,5);
\draw (0.75,4.25) -- (0.5,4.5) -- (0.5,5);
\draw[blue] (0.75,4.25) -- (1,4.5) -- (1,5) -- (2.75,5.5);
\draw[blue] (0.75,4.25) -- (1.5,4.5) -- (1.5,5) -- (2.75,5.357);

\draw[green] (2.75,5.071) -- (3,5) -- (3,0.5) -- (4.75,0);
\draw[green] (2.75,5.214) -- (3.5,5) -- (3.5,0.5) -- (4.75,0.143);
\draw[green] (2.75,5.357) -- (4,5) -- (4,0.5) -- (4.75,0.286);
\draw[green] (2.75,5.5) -- (4.5,5) -- (4.5,0.5) -- (4.75,0.429);

\draw[blue] (4.75,0.429) -- (5,0.5) -- (5,5);
\draw[blue] (4.75,0.286) -- (5.5,0.5) -- (5.5,5);
\draw[blue] (4.75,0.143) -- (6,0.5) -- (6,5);
\draw[blue] (4.75,0) -- (6.5,0.5) -- (6.5,5);

\draw[decorate,decoration={brace,amplitude=5pt}] (-0.1,5.1) -- (0.6,5.1);
\node at (0.25,5.6) {$|\textcolor{darkgreen}{\uparrow}\rangle_B$};

\draw[decorate,decoration={brace,amplitude=5pt}] (4.9,5.1) -- (6.6,5.1);
\node at (5.75,5.6) {$|\textcolor{red}{\downarrow}\rangle_B$};


\filldraw[red] (1.75,0.75) circle (2pt);
\filldraw[red] (1.75,1.25) circle (2pt);

\filldraw[red] (1.5,2.25) circle (2pt);
\filldraw[red] (1.5,2.75) circle (2pt);

\filldraw[red] (0.75,3.75) circle (2pt);
\filldraw[red] (0.75,4.25) circle (2pt);
    
\end{scope}
    
\end{tikzpicture}
}
    \caption{The two dominant contributions in the calculation of the fine-grained entropy of $R_\text{out}$ using the BFP holographic map. \textcolor{darkgreen}{Green} dots indicate the insertion of a $\textcolor{darkgreen}{\uparrow}$ state; \textcolor{red}{red} dots indicated the insertion of a $\textcolor{red}{\downarrow}$ state.}
    \label{fig:Page_curve}
\end{figure}
Assuming for the moment that $\Psi_{R_\text{in}}$ is pure, the first term in (\ref{eq:QES}) dominates for $|R_\text{out}| < |B|$ and the second dominates for $|R_\text{out}| > |B|$. The dominance of the terms will switch when $|R_\text{out}| \sim |B|$, and we define this to be the Page time,
\begin{equation} \label{eq:page_time}
    t_\text{Page} \sim \frac{n_0}{3}.
\end{equation}
Using a mixed state for $\Psi_{R_\text{in}}$ will only push the Page time later. If we also consider a more general set of effective dynamics where $\mathcal{I}$ qudits fall in and $\mathcal{O}$ qudits are radiated at each time step,\ the Page time is given by\footnote{In \cite{akers_black_2022}, the Page time was identified as the time when $|\ell|\cdot|r| \sim |B|$; at this time, the holographic map transitions from an approximate isometry to non-isometric. For general $\mathcal{I}$ and $\mathcal{O}$, this time is $t \sim (n_0 - m_0)/(2\mathcal{O})$. Since both Page times are linear in $n_0$, we take their difference to be insignificant.}
\begin{equation}
    t_\text{Page} \sim \frac{n_0 + S_2(\Psi_{R_\text{in}})}{2\mathcal{O} - \mathcal{I}}.
\end{equation}
Choosing $\mathcal{I}=1$ and $\mathcal{O}=2$ gives the dynamics shown in Fig.~\ref{fig:fun_eff_dyn} and reproduces (\ref{eq:page_time}) above. Had we instead chosen $\mathcal{I}=0$ and $\mathcal{O}=1$ (letting one qudit escape as radiation per time step and none fall in) we would've found the usual $t_\text{Page} \sim n_0/2$.

\section{Action on generic states} \label{sec:generic}

Now that we have reviewed the successes of the backwards-forwards and BFP maps on the dynamically generated subspace, we turn to generic states of the effective Hilbert space. The first dynamical holographic map proposed by PHEVA was shown to satisfy two important criteria demonstrating its effectiveness on generic states: (1) it acts isometrically on average with exponentially small fluctuations, and  (2) it provides a good state-dependent reconstruction of unitary operators in the effective description on $BR_\text{out}$ in the fundamental description. We show here that the BFP map also satisfies these criteria, continuing to demonstrate that it is a proper generalization of PHEVA's work to include non-trivial interior interactions, while the BF map does not.

For the remainder of this work, we drop the assumption that black hole states are in the dynamically generated subspaces of either description. In particular, we will not assume $r$ and $R_\text{out}$ will return to a maximally entangled state after backwards evolution in the effective description. Similarly, we will not assume that a state in the fundamental description was obtained by fundamental dynamics applied to a particular fixed state on $f$. Any state on the full effective or fundamental Hilbert spaces can be used as an input to the backwards-forwards or BFP maps in what follows.

\subsection{Failure of the BF map} \label{sec:BF_fail}
We begin by showing that the backwards-forwards map does not even satisfy the first criterion for acting on generic states: it does not act isometrically on average. Performing an average over $U$ and $\hat{U}$ of $V_\text{BF}^\dagger V_\text{BF}$ acting on generic states, we find
\begin{equation}
    \int d\hat{U} dU \, \langle\hat{\Psi}_j|_{\ell rR_\text{out}} V_\text{BF}^\dagger V_\text{BF} |\hat{\Psi}_i\rangle_{\ell rR_\text{out}} = \int d\hat{U} dU \, 
    \vcenter{
    \hbox{
    \scalebox{0.7}{
\begin{tikzpicture}[line width=1.1pt]

\node[scale=1.4] at (1.5,-0.75) {$|\hat{\Psi}_i\rangle_{\ell r R_\text{out}}$};

\draw[orange] (0,-0.25) -- (0,0.5);
\draw[green] (2,-0.25) -- (2,0.5);
\draw[blue] (3,-0.25) -- (3,2) -- (2.5,2.25);

\draw[brown,fill=brown!20] (-0.25,0.5) rectangle (2.25,1.25);
\node[font=\large,brown] at (1,0.875) {$\hat{U}^\dagger$};

\draw[red] (0,1.25) -- (0,2.75);
\draw[green] (2,1.25) -- (2,2) -- (2.5,2.25);
\node[font=\small] at (2.5,2.5) {$\langle\text{MAX}|$};
\draw (1,2.25) -- (1,2.75);
\node[font=\small] at (1,1.95) {$|\psi\rangle_f$};

\draw[fill=gray!30] (-0.25,2.75) rectangle (1.25,3.5);
\node[font=\large] at (0.5,3.125) {$U$};

\draw (0,3.5) -- (0,4.25);
\draw[blue] (1,3.5) -- (1,4.25);


\begin{scope}[yscale=-1,shift={(0,-7.75)}]

\node[scale=1.4] at (1.5,-0.75) {$\langle\hat{\Psi}_j|_{\ell r R_\text{out}}$};

\draw[orange] (0,-0.25) -- (0,0.5);
\draw[green] (2,-0.25) -- (2,0.5);
\draw[blue] (3,-0.25) -- (3,2) -- (2.5,2.25);

\draw[brown,fill=brown!20] (-0.25,0.5) rectangle (2.25,1.25);
\node[font=\large,brown] at (1,0.875) {$\hat{U}$};

\draw[red] (0,1.25) -- (0,2.75);
\draw[green] (2,1.25) -- (2,2) -- (2.5,2.25);
\node[font=\small] at (2.5,2.5) {$|\text{MAX}\rangle$};
\draw (1,2.25) -- (1,2.75);
\node[font=\small] at (1,1.95) {$\langle\psi|_f$};

\draw[fill=gray!30] (-0.25,2.75) rectangle (1.25,3.5);
\node[font=\large] at (0.5,3.125) {$U^\dagger$};
    
\end{scope}

\end{tikzpicture}
}
    }
    }
    = \int d\hat{U}
    \vcenter{
    \hbox{
    \scalebox{0.7}{
\begin{tikzpicture}[line width=1.1pt]

\node[scale=1.4] at (1,-0.75) {$|\hat{\Psi}_i\rangle_{\ell rR_\text{out}}$};

\draw[orange] (0,-0.25) -- (0,0.5);
\draw[green] (1,-0.25) -- (1,0.5);

\draw[brown,fill=brown!20] (-0.25,0.5) rectangle (1.25,1.25);
\node[font=\large,brown] at (0.5,0.875) {$\hat{U}^\dagger$};

\draw[red] (0,1.25) -- (0,4.25);
\draw[green] (1,1.25) -- (1,2) -- (1.5,2.25);

\node[font=\small] at (1.5,2.5) {$\langle\text{MAX}|$};

\draw[blue]  (1.5,2.25) -- (2,2) -- (2,-0.25);
\node[font=\small] at (1.5,3) {$|\text{MAX}\rangle$};
\draw[green] (1.5,3.25) -- (1,3.5) -- (1,4.25);
\draw[blue] (1.5,3.25) -- (2,3.5) -- (2,5.75);

\draw[brown,fill=brown!20] (-0.25,4.25) rectangle (1.25,5);
\node[font=\large,brown] at (0.5,4.625) {$\hat{U}$};

\draw[orange] (0,5) -- (0,5.75);
\draw[green] (1,5) -- (1,5.75);

\node[scale=1.4] at (1,6.25) {$\langle\hat{\Psi}_j|_{\ell rR_\text{out}}$};

\end{tikzpicture}
}
    }
    }
\end{equation}
where we used the unitarity of $U$ to remove it in the second equality without a need for averaging. Averaging over $\hat{U}$ using (\ref{eq:dU1}) gives a factor of $1/|r|$, and straightening the now connected radiation lines gives a second factor of $1/|r|$. Together, we find
\begin{equation}
    \int d\hat{U} \langle\hat{\Psi}_j|_{\ell r R_\text{out}} V_\text{BF}^\dagger V_\text{BF} |\hat{\Psi}_i\rangle_{\ell r R_\text{out}} = \frac{1}{|r|^2} \delta_{ij}.
\end{equation}
Therefore $V_\text{BF}$ does not preserve the inner product of generic effective description states on average. Instead, it tends to decrease the inner product, indicating that the backwards-forwards map is annihilating too many generic states. 

We can intuitively understand this failure of the backwards-forwards map as follows. Consider a state $|\Psi\rangle_{R_\text{in}}$ that collapses unitarily into a black hole and begins evaporating by the effective dynamics. A naive measurement of the interior will perturb the state in a way that cannot be undone by the backwards evolution in the backwards-forwards map. Therefore, the state that comes out of the black hole after backwards evolution could be in some state $|\Psi'\rangle_{R_\text{in}}$ not equal to the original. Instead, the original state $|\Psi\rangle$ would now be shared between the $R_\text{in}$ and radiation $r$ modes stored after the backwards evolution. If we were to follow the backwards-forwards map and post-select on $rR_\text{out}$ at this point, the information now stored in the radiation would be lost. In this way, the backwards-forwards map annihilates too many generic states.

\subsection{Deviation of BFP from isometry} \label{sec:deviation}
Following the failure of the backwards-forwards map on generic states, we test the BFP map against the two criteria introduced in \cite{akers_black_2022}. First, we check the action of $V_\text{BFP}^\dagger V_\text{BFP}$ on generic states of the effective Hilbert space,
\begin{equation} \label{eq:V_daggerV_BFP}
    \int d\hat{U} dU \, \langle\hat{\Psi}_j|_{\ell r} V_\text{BFP}^\dagger V_\text{BFP} |\hat{\Psi}_i\rangle_{\ell r} = \int d\hat{U} dU \, |r|^2 
    \vcenter{
    \hbox{
    \scalebox{0.7}{
\begin{tikzpicture}[line width=1.1pt]

\node[scale=1.4] at (1,-0.75) {$|\hat{\Psi}_i\rangle_{\ell r}$};

\draw[orange] (0,-0.25) -- (0,0.5);
\draw[green] (2,-0.25) -- (2,0.5);

\draw[brown,fill=brown!20] (-0.25,0.5) rectangle (2.25,1.25);
\node[font=\large,brown] at (1,0.875) {$\hat{U}^\dagger$};

\draw[red] (0,1.25) -- (0,2.75);
\draw[green] (2,1.25) -- (2,4.25) -- (1.5,4.5);
\draw (1,2.25) -- (1,2.75);
\node[font=\small] at (1,1.95) {$|\psi\rangle_f$};

\draw[fill=gray!30] (-0.25,2.75) rectangle (1.25,3.5);
\node[font=\large] at (0.5,3.125) {$U$};

\draw (0,3.5) -- (0,6.5);
\draw[blue] (1,3.5) -- (1,4.25) -- (1.5,4.5);
\node[font=\small] at (1.5,4.75) {$\langle\text{MAX}|$};


\begin{scope}[yscale=-1,shift={(0,-10)}]

\node[scale=1.4] at (1,-0.75) {$\langle\hat{\Psi}_j|_{\ell r}$};

\draw[orange] (0,-0.25) -- (0,0.5);
\draw[green] (2,-0.25) -- (2,0.5);

\draw[brown,fill=brown!20] (-0.25,0.5) rectangle (2.25,1.25);
\node[font=\large,brown] at (1,0.875) {$\hat{U}$};

\draw[red] (0,1.25) -- (0,2.75);
\draw[green] (2,1.25) -- (2,4.25) -- (1.5,4.5);
\draw (1,2.25) -- (1,2.75);
\node[font=\small] at (1,1.95) {$\langle\psi|_f$};

\draw[fill=gray!30] (-0.25,2.75) rectangle (1.25,3.5);
\node[font=\large] at (0.5,3.125) {$U^\dagger$};

\draw[blue] (1,3.5) -- (1,4.25) -- (1.5,4.5);
\node[font=\small] at (1.5,4.75) {$|\text{MAX}\rangle$};
    
\end{scope}

\end{tikzpicture}
}
    }
    }
    = \int d\hat{U} \, |r|^2 \frac{|B|}{|B|\cdot|R'_\text{out}|} 
    \vcenter{
    \hbox{
    \scalebox{0.7}{
\begin{tikzpicture}[line width=1.1pt]

\node[scale=1.4] at (1,-0.75) {$|\hat{\Psi}_i\rangle_{\ell r}$};

\draw[orange] (0,-0.25) -- (0,0.5);
\draw[green] (2,-0.25) -- (2,0.5);

\draw[brown,fill=brown!20] (-0.25,0.5) rectangle (2.25,1.25);
\node[font=\large,brown] at (1,0.875) {$\hat{U}^\dagger$};

\draw[red] (0,1.25) -- (0,5.75);
\draw[green] (2,1.25) -- (2,2.75) -- (1.5,3);

\node[font=\small] at (1.5,3.25) {$\langle\text{MAX}|$};

\draw[blue]  (1.5,3) -- (1,2.75) -- (1,2) -- (0.75,2) -- (0.75,5) -- (1,5) -- (1,4.25) -- (1.5,4);
\node[font=\small] at (1.5,3.75) {$|\text{MAX}\rangle$};
\draw[green] (1.5,4) -- (2,4.25) -- (2,5.75);

\draw[brown,fill=brown!20] (-0.25,5.75) rectangle (2.25,6.5);
\node[font=\large,brown] at (1,6.125) {$\hat{U}$};

\draw[orange] (0,6.5) -- (0,7.25);
\draw[green] (2,6.5) -- (2,7.25);

\node[scale=1.4] at (1,7.75) {$\langle\hat{\Psi}_j|_{\ell r}$};

\end{tikzpicture}
}
    }
    },
\end{equation}
where we've temporarily dropped the $R_\text{out}$ factor since $V_\text{BFP}$ acts trivially on the outgoing radiation. Since we aren't assuming that $rR_\text{out}$ began in the maximally entangled state, we cannot simply get rid of post-selection by straightening lines as we did in (\ref{eq:V_daggerV}). Instead, we integrate $U$ over the Haar measure using equation (\ref{eq:dU1}); the result of this integration is shown in the second equality of (\ref{eq:V_daggerV_BFP}). We can now straighten the connected radiation lines, contributing another factor of $1/|r|$; this allows the effective dynamics to reduce to the identity without any need for integration. Recalling that $|R'_\text{out}|=|R_\text{out}|=|r|$, we find
\begin{equation}
    \int dU \, \langle\hat{\Psi}_j|_{\ell r} V_\text{BFP}^\dagger V_\text{BFP} | \hat{\Psi}_i\rangle_{\ell r} = \delta_{ij}.
\end{equation}
For Haar average fundamental dynamics $U$, the BFP map acts isometrically on generic states of the effective Hilbert space.

\begin{figure} 
    \centering
    \begin{subfigure}{.5\textwidth}
        \centering
        \includegraphics[width=\linewidth]{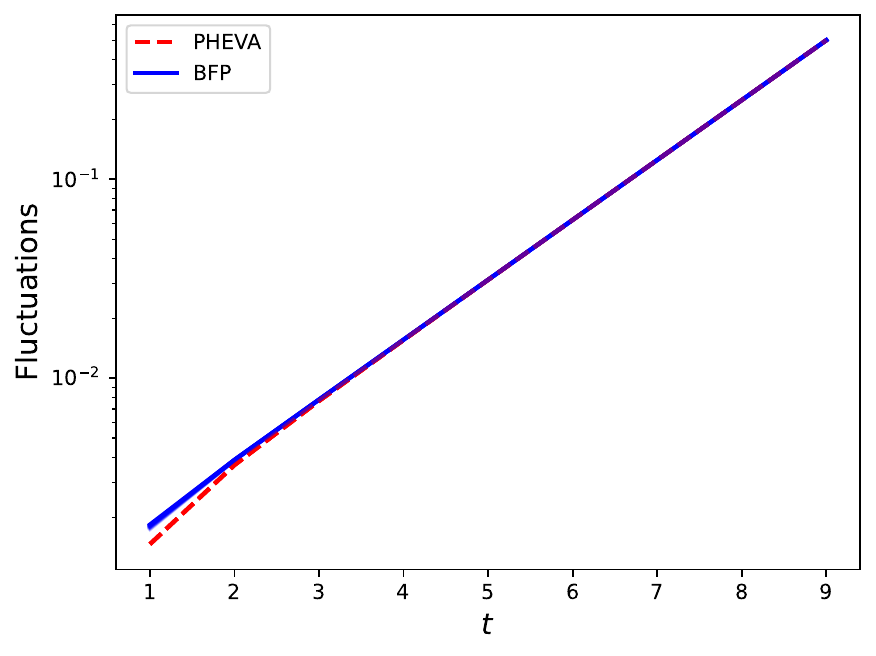}
    \end{subfigure}%
    \begin{subfigure}{.5\textwidth}
        \centering
        \includegraphics[width=\linewidth]{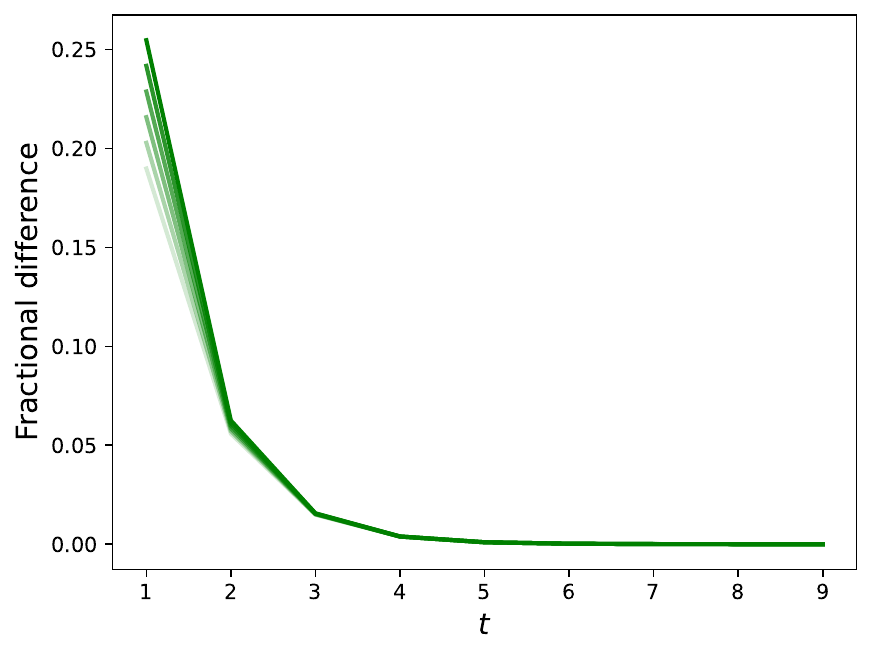}
    \end{subfigure}
    \caption{Plots comparing the fluctuations around isometry as a function of time for $V_\text{BFP}$ and $V_\text{PHEVA}$ using $q=2$, $n_0=10$, and $m_0=1$. The left panel plots the exact expressions for the BFP map (\ref{eq:fluctuation_exact}) in blue and PHEVA's dynamical map (\ref{eq:fluctuation_PHEVA}) in dashed red. The right panel shows the fractional difference between the two, $(\text{BFP} - \text{PHEVA})/\text{PHEVA}$. $\tr ( (\hat{\Psi}_\alpha)_{R_\text{out}} (\hat{\Psi}_\beta)_{R_\text{out}} )$ was set to 1 in all plots; the color shade indicates the value of $|\langle\hat{\Psi}_\beta|\hat{\Psi}_\alpha\rangle_{\ell r R_\text{out}}|^2$, with 0 being the darkest, and 1 being the lightest.}
    \label{fig:fluctuations_compare}
\end{figure}

It is not enough to only act isometrically on average; PHEVA demonstrated that holographic maps for the black hole interior must have fluctuations around this average that are exponentially small in the black hole's entropy. In calculating the fluctuations, it will be useful to consider overlaps between non-orthogonal states; for this we define
\begin{equation}
    |\hat{\Psi}_\alpha\rangle_{\ell rR_\text{out}} = \sum_i \alpha_i |\hat{\Psi}_i\rangle_{\ell rR_\text{out}},
\end{equation}
such that $0 \leq \langle \hat{\Psi}_\beta| \hat{\Psi}_\alpha\rangle \leq 1$. As in \cite{akers_black_2022}, we quantify the fluctuations of $V_\text{BFP}$ about isometry using
\begin{align*}
    \int d\hat{U} dU \, &\Big| \langle \hat{\Psi}_\beta|_{\ell r R_\text{out}} V_\text{BFP}^\dagger V_\text{BFP} |\hat{\Psi}_\alpha\rangle_{\ell r R_\text{out}} - \langle\hat{\Psi}_\beta|\hat{\Psi}_\alpha\rangle_{\ell r R_\text{out}} \Big|^2 \\
    &\quad= \int d\hat{U} dU \, \langle\hat{\Psi}_\beta|\langle\hat{\Psi}_\alpha| \big(V_\text{BFP}^\dagger V_\text{BFP}\big)^{\otimes2} |\hat{\Psi}_\alpha\rangle|\hat{\Psi}_\beta\rangle - \big|\langle\hat{\Psi}_\beta|\hat{\Psi}_\alpha\rangle\big|^2 \numberthis \label{eq:fluctuation_formula},
\end{align*}
where we have dropped the $\ell r R_\text{out}$ subscripts in the second line for ease of reading. Equation (\ref{eq:dU2}) is required to integrate over the two copies of $V_\text{BFP}^\dagger V_\text{BFP}$ in the first term. Integrating over $U$ and straightening any bent radiation lines gives
\begin{flalign*} 
    \int d\hat{U} dU \, |r|^4
    \vcenter{
    \hbox{
    \scalebox{0.7}{
\begin{tikzpicture}[line width=1.1pt]

\node[scale=1.4] at (1.5,-0.75) {$|\hat{\Psi}_\alpha\rangle_{\ell rR_\text{out}}$};

\draw[orange] (0,-0.25) -- (0,0.5);
\draw[green] (2,-0.25) -- (2,0.5);

\draw[brown,fill=brown!20] (-0.25,0.5) rectangle (2.25,1.25);
\node[font=\large,brown] at (1,0.875) {$\hat{U}^\dagger$};

\draw[red] (0,1.25) -- (0,2.75);
\draw[green] (2,1.25) -- (2,4.25) -- (1.5,4.5);
\draw (1,2.25) -- (1,2.75);
\node[font=\small] at (1,1.95) {$|\psi\rangle_f$};

\draw[fill=gray!30] (-0.25,2.75) rectangle (1.25,3.5);
\node[font=\large] at (0.5,3.125) {$U$};

\draw (0,3.5) -- (0,6.5);
\draw[blue] (1,3.5) -- (1,4.25) -- (1.5,4.5);
\node[font=\small] at (1.5,4.75) {$\langle\text{MAX}|$};


\begin{scope}[yscale=-1,shift={(0,-10)}]

\node[scale=1.4] at (1.5,-0.75) {$\langle\hat{\Psi}_\beta|_{\ell rR_\text{out}}$};

\draw[orange] (0,-0.25) -- (0,0.5);
\draw[green] (2,-0.25) -- (2,0.5);

\draw[brown,fill=brown!20] (-0.25,0.5) rectangle (2.25,1.25);
\node[font=\large,brown] at (1,0.875) {$\hat{U}$};

\draw[red] (0,1.25) -- (0,2.75);
\draw[green] (2,1.25) -- (2,4.25) -- (1.5,4.5);
\draw (1,2.25) -- (1,2.75);
\node[font=\small] at (1,1.95) {$\langle\psi|_f$};

\draw[fill=gray!30] (-0.25,2.75) rectangle (1.25,3.5);
\node[font=\large] at (0.5,3.125) {$U^\dagger$};

\draw[blue] (1,3.5) -- (1,4.25) -- (1.5,4.5);
\node[font=\small] at (1.5,4.75) {$|\text{MAX}\rangle$};
    
\end{scope}

\draw[blue] (3,-0.25) -- (3,10.25);


\begin{scope}[shift={(5,0)}]

\node[scale=1.4] at (1.5,-0.75) {$|\hat{\Psi}_\beta\rangle_{\ell rR_\text{out}}$};

\draw[orange] (0,-0.25) -- (0,0.5);
\draw[green] (2,-0.25) -- (2,0.5);

\draw[brown,fill=brown!20] (-0.25,0.5) rectangle (2.25,1.25);
\node[font=\large,brown] at (1,0.875) {$\hat{U}^\dagger$};

\draw[red] (0,1.25) -- (0,2.75);
\draw[green] (2,1.25) -- (2,4.25) -- (1.5,4.5);
\draw (1,2.25) -- (1,2.75);
\node[font=\small] at (1,1.95) {$|\psi\rangle_f$};

\draw[fill=gray!30] (-0.25,2.75) rectangle (1.25,3.5);
\node[font=\large] at (0.5,3.125) {$U$};

\draw (0,3.5) -- (0,6.5);
\draw[blue] (1,3.5) -- (1,4.25) -- (1.5,4.5);
\node[font=\small] at (1.5,4.75) {$\langle\text{MAX}|$};


\begin{scope}[yscale=-1,shift={(0,-10)}]

\node[scale=1.4] at (1.5,-0.75) {$\langle\hat{\Psi}_\alpha|_{\ell rR_\text{out}}$};

\draw[orange] (0,-0.25) -- (0,0.5);
\draw[green] (2,-0.25) -- (2,0.5);

\draw[brown,fill=brown!20] (-0.25,0.5) rectangle (2.25,1.25);
\node[font=\large,brown] at (1,0.875) {$\hat{U}$};

\draw[red] (0,1.25) -- (0,2.75);
\draw[green] (2,1.25) -- (2,4.25) -- (1.5,4.5);
\draw (1,2.25) -- (1,2.75);
\node[font=\small] at (1,1.95) {$\langle\psi|_f$};

\draw[fill=gray!30] (-0.25,2.75) rectangle (1.25,3.5);
\node[font=\large] at (0.5,3.125) {$U^\dagger$};

\draw[blue] (1,3.5) -- (1,4.25) -- (1.5,4.5);
\node[font=\small] at (1.5,4.75) {$|\text{MAX}\rangle$};
    
\end{scope}

\draw[blue] (3,-0.25) -- (3,10.25);

\end{scope}

\end{tikzpicture}
}
    }
    } && 
\end{flalign*}
\begin{flalign*} 
    \qquad\quad&= \int d\hat{U} \, \frac{|r|^2}{|B|^2|r|^2 - 1} \Big[ |B|^2 |\langle\hat{\Psi}_\alpha|\hat{\Psi}_\beta\rangle_{\ell rR_\text{out}}|^2 + |B|\tr\big((\hat{\Psi}_\alpha)_{R_\text{out}} (\hat{\Psi}_\beta)_{R_\text{out}} \big) \Big] && \\
    &\qquad\quad -\frac{|r|^2}{|B|\cdot|r| ( |B|^2|r|^2 - 1 )} \lb
    \vcenter{
    \hbox{
    \scalebox{0.7}{
\begin{tikzpicture}[line width=1.1pt]

\node[scale=1.4] at (-1,1.875) {$|B|$};

\node[scale=1.1] at (0.5,-0.3) {$|\hat{\Psi}_\alpha\rangle_{\ell r}$};
\node[scale=1.1] at (2.5,-0.3) {$|\hat{\Psi}_\beta\rangle_{\ell r}$};

\draw[orange] (0,0) -- (0,0.75);
\draw[green] (1,0) -- (1,0.75);
\draw[orange] (2,0) -- (2,0.75);
\draw[green] (3,0) -- (3,0.75);

\draw[brown,fill=brown!20] (-0.25,0.75) rectangle (1.25,1.5);
\node[font=\large,brown] at (0.5,1.125) {$\hat{U}^\dagger$};

\draw[brown,fill=brown!20] (1.75,0.75) rectangle (3.25,1.5);
\node[font=\large,brown] at (2.5,1.125) {$\hat{U}^\dagger$};

\draw[red] (0,1.5) -- (0,2.25);
\draw[red] (2,1.5) -- (2,2.25);
\draw[green] (1,1.5) -- (3,2.25);
\draw[green] (3,1.5) -- (1,2.25);

\draw[brown,fill=brown!20] (-0.25,2.25) rectangle (1.25,3);
\node[font=\large,brown] at (0.5,2.625) {$\hat{U}$};

\draw[brown,fill=brown!20] (1.75,2.25) rectangle (3.25,3);
\node[font=\large,brown] at (2.5,2.625) {$\hat{U}$};

\draw[orange] (0,3) -- (0,3.75);
\draw[green] (1,3) -- (1,3.75);
\draw[orange] (2,3) -- (2,3.75);
\draw[green] (3,3) -- (3,3.75);

\node[scale=1.1] at (0.5,4.05) {$\langle\hat{\Psi}_\beta|_{\ell r}$};
\node[scale=1.1] at (2.5,4.05) {$\langle\hat{\Psi}_\alpha|_{\ell r}$};


\begin{scope}[shift={(5.75,0)}]

\node[scale=1.4] at (-1.25,1.875) {$+ \,|B|^2$};

\node[scale=1.1] at (0.5,-0.3) {$|\hat{\Psi}_\alpha\rangle_{\ell r}$};
\node[scale=1.1] at (2.5,-0.3) {$|\hat{\Psi}_\beta\rangle_{\ell r}$};

\draw[orange] (0,0) -- (0,0.75);
\draw[green] (1,0) -- (1,0.75);
\draw[orange] (2,0) -- (2,0.75);
\draw[green] (3,0) -- (3,0.75);

\draw[brown,fill=brown!20] (-0.25,0.75) rectangle (1.25,1.5);
\node[font=\large,brown] at (0.5,1.125) {$\hat{U}^\dagger$};

\draw[brown,fill=brown!20] (1.75,0.75) rectangle (3.25,1.5);
\node[font=\large,brown] at (2.5,1.125) {$\hat{U}^\dagger$};

\draw[red] (0,1.5) -- (2,2.25);
\draw[red] (2,1.5) -- (0,2.25);
\draw[green] (1,1.5) -- (1,2.25);
\draw[green] (3,1.5) -- (3,2.25);

\draw[brown,fill=brown!20] (-0.25,2.25) rectangle (1.25,3);
\node[font=\large,brown] at (0.5,2.625) {$\hat{U}$};

\draw[brown,fill=brown!20] (1.75,2.25) rectangle (3.25,3);
\node[font=\large,brown] at (2.5,2.625) {$\hat{U}$};

\draw[orange] (0,3) -- (0,3.75);
\draw[green] (1,3) -- (1,3.75);
\draw[orange] (2,3) -- (2,3.75);
\draw[green] (3,3) -- (3,3.75);

\node[scale=1.1] at (0.5,4.05) {$\langle\hat{\Psi}_\beta|_{\ell r}$};
\node[scale=1.1] at (2.5,4.05) {$\langle\hat{\Psi}_\alpha|_{\ell r}$};
    
\end{scope}

\end{tikzpicture}
}
    }
    } \rb && \numberthis \label{eq:fluctuation}
\end{flalign*}
where $(\hat{\Psi}_\alpha)_{R_\text{out}} = \tr_{\ell r} |\hat{\Psi}_\alpha\rangle\langle\hat{\Psi}_\alpha|_{\ell rR_\text{out}}$. Integrating the last two terms of (\ref{eq:fluctuation}) over $\hat{U}$ using (\ref{eq:dU2}) and plugging the result in (\ref{eq:fluctuation_formula}), we find the fluctuations of the BFP map around isometry:
\begin{align*}
    \int d\hat{U} dU \, &\Big| \langle \hat{\Psi}_\beta|_{\ell r R_\text{out}} V_\text{BFP}^\dagger V_\text{BFP} |\hat{\Psi}_\alpha\rangle_{\ell r R_\text{out}} - \langle\hat{\Psi}_\beta|\hat{\Psi}_\alpha\rangle_{\ell r R_\text{out}} \Big|^2 \\
    & = \frac{(|B|\cdot|\ell|\cdot|r|-1)(|r|^2-1)}{(|B|^2 |r|^2 -1)(|\ell|^2 |r|^2 - 1)} \Bigg[ |\ell|\cdot|r| \tr \Big( (\hat{\Psi}_\alpha)_{R_\text{out}} (\hat{\Psi}_\beta)_{R_\text{out}} \Big) - \big|\langle\hat{\Psi}_\beta|\hat{\Psi}_\alpha\rangle_{\ell r R_\text{out}}\big|^2 \Bigg]. \numberthis \label{eq:fluctuation_exact}
\end{align*}
To determine the strength of these fluctuations, we find an upper bound of (\ref{eq:fluctuation_exact}). We can maximize over the choice of $\hat{\Psi}_\alpha$ and $\hat{\Psi}_\beta$ by setting
\begin{equation}
    \tr \Big( (\hat{\Psi}_\alpha)_{R_\text{out}} (\hat{\Psi}_\beta)_{R_\text{out}} \Big) = 1, \qquad \big|\langle\hat{\Psi}_\beta|\hat{\Psi}_\alpha\rangle_{\ell r R_\text{out}}\big|^2 = 0.
\end{equation}
Applying the inequality
\begin{equation} \label{eq:bounder}
    \frac{1}{x-1} \leq \frac{2}{x}, \quad x\geq2,
\end{equation}
on the overall prefactor twice gives an upper bound on the fluctuations of $V_\text{BFP}^\dagger V_\text{BFP}$ as
\begin{equation}
    \int d\hat{U} dU \, \Big| \langle \hat{\Psi}_\beta|_{\ell r R_\text{out}} V_\text{BFP}^\dagger V_\text{BFP} |\hat{\Psi}_\alpha\rangle_{\ell r R_\text{out}} - \langle\hat{\Psi}_\beta|\hat{\Psi}_\alpha\rangle_{\ell r R_\text{out}} \Big|^2 \leq \frac{4}{|B|}.
\end{equation}
For average $U$ and $\hat{U}$, the fluctuations of the BFP map around isometry are indeed exponentially small in the black hole's entropy, thanks to the factor of $|B|$ in the denominator. Therefore, the backwards-forwards map preserves the inner product of all generic states on the effective Hilbert space with exponentially small errors. Furthermore, this upper bound differs from the upper bound on fluctuations of PHEVA's dynamical model by only a factor of 2; for convenience, we reproduce their exact result and bound here,\footnote{PHEVA found fluctuations in equation (2.7) of \cite{akers_black_2022} for their generic holographic map defined in equation (\ref{eq:PHEVA_gen}) of this work. The result quoted here applies to their dynamical holographic map defined here in (\ref{eq:V_PHEVA}).}
\begin{align*} 
    \int d\hat{U} dU \, &\Big| \langle \hat{\Psi}_\beta|_{\ell r R_\text{out}} V_\text{PHEVA}^\dagger V_\text{PHEVA} |\hat{\Psi}_\alpha\rangle_{\ell r R_\text{out}} - \langle\hat{\Psi}_\beta|\hat{\Psi}_\alpha\rangle_{\ell r R_\text{out}} \Big|^2 \\
    &=\frac{|r|-1}{|B|^2 |r|^2 - 1} \Bigg[ |B|\cdot|r| \tr \Big( (\hat{\Psi}_\alpha)_{R_\text{out}} (\hat{\Psi}_\beta)_{R_\text{out}} \Big) - \big|\langle\hat{\Psi}_\beta|\hat{\Psi}_\alpha\rangle_{\ell r R_\text{out}}\big|^2 \Bigg] \leq \frac{2}{|B|}. \numberthis \label{eq:fluctuation_PHEVA}
\end{align*}
Fig.~\ref{fig:fluctuations_compare} compares the exact fluctuations for the BFP map in (\ref{eq:fluctuation_exact}) with those for PHEVA's dynamical map in (\ref{eq:fluctuation_PHEVA}) as a function of time using
\begin{equation}
    |B| = q^{n_0 - t}, \quad |r| = q^{2t}, \quad |\ell| = q^{m_0 + t}.
\end{equation}
The parameters $q=2$, $n_0=10$, $m_0=1$, and $\tr ( (\hat{\Psi}_\alpha)_{R_\text{out}} (\hat{\Psi}_\beta)_{R_\text{out}} ) = 1$ were used for both plots. We see that fluctuations of both models behave similarly, getting larger at late times. With these choices of parameters, the fluctuations of the BFP map are slightly larger than PHEVA's dynamical map, but their agreement strengthens as the black hole continues to evaporate. We note that the BFP map does have smaller fluctuations than PHEVA's map for other choices of parameters. Agreement between the two strengthens as $q$ is increased since both expressions have the same large $q$ approximation,
\begin{equation}
    \int d\hat{U} dU \, \Big| \langle \hat{\Psi}_\beta|_{\ell r R_\text{out}} V^\dagger V |\hat{\Psi}_\alpha\rangle_{\ell r R_\text{out}} - \langle\hat{\Psi}_\beta|\hat{\Psi}_\alpha\rangle_{\ell r R_\text{out}} \Big|^2 \approx q^{-n_0 + t} \tr \Big( (\hat{\Psi}_\alpha)_{R_\text{out}} (\hat{\Psi}_\beta)_{R_\text{out}} \Big), \quad \text{large } q.
\end{equation}

We've now seen that the BFP map satisfies the first criterion for a holographic map acting on generic states: it preserves the inner product of generic states on average with exponentially suppressed fluctuations. Furthermore, it does so in a way comparable to the dynamical holographic maps proposed by PHEVA in \cite{akers_black_2022}. We note that unlike previous results, this did require averaging $\hat{U}$ over the Haar measure. We do not expect each $\hat{U}_t$, which are determined by the dynamics of semiclassical gravity, to be strongly scrambling or complex enough to be well modeled by Haar random unitaries. Instead, we expect unitary $k$-designs could be used to achieve the above integrals without making the dynamics too scrambling or complex.

Before continuing to reconstruction using the BFP map, we comment on the relationship between the BFP map and the backwards-forwards map. As described in Sec.~\ref{sec:relationship}, the transformations relating the two allowed us to argue that the unitarity of black hole evaporation was preserved by the map, since they connect the BFP map to the unitarity of the backwards-forwards map on dynamically generated states. However, these transformations are no longer valid for generic states. We can no longer guarantee that radiation modes in the effective description will backwards evolve to the maximally entangled state, so we cannot replace the projector $|\text{MAX}\rangle\langle\text{MAX}|_{r,R_\text{out}}$ with the identity in the second step of Fig.~\ref{fig:transformations}. Even so, we've seen that the backwards-forwards map no longer acts isometrically on generic states, let alone unitarily. Therefore, we cannot claim unitarity of black hole evaporation for the full effective Hilbert space.

However, there is no issue here -- generic states can only be reached by a measurement performed by an interior observer. Since an observer cannot force themselves to forget the result of their measurement, there is no reason to expect unitarity from their perspective. A second observer who hasn't learned the result of the measurement would see no violation of unitarity; all that's occurred from their perspective is a dynamical unitary entangling the first observer with the black hole interior. This keeps the interior state in the dynamically generated subspace, where both the transformations of Fig.~\ref{fig:transformations} and unitarity still hold.

\subsection{Reconstruction using BFP} \label{sec:recon}
We now check the second criterion for a holographic map to work properly on generic states: the BFP map must be able to reconstruct non-dynamical effective description unitary operators in the fundamental description. It was shown in \cite{akers_black_2022} that PHEVA's holographic maps provided a state-dependent reconstruction somewhere on $BR_\text{out}$ at all times. Furthermore, reconstruction on $B$ and $R_\text{out}$ separately implied the appropriate bounds for entanglement wedge reconstruction:
\begin{alignat}{2}
    &\text{Reconstruction on } B \, &&\implies \, S(\hat{\Psi}_{R_\text{out}}) \ll S(\hat{\Psi}_L) + \log|B| \\
    &\text{Reconstruction on } R_\text{out} \, &&\implies \, S(\hat{\Psi}_{R_\text{out}}) \gg S(\hat{\Psi}_L) + \log|B|.
\end{alignat}
The ability of their holographic maps to reconstruct measurements performed by an observer in the black hole interior then follows from their ability to reconstruct unitary operators entangling the observer and their measurement apparatus with the interior state.

Here, we demonstrate that the BFP map provides an equally good state-dependent reconstruction of interior unitaries $W$ in the fundamental description. To do so, we follow the ideas laid out by theorem 3.7 of \cite{akers_quantum_2022} and theorem 5.2 of \cite{akers_black_2022}. In our current notation, these state that if the action of $W \in \hs_\text{eff}$ on the effective Hilbert space decouples from a reference system $L$,\footnote{Recall that $L$ keeps a record of and purifies what fell into the black hole as $\ell$. Since it remains outside of the black hole, it is shared by both the effective and fundamental descriptions.}
\begin{equation} \label{eq:decouple}
    \big\| \tr_{BR_\text{out}} \big( V_\text{BFP} W |\hat{\Psi}\rangle\langle\hat{\Psi}| W^\dagger V_\text{BFP}^\dagger \big) - \tr_{BR_\text{out}} \big( V_\text{BFP} |\hat{\Psi}\rangle\langle\hat{\Psi}| V_\text{BFP}^\dagger \big) \big\|_1 \leq \epsilon_2,
\end{equation}
then there exists some unitary operator $\tilde W \in \hs_\text{fun}$ acting on the complement of $L$ in the fundamental Hilbert space such that
\begin{equation}
    \big\| \tilde{W} V_\text{BFP} |\hat{\Psi}\rangle - V_\text{BFP} W |\hat{\Psi}\rangle \big\| \leq \epsilon_1
\end{equation}
with $\epsilon_1 \leq \sqrt{\epsilon_2}$. In other words, if the reference system $L$ cannot be used to discern whether or not $W$ was applied inside the interior, then there exists some unitary acting somewhere on $BR_\text{out}$ in the fundamental Hilbert space that serves as the reconstruction of $W$. We will consider this reconstruction to be ``good'' if $\epsilon_2$ (and therefore $\epsilon_1$) is exponentially small in the black hole's entropy.

We emphasize that this state-dependent reconstruction is only needed for non-dynamical $W$. Because the BFP map acts isometrically on dynamically generated states, reconstruction in the dynamical subspace is exact and state-independent. Therefore, if $W$ were a dynamical unitary, it would be undone by the backwards evolution and there would exist a state-independent reconstruction in the fundamental description. However, these nice properties do not apply to non-dynamical unitaries since they cannot be undone by backwards dynamics. In this case, $V_\text{BFP}$ acts only isometrically on average; therefore no state-independent reconstruction is guaranteed to exist. (See theorem~5.1 of \cite{akers_black_2022} for more details.) In what follows, we will always assume $W$ is a non-dynamical unitary.

We begin by checking the reconstruction of $W_{\ell rR_\text{out}}$ on $BR_\text{out}$ through decoupling from $L$. This calculation is essentially a repeat of that done by PHEVA in \cite{akers_black_2022} for their generic holographic map, here applied to the BFP map. As in \cite{akers_black_2022}, we first define a $W$-dependent state in the fundamental description for notational convenience:
\begin{equation}
    \Psi_{LBR_\text{out}} (W) \equiv \big( V_\text{BFP} \otimes \id_L \big) W |\hat{\Psi}\rangle\langle\hat{\Psi}|_{L\ell rR_\text{out}} W^\dagger \big( V_\text{BFP}^\dagger \otimes \id_L \big)
\end{equation}
Introducing Haar integration over fundamental dynamics $U$ and effective dynamics $\hat{U}$, the left hand side of the decoupling inequality (\ref{eq:decouple}) becomes
\begin{equation} \label{eq:bounded_recon_BR}
    \int d\hat{U} dU \, \big\| \Psi_L (W_{\ell rR_\text{out}}) - \Psi_L (\id) \big\|_1 \leq \sqrt{ |L| \int d\hat{U} dU \, \big\| \Psi_L (W_{\ell rR_\text{out}}) - \Psi_L (\id) \big\|_2^2 },
\end{equation}
where the right hand side of (\ref{eq:bounded_recon_BR}) is the upper bound of the $L_1$ norm given (as in \cite{akers_black_2022}) by the $L_2$ norm and Jensen's inequality. Our task is then to compute the radicand of (\ref{eq:bounded_recon_BR}),
\begin{equation} \label{eq:integral_L2}
    \int d\hat{U} dU \, \big\| \Psi_L (W_{\ell rR_\text{out}}) - \Psi_L (\id) \big\|_2^2 = \int d\hat{U} dU \, \tr \Big[ \Psi_L(W_{\ell rR_\text{out}})^2 - 2 \Psi_L(W_{\ell rR_\text{out}}) \Psi_L(\id) + \Psi_L(\id)^2 \Big].
\end{equation}
All three terms in the right hand side of (\ref{eq:integral_L2}) involve essentially the same integration, differing only by whether or not $W$ is acting on $\ell rR_\text{out}$. We will demonstrate the calculation of the first term explicitly; the remaining two will follow naturally from the first. Represented as a circuit diagram, this term is
\begin{equation} \label{eq:Psi_W_2}
    \int d\hat{U} dU \, \tr \Psi_L(W_{\ell rR_\text{out}})^2 = |r|^4
    \vcenter{
    \hbox{
    \scalebox{0.7}{
\begin{tikzpicture}[line width=1.1pt]

\node[scale=1.3] at (1.5,-0.375) {$|\hat{\Psi}\rangle\langle\hat{\Psi}|_{\ell r R_\text{out} L}$};

\draw[orange] (0,0) -- (0,0.75);
\draw[green] (1,0) -- (1,0.75);
\draw[blue] (2,0) -- (2,0.75);
\draw[brown] (3,0) -- (3,5.25);

\draw[darkblue,fill=darkblue!20] (-0.25,0.75) rectangle (2.25,1.5);
\node[font=\large,darkblue] at (1,1.125) {$W$};

\draw[orange] (0,1.5) -- (0,2.25);
\draw[green] (1,1.5) -- (1,2.25);
\draw[blue] (2,1.5) -- (2,6) -- (-3,6) -- (-3,-0.375);

\draw[brown,fill=brown!20] (-0.25,2.25) rectangle (1.25,3);
\node[font=\large,brown] at (0.5,2.625) {$\hat{U}^\dagger$};

\draw (-1,3.25) -- (-1,3.75);
\draw[red] (0,3) -- (0,3.75);
\draw[green] (1,3) -- (1,5.25) -- (0.5,5.5);
\draw[blue] (0.5,5.5) -- (0,5.25) -- (0,4.5);

\draw[fill=gray!30] (-1.25,3.75) rectangle (0.25,4.5);
\node[font=\large] at (-0.5,4.125) {$U$};
\node[font=\small] at (-1,2.95) {$|\psi\rangle_f$};
\node[font=\small] at (0.5,5.7) {$\langle\text{MAX}|$};

\draw (-1,4.5) -- (-1,5.25) -- (-2,5.25) -- (-2,-0.375);


\begin{scope}[yscale=-1,shift={(0,0.75)}]

\draw[orange] (0,0) -- (0,0.75);
\draw[green] (1,0) -- (1,0.75);
\draw[blue] (2,0) -- (2,0.75);
\draw[brown] (3,0) -- (3,5.25);

\draw[darkblue,fill=darkblue!20] (-0.25,0.75) rectangle (2.25,1.5);
\node[font=\large,darkblue] at (1,1.125) {$W^\dagger$};

\draw[orange] (0,1.5) -- (0,2.25);
\draw[green] (1,1.5) -- (1,2.25);
\draw[blue] (2,1.5) -- (2,6) -- (-3,6) -- (-3,-0.375);

\draw[brown,fill=brown!20] (-0.25,2.25) rectangle (1.25,3);
\node[font=\large,brown] at (0.5,2.625) {$\hat{U}$};

\draw (-1,3.25) -- (-1,3.75);
\draw[red] (0,3) -- (0,3.75);
\draw[green] (1,3) -- (1,5.25) -- (0.5,5.5);
\draw[blue] (0.5,5.5) -- (0,5.25) -- (0,4.5);

\draw[fill=gray!30] (-1.25,3.75) rectangle (0.25,4.5);
\node[font=\large] at (-0.5,4.125) {$U^\dagger$};
\node[font=\small] at (-1,2.95) {$\langle\psi|_f$};
\node[font=\small] at (0.5,5.7) {$|\text{MAX}\rangle$};

\draw (-1,4.5) -- (-1,5.25) -- (-2,5.25) -- (-2,-0.375);
    
\end{scope}


\begin{scope}[shift={(7,0)}]

\node[scale=1.3] at (1.5,-0.375) {$|\hat{\Psi}\rangle\langle\hat{\Psi}|_{\ell r R_\text{out} L}$};

\draw[orange] (0,0) -- (0,0.75);
\draw[green] (1,0) -- (1,0.75);
\draw[blue] (2,0) -- (2,0.75);
\draw[brown] (3,0) -- (3,5.25);

\draw[darkblue,fill=darkblue!20] (-0.25,0.75) rectangle (2.25,1.5);
\node[font=\large,darkblue] at (1,1.125) {$W$};

\draw[orange] (0,1.5) -- (0,2.25);
\draw[green] (1,1.5) -- (1,2.25);
\draw[blue] (2,1.5) -- (2,6) -- (-3,6) -- (-3,-0.375);

\draw[brown,fill=brown!20] (-0.25,2.25) rectangle (1.25,3);
\node[font=\large,brown] at (0.5,2.625) {$\hat{U}^\dagger$};

\draw (-1,3.25) -- (-1,3.75);
\draw[red] (0,3) -- (0,3.75);
\draw[green] (1,3) -- (1,5.25) -- (0.5,5.5);
\draw[blue] (0.5,5.5) -- (0,5.25) -- (0,4.5);

\draw[fill=gray!30] (-1.25,3.75) rectangle (0.25,4.5);
\node[font=\large] at (-0.5,4.125) {$U$};
\node[font=\small] at (-1,2.95) {$|\psi\rangle_f$};
\node[font=\small] at (0.5,5.7) {$\langle\text{MAX}|$};

\draw (-1,4.5) -- (-1,5.25) -- (-2,5.25) -- (-2,-0.375);


\begin{scope}[yscale=-1,shift={(0,0.75)}]

\draw[orange] (0,0) -- (0,0.75);
\draw[green] (1,0) -- (1,0.75);
\draw[blue] (2,0) -- (2,0.75);
\draw[brown] (3,0) -- (3,5.25);

\draw[darkblue,fill=darkblue!20] (-0.25,0.75) rectangle (2.25,1.5);
\node[font=\large,darkblue] at (1,1.125) {$W^\dagger$};

\draw[orange] (0,1.5) -- (0,2.25);
\draw[green] (1,1.5) -- (1,2.25);
\draw[blue] (2,1.5) -- (2,6) -- (-3,6) -- (-3,-0.375);

\draw[brown,fill=brown!20] (-0.25,2.25) rectangle (1.25,3);
\node[font=\large,brown] at (0.5,2.625) {$\hat{U}$};

\draw (-1,3.25) -- (-1,3.75);
\draw[red] (0,3) -- (0,3.75);
\draw[green] (1,3) -- (1,5.25) -- (0.5,5.5);
\draw[blue] (0.5,5.5) -- (0,5.25) -- (0,4.5);

\draw[fill=gray!30] (-1.25,3.75) rectangle (0.25,4.5);
\node[font=\large] at (-0.5,4.125) {$U^\dagger$};
\node[font=\small] at (-1,2.95) {$\langle\psi|_f$};
\node[font=\small] at (0.5,5.7) {$|\text{MAX}\rangle$};

\draw (-1,4.5) -- (-1,5.25) -- (-2,5.25) -- (-2,-0.375);
    
\end{scope}

\end{scope}

\draw[brown] (3,5.25) -- (10,-6);
\draw[brown] (3,-6) -- (10,5.25);
    
\end{tikzpicture}
}
    }
    }
\end{equation}
We first integrate over the four factors of fundamental dynamics $U$ using equation (\ref{eq:dU2}) from Appendix~\ref{app:int}. This generates four terms,
\begin{align*} 
    \frac{|r|^2}{|B|^2|r|^2 - 2} &\lb |B|^2
    \vcenter{
    \hbox{
    \scalebox{0.5}{
\begin{tikzpicture}[line width=1.1pt]

\begin{scope}[xscale=0.8]

\node[scale=1.3] at (1.5,-0.375) {$|\hat{\Psi}\rangle\langle\hat{\Psi}|_{\ell r R_\text{out} L}$};

\draw[orange] (0,0) -- (0,0.75);
\draw[green] (1,0) -- (1,0.75);
\draw[blue] (2,0) -- (2,0.75);
\draw[brown] (3,0) -- (3,3.75);

\draw[darkblue,fill=darkblue!20] (-0.25,0.75) rectangle (2.25,1.5);
\node[font=\large,darkblue] at (1,1.125) {$W$};

\draw[orange] (0,1.5) -- (0,2.25);
\draw[green] (1,1.5) -- (1,2.25);
\draw[blue] (2,1.5) -- (2,5.25) -- (-3,5.25) -- (-3,-0.375);

\draw[brown,fill=brown!20] (-0.25,2.25) rectangle (1.25,3);
\node[font=\large,brown] at (0.5,2.625) {$\hat{U}^\dagger$};

\draw[red] (0,3) -- (0,3.75) -- (-1,3.75) -- (-1,-0.375);
\draw[green] (1,3) -- (1,4.5) -- (-2,4.5) -- (-2,-0.375);


\begin{scope}[yscale=-1,shift={(0,0.75)}]

\draw[orange] (0,0) -- (0,0.75);
\draw[green] (1,0) -- (1,0.75);
\draw[blue] (2,0) -- (2,0.75);
\draw[brown] (3,0) -- (3,3.75);

\draw[darkblue,fill=darkblue!20] (-0.25,0.75) rectangle (2.25,1.5);
\node[font=\large,darkblue] at (1,1.125) {$W^\dagger$};

\draw[orange] (0,1.5) -- (0,2.25);
\draw[green] (1,1.5) -- (1,2.25);
\draw[blue] (2,1.5) -- (2,5.25) -- (-3,5.25) -- (-3,-0.375);

\draw[brown,fill=brown!20] (-0.25,2.25) rectangle (1.25,3);
\node[font=\large,brown] at (0.5,2.625) {$\hat{U}$};

\draw[red] (0,3) -- (0,3.75) -- (-1,3.75) -- (-1,-0.375);
\draw[green] (1,3) -- (1,4.5) -- (-2,4.5) -- (-2,-0.375);
    
\end{scope}


\begin{scope}[shift={(7,0)}]

\node[scale=1.3] at (1.5,-0.375) {$|\hat{\Psi}\rangle\langle\hat{\Psi}|_{\ell r R_\text{out} L}$};

\draw[orange] (0,0) -- (0,0.75);
\draw[green] (1,0) -- (1,0.75);
\draw[blue] (2,0) -- (2,0.75);
\draw[brown] (3,0) -- (3,3.75);

\draw[darkblue,fill=darkblue!20] (-0.25,0.75) rectangle (2.25,1.5);
\node[font=\large,darkblue] at (1,1.125) {$W$};

\draw[orange] (0,1.5) -- (0,2.25);
\draw[green] (1,1.5) -- (1,2.25);
\draw[blue] (2,1.5) -- (2,5.25) -- (-3,5.25) -- (-3,-0.375);

\draw[brown,fill=brown!20] (-0.25,2.25) rectangle (1.25,3);
\node[font=\large,brown] at (0.5,2.625) {$\hat{U}^\dagger$};

\draw[red] (0,3) -- (0,3.75) -- (-1,3.75) -- (-1,-0.375);
\draw[green] (1,3) -- (1,4.5) -- (-2,4.5) -- (-2,-0.375);


\begin{scope}[yscale=-1,shift={(0,0.75)}]

\draw[orange] (0,0) -- (0,0.75);
\draw[green] (1,0) -- (1,0.75);
\draw[blue] (2,0) -- (2,0.75);
\draw[brown] (3,0) -- (3,3.75);

\draw[darkblue,fill=darkblue!20] (-0.25,0.75) rectangle (2.25,1.5);
\node[font=\large,darkblue] at (1,1.125) {$W^\dagger$};

\draw[orange] (0,1.5) -- (0,2.25);
\draw[green] (1,1.5) -- (1,2.25);
\draw[blue] (2,1.5) -- (2,5.25) -- (-3,5.25) -- (-3,-0.375);

\draw[brown,fill=brown!20] (-0.25,2.25) rectangle (1.25,3);
\node[font=\large,brown] at (0.5,2.625) {$\hat{U}$};

\draw[red] (0,3) -- (0,3.75) -- (-1,3.75) -- (-1,-0.375);
\draw[green] (1,3) -- (1,4.5) -- (-2,4.5) -- (-2,-0.375);
    
\end{scope}

\end{scope}

\draw[brown] (3,3.75) -- (10,-4.5);
\draw[brown] (3,-4.5) -- (10,3.75);

\end{scope}
    
\end{tikzpicture}
}
    }
    } + |B|
    \vcenter{
    \hbox{
    \scalebox{0.5}{
\begin{tikzpicture}[line width=1.1pt]

\begin{scope}[xscale=0.8]

\node[scale=1.3] at (1.5,-0.375) {$|\hat{\Psi}\rangle\langle\hat{\Psi}|_{\ell r R_\text{out} L}$};

\draw[orange] (0,0) -- (0,0.75);
\draw[green] (1,0) -- (1,0.75);
\draw[blue] (2,0) -- (2,0.75);
\draw[brown] (3,0) -- (3,3.75);

\draw[darkblue,fill=darkblue!20] (-0.25,0.75) rectangle (2.25,1.5);
\node[font=\large,darkblue] at (1,1.125) {$W$};

\draw[orange] (0,1.5) -- (0,2.25);
\draw[green] (1,1.5) -- (1,2.25);
\draw[blue] (2,1.5) -- (2,4.5) -- (-1,4.5) -- (-1,-0.375);

\draw[brown,fill=brown!20] (-0.25,2.25) rectangle (1.25,3);
\node[font=\large,brown] at (0.5,2.625) {$\hat{U}^\dagger$};

\draw[red] (0,3) -- (0,3.75);
\draw[green] (1,3) -- (1,3.75);


\begin{scope}[yscale=-1,shift={(0,0.75)}]

\draw[orange] (0,0) -- (0,0.75);
\draw[green] (1,0) -- (1,0.75);
\draw[blue] (2,0) -- (2,0.75);
\draw[brown] (3,0) -- (3,3.75);

\draw[darkblue,fill=darkblue!20] (-0.25,0.75) rectangle (2.25,1.5);
\node[font=\large,darkblue] at (1,1.125) {$W^\dagger$};

\draw[orange] (0,1.5) -- (0,2.25);
\draw[green] (1,1.5) -- (1,2.25);
\draw[blue] (2,1.5) -- (2,4.5) -- (-1,4.5) -- (-1,-0.375);

\draw[brown,fill=brown!20] (-0.25,2.25) rectangle (1.25,3);
\node[font=\large,brown] at (0.5,2.625) {$\hat{U}$};

\draw[red] (0,3) -- (0,3.75);
\draw[green] (1,3) -- (1,3.75);
    
\end{scope}


\begin{scope}[shift={(5,0)}]

\node[scale=1.3] at (1.5,-0.375) {$|\hat{\Psi}\rangle\langle\hat{\Psi}|_{\ell r R_\text{out} L}$};

\draw[orange] (0,0) -- (0,0.75);
\draw[green] (1,0) -- (1,0.75);
\draw[blue] (2,0) -- (2,0.75);
\draw[brown] (3,0) -- (3,3.75);

\draw[darkblue,fill=darkblue!20] (-0.25,0.75) rectangle (2.25,1.5);
\node[font=\large,darkblue] at (1,1.125) {$W$};

\draw[orange] (0,1.5) -- (0,2.25);
\draw[green] (1,1.5) -- (1,2.25);
\draw[blue] (2,1.5) -- (2,4.5) -- (-1,4.5) -- (-1,-0.375);

\draw[brown,fill=brown!20] (-0.25,2.25) rectangle (1.25,3);
\node[font=\large,brown] at (0.5,2.625) {$\hat{U}^\dagger$};

\draw[red] (0,3) -- (0,3.75);
\draw[green] (1,3) -- (1,3.75);


\begin{scope}[yscale=-1,shift={(0,0.75)}]

\draw[orange] (0,0) -- (0,0.75);
\draw[green] (1,0) -- (1,0.75);
\draw[blue] (2,0) -- (2,0.75);
\draw[brown] (3,0) -- (3,3.75);

\draw[darkblue,fill=darkblue!20] (-0.25,0.75) rectangle (2.25,1.5);
\node[font=\large,darkblue] at (1,1.125) {$W^\dagger$};

\draw[orange] (0,1.5) -- (0,2.25);
\draw[green] (1,1.5) -- (1,2.25);
\draw[blue] (2,1.5) -- (2,4.5) -- (-1,4.5) -- (-1,-0.375);

\draw[brown,fill=brown!20] (-0.25,2.25) rectangle (1.25,3);
\node[font=\large,brown] at (0.5,2.625) {$\hat{U}$};

\draw[red] (0,3) -- (0,3.75);
\draw[green] (1,3) -- (1,3.75);
    
\end{scope}

\end{scope}

\draw[brown] (3,3.75) -- (8,-4.5);
\draw[brown] (3,-4.5) -- (8,3.75);

\draw[red] (0,3.75) -- (5,-4.5);
\draw[red] (0,-4.5) -- (5,3.75);

\draw[green] (1,3.75) -- (6,-4.5);
\draw[green] (1,-4.5) -- (6,3.75);

\end{scope}
    
\end{tikzpicture}
}
    }
    } \rb \\[0.8cm]
    -\frac{|r|^2}{|B|\cdot|r| ( |B|^2|r|^2 - 1 ) } & \lb |B|^2 
    \vcenter{
    \hbox{
    \scalebox{0.5}{
\begin{tikzpicture}[line width=1.1pt]

\begin{scope}[xscale=0.8]

\node[scale=1.3] at (1.5,-0.375) {$|\hat{\Psi}\rangle\langle\hat{\Psi}|_{\ell r R_\text{out} L}$};

\draw[orange] (0,0) -- (0,0.75);
\draw[green] (1,0) -- (1,0.75);
\draw[blue] (2,0) -- (2,0.75);
\draw[brown] (3,0) -- (3,3.75);

\draw[darkblue,fill=darkblue!20] (-0.25,0.75) rectangle (2.25,1.5);
\node[font=\large,darkblue] at (1,1.125) {$W$};

\draw[orange] (0,1.5) -- (0,2.25);
\draw[green] (1,1.5) -- (1,2.25);
\draw[blue] (2,1.5) -- (2,5.25) -- (-2,5.25) -- (-2,-0.375);

\draw[brown,fill=brown!20] (-0.25,2.25) rectangle (1.25,3);
\node[font=\large,brown] at (0.5,2.625) {$\hat{U}^\dagger$};

\draw[red] (0,3) -- (0,3.75);
\draw[green] (1,3) -- (1,4.5) -- (-1,4.5) -- (-1,-0.375);


\begin{scope}[yscale=-1,shift={(0,0.75)}]

\draw[orange] (0,0) -- (0,0.75);
\draw[green] (1,0) -- (1,0.75);
\draw[blue] (2,0) -- (2,0.75);
\draw[brown] (3,0) -- (3,3.75);

\draw[darkblue,fill=darkblue!20] (-0.25,0.75) rectangle (2.25,1.5);
\node[font=\large,darkblue] at (1,1.125) {$W^\dagger$};

\draw[orange] (0,1.5) -- (0,2.25);
\draw[green] (1,1.5) -- (1,2.25);
\draw[blue] (2,1.5) -- (2,5.25) -- (-2,5.25) -- (-2,-0.375);

\draw[brown,fill=brown!20] (-0.25,2.25) rectangle (1.25,3);
\node[font=\large,brown] at (0.5,2.625) {$\hat{U}$};

\draw[red] (0,3) -- (0,3.75);
\draw[green] (1,3) -- (1,4.5) -- (-1,4.5) -- (-1,-0.375);
    
\end{scope}


\begin{scope}[shift={(6,0)}]

\node[scale=1.3] at (1.5,-0.375) {$|\hat{\Psi}\rangle\langle\hat{\Psi}|_{\ell r R_\text{out} L}$};

\draw[orange] (0,0) -- (0,0.75);
\draw[green] (1,0) -- (1,0.75);
\draw[blue] (2,0) -- (2,0.75);
\draw[brown] (3,0) -- (3,3.75);

\draw[darkblue,fill=darkblue!20] (-0.25,0.75) rectangle (2.25,1.5);
\node[font=\large,darkblue] at (1,1.125) {$W$};

\draw[orange] (0,1.5) -- (0,2.25);
\draw[green] (1,1.5) -- (1,2.25);
\draw[blue] (2,1.5) -- (2,5.25) -- (-2,5.25) -- (-2,-0.375);

\draw[brown,fill=brown!20] (-0.25,2.25) rectangle (1.25,3);
\node[font=\large,brown] at (0.5,2.625) {$\hat{U}^\dagger$};

\draw[red] (0,3) -- (0,3.75);
\draw[green] (1,3) -- (1,4.5) -- (-1,4.5) -- (-1,-0.375);


\begin{scope}[yscale=-1,shift={(0,0.75)}]

\draw[orange] (0,0) -- (0,0.75);
\draw[green] (1,0) -- (1,0.75);
\draw[blue] (2,0) -- (2,0.75);
\draw[brown] (3,0) -- (3,3.75);

\draw[darkblue,fill=darkblue!20] (-0.25,0.75) rectangle (2.25,1.5);
\node[font=\large,darkblue] at (1,1.125) {$W^\dagger$};

\draw[orange] (0,1.5) -- (0,2.25);
\draw[green] (1,1.5) -- (1,2.25);
\draw[blue] (2,1.5) -- (2,5.25) -- (-2,5.25) -- (-2,-0.375);

\draw[brown,fill=brown!20] (-0.25,2.25) rectangle (1.25,3);
\node[font=\large,brown] at (0.5,2.625) {$\hat{U}$};

\draw[red] (0,3) -- (0,3.75);
\draw[green] (1,3) -- (1,4.5) -- (-1,4.5) -- (-1,-0.375);
    
\end{scope}

\end{scope}

\draw[brown] (3,3.75) -- (9,-4.5);
\draw[brown] (3,-4.5) -- (9,3.75);

\draw[red] (0,3.75) -- (6,-4.5);
\draw[red] (0,-4.5) -- (6,3.75);

\end{scope}
    
\end{tikzpicture}
}
    }
    } + |B|
    \vcenter{
    \hbox{
    \scalebox{0.5}{
\begin{tikzpicture}[line width=1.1pt]

\begin{scope}[xscale=0.8]

\node[scale=1.3] at (1.5,-0.375) {$|\hat{\Psi}\rangle\langle\hat{\Psi}|_{\ell r R_\text{out} L}$};

\draw[orange] (0,0) -- (0,0.75);
\draw[green] (1,0) -- (1,0.75);
\draw[blue] (2,0) -- (2,0.75);
\draw[brown] (3,0) -- (3,3.75);

\draw[darkblue,fill=darkblue!20] (-0.25,0.75) rectangle (2.25,1.5);
\node[font=\large,darkblue] at (1,1.125) {$W$};

\draw[orange] (0,1.5) -- (0,2.25);
\draw[green] (1,1.5) -- (1,2.25);
\draw[blue] (2,1.5) -- (2,4.5) -- (-2,4.5) -- (-2,-0.375);

\draw[brown,fill=brown!20] (-0.25,2.25) rectangle (1.25,3);
\node[font=\large,brown] at (0.5,2.625) {$\hat{U}^\dagger$};

\draw[red] (0,3) -- (0,3.75) -- (-1,3.75) -- (-1,-0.375);
\draw[green] (1,3) -- (1,3.75);


\begin{scope}[yscale=-1,shift={(0,0.75)}]

\draw[orange] (0,0) -- (0,0.75);
\draw[green] (1,0) -- (1,0.75);
\draw[blue] (2,0) -- (2,0.75);
\draw[brown] (3,0) -- (3,3.75);

\draw[darkblue,fill=darkblue!20] (-0.25,0.75) rectangle (2.25,1.5);
\node[font=\large,darkblue] at (1,1.125) {$W^\dagger$};

\draw[orange] (0,1.5) -- (0,2.25);
\draw[green] (1,1.5) -- (1,2.25);
\draw[blue] (2,1.5) -- (2,4.5) -- (-2,4.5) -- (-2,-0.375);

\draw[brown,fill=brown!20] (-0.25,2.25) rectangle (1.25,3);
\node[font=\large,brown] at (0.5,2.625) {$\hat{U}$};

\draw[red] (0,3) -- (0,3.75) -- (-1,3.75) -- (-1,-0.375);
\draw[green] (1,3) -- (1,3.75);
    
\end{scope}


\begin{scope}[shift={(6,0)}]

\node[scale=1.3] at (1.5,-0.375) {$|\hat{\Psi}\rangle\langle\hat{\Psi}|_{\ell r R_\text{out} L}$};

\draw[orange] (0,0) -- (0,0.75);
\draw[green] (1,0) -- (1,0.75);
\draw[blue] (2,0) -- (2,0.75);
\draw[brown] (3,0) -- (3,3.75);

\draw[darkblue,fill=darkblue!20] (-0.25,0.75) rectangle (2.25,1.5);
\node[font=\large,darkblue] at (1,1.125) {$W$};

\draw[orange] (0,1.5) -- (0,2.25);
\draw[green] (1,1.5) -- (1,2.25);
\draw[blue] (2,1.5) -- (2,4.5) -- (-2,4.5) -- (-2,-0.375);

\draw[brown,fill=brown!20] (-0.25,2.25) rectangle (1.25,3);
\node[font=\large,brown] at (0.5,2.625) {$\hat{U}^\dagger$};

\draw[red] (0,3) -- (0,3.75) -- (-1,3.75) -- (-1,-0.375);
\draw[green] (1,3) -- (1,3.75);


\begin{scope}[yscale=-1,shift={(0,0.75)}]

\draw[orange] (0,0) -- (0,0.75);
\draw[green] (1,0) -- (1,0.75);
\draw[blue] (2,0) -- (2,0.75);
\draw[brown] (3,0) -- (3,3.75);

\draw[darkblue,fill=darkblue!20] (-0.25,0.75) rectangle (2.25,1.5);
\node[font=\large,darkblue] at (1,1.125) {$W^\dagger$};

\draw[orange] (0,1.5) -- (0,2.25);
\draw[green] (1,1.5) -- (1,2.25);
\draw[blue] (2,1.5) -- (2,4.5) -- (-2,4.5) -- (-2,-0.375);

\draw[brown,fill=brown!20] (-0.25,2.25) rectangle (1.25,3);
\node[font=\large,brown] at (0.5,2.625) {$\hat{U}$};

\draw[red] (0,3) -- (0,3.75) -- (-1,3.75) -- (-1,-0.375);
\draw[green] (1,3) -- (1,3.75);
    
\end{scope}

\end{scope}

\draw[brown] (3,3.75) -- (9,-4.5);
\draw[brown] (3,-4.5) -- (9,3.75);

\draw[green] (1,3.75) -- (7,-4.5);
\draw[green] (1,-4.5) -- (7,3.75);

\end{scope}
    
\end{tikzpicture}
}
    }
    }
    \rb \numberthis \label{eq:Int_dU_Psi_W_2}
\end{align*}
Thanks to the cyclic property of the trace, $\hat{U}$ disappears from the first and second terms of (\ref{eq:Int_dU_Psi_W_2}) without the need for integration. $W$ then also disappears in the first term to give $\tr \hat{\Psi}_L^2$, while the second reduces to $\tr(\tr_{R_\text{out}} W|\hat{\Psi}\rangle\langle\hat{\Psi}|W^\dagger)^2$. However, we are not so fortunate with the third and fourth terms, and we must integrate over the effective dynamics using (\ref{eq:dU2}) from Appendix~\ref{app:int}. The resulting integrations also give contributions of $\tr \hat{\Psi}_L^2$ and $\tr(\tr_{R_\text{out}} W|\hat{\Psi}\rangle\langle\hat{\Psi}|W^\dagger)^2$, but with additional factors of $|\ell|$ and $|r|$; combining with the first two, we find
\begin{align*} 
    \int d\hat{U} dU \, &\tr \Psi_L(W_{\ell rR_\text{out}})^2 \\
    &= \Bigg[ \frac{|B|^2|r|^2}{|B|^2|r|^2 - 1} - \frac{|B|\cdot|\ell|\cdot|r|(|r|^2 - 1) + |r|^2(|\ell|^2 - 1)}{(|B|^2|r|^2 - 1)(|\ell|^2|r|^2 - 1)} \Bigg] \tr \hat{\Psi}_L^2 \\
    &+ \Bigg[ \frac{|B|\cdot|r|^2}{|B|^2|r|^2 - 1} - \frac{|B|\cdot|r|^2(|\ell|^2 - 1) + |\ell|\cdot|r|(|r|^2 - 1)}{(|B|^2|r|^2 - 1)(|\ell|^2|r|^2 - 1)} \Bigg] \tr \big( \tr_{R_\text{out}} W|\hat{\Psi}\rangle\langle\hat{\Psi}|W^\dagger \big)^2. \numberthis \label{eq:Int_dU_dHat_Psi_W_2}
\end{align*}
This completes the integration over the first term in the right hand side of (\ref{eq:integral_L2}). The result for the second and third terms are similar -- we only need to replace $\tr(\tr_{R_\text{out}} W|\hat{\Psi}\rangle\langle\hat{\Psi}|W^\dagger)^2$ with \begin{equation}
    \tr(\tr_{R_\text{out}} W|\hat{\Psi}\rangle\langle\hat{\Psi}|W^\dagger)(\tr_{R_\text{out}} |\hat{\Psi}\rangle\langle\hat{\Psi}|) \qquad \text{and} \qquad \tr(\tr_{R_\text{out}} |\hat{\Psi}\rangle\langle\hat{\Psi}|)^2,
\end{equation}
respectively. The $\tr\hat{\Psi}_L^2$ term in (\ref{eq:Int_dU_dHat_Psi_W_2}) has no $W$ dependence and is the same for all three terms in (\ref{eq:integral_L2}); it therefore cancels in the sum. The final result for the integration over the squared $L_2$ norm is then given by
\begin{align*}
    \int d\hat{U} &dU \, \big|\big| \Psi_L(W_{\ell rR_\text{out}}) - \Psi_L(\id) \big|\big|_2^2 \\
        &= \Bigg[ \frac{|B|\cdot|r|^2}{|B|^2|r|^2 - 1} - \frac{|B|\cdot|r|^2(|\ell|^2 - 1) + |\ell|\cdot|r|(|r|^2 - 1)}{(|B|^2|r|^2 - 1)(|\ell|^2|r|^2 - 1)} \Bigg] \\
        &\times \tr \Big[ \big( \tr_{R_\text{out}} W|\hat{\Psi}\rangle\langle\hat{\Psi}|W^\dagger \big)^2 - 2 \big( \tr_{R_\text{out}} W|\hat{\Psi}\rangle\langle\hat{\Psi}|W^\dagger \big)\big( \tr_{R_\text{out}} |\hat{\Psi}\rangle\langle\hat{\Psi}| \big) + \big( \tr_{R_\text{out}} |\hat{\Psi}\rangle\langle\hat{\Psi}| \big)^2 \Big]. \numberthis \label{eq:integral_L2_exact}
\end{align*}
To check that the action of $W$ decouples to exponential precision in the black hole's entropy, we upper bound this function. We first maximize over choices of the initial state $|\hat{\Psi}\rangle$ in the effective description; since $\tr(\rho)^2 \leq 1$ for any density matrix $\rho$, the state dependent portion of (\ref{eq:integral_L2_exact}) takes a maximum value of 2. Applying (\ref{eq:bounder}) multiple times, we find
\begin{equation}
    \int d\hat{U} dU \, \big|\big| \Psi_L(W_{\ell rR_\text{out}}) - \Psi_L(\id) \big|\big|_2^2 \leq \frac{4}{|B|} \lp 1 + \frac{2}{|\ell|^2|r|^2} + \frac{2}{|B|\cdot|\ell|\cdot|r|^3} \rp.
\end{equation}
Plugging into (\ref{eq:bounded_recon_BR}), we find the upper bound for the average decoupling from $L$,
\begin{equation} \label{eq:recon_BR}
    \int d\hat{U} dU \, \big|\big| \Psi_L(W_{\ell rR_\text{out}}) - \Psi_L(\id) \big|\big|_1 \leq 2 \sqrt{ \frac{|L|}{|B|} \lp 1 + \frac{2}{|\ell|^2 |r|^2} + \frac{2}{|B|\cdot|\ell|\cdot|r|^3} \rp}.
\end{equation}
Indeed, the factor of $1/\sqrt{|B|}$ ensures that the information about $W$ decouples from $L$ to a precision exponentially small in the black hole's entropy. According to the theorems discussed earlier, this implies that the BFP map does indeed provide a good reconstruction of interior operators $W$ in the effective description on $BR_\text{out}$ in the fundamental description. In fact, the first term in the upper bound (\ref{eq:recon_BR}) is exactly the same upper bound found by PHEVA in \cite{akers_black_2022} for their generic holographic map (\ref{eq:PHEVA_gen}). The second two terms arise from the inclusion of non-trivial effective dynamics. They are small compared to the first term (particularly at late times) and we consider them to be ``corrections'' to the PHEVA result. 

We also compare our reconstruction with the BFP map against reconstruction using PHEVA's dynamical map. The exact result for their squared $L_2$ norm is given by,
\begin{equation} \label{eq:PHEVA_integral_L2_exact}
    \int dU \, \big|\big| \Psi_L^\text{PHEVA}(W_{\ell rR_\text{out}}) - \Psi_L^\text{PHEVA}(\id) \big|\big|_2^2 = \frac{|r|^2}{|B|^2 |r|^2 - 1} \lb |B| D_{R_\text{out}} - \frac{|B|}{|r|} D_{rR_\text{out}} - \frac{1}{|r|} D_{\ell R_\text{out}} \rb,
\end{equation}
where
\begin{equation}
    D_{A} \equiv \tr \Big[ \big( \tr_A W|\hat{\Psi}\rangle\langle\hat{\Psi}|W^\dagger \big)^2 - 2 \big( \tr_A W|\hat{\Psi}\rangle\langle\hat{\Psi}|W^\dagger \big)\big( \tr_A |\hat{\Psi}\rangle\langle\hat{\Psi}| \big) + \big( \tr_A |\hat{\Psi}\rangle\langle\hat{\Psi}| \big)^2 \Big].
\end{equation}
We note that our result for the BFP map (\ref{eq:integral_L2_exact}) also depends on $D_{R_\text{out}}$, but $D_{rR_\text{out}}$ and $D_{\ell R_\text{out}}$ contribute to the decoupling for PHEVA's dynamical map only. The upper bound on the $L_1$ norm for PHEVA's dynamical map is then
\begin{equation}
     \int dU \, \big|\big| \Psi_L^\text{PHEVA}(W_{\ell rR_\text{out}}) - \Psi_L^\text{PHEVA}(\id) \big|\big|_1 \leq 2 \sqrt{ \frac{|L|}{|B|} \lp 1 - \frac{1}{|r|} - \frac{1}{|B|\cdot|r|} \rp },
\end{equation}
which also has the same leading term as the generic map accompanied by small corrections from the fundamental dynamics alone. We note that the corrections appearing in (\ref{eq:recon_BR}) involve additional factors of $|\ell|$ and $|r|$ thanks to the effective dynamics.

\begin{figure} 
    \centering
    \begin{subfigure}{.5\textwidth}
        \centering
        \includegraphics[width=\linewidth]{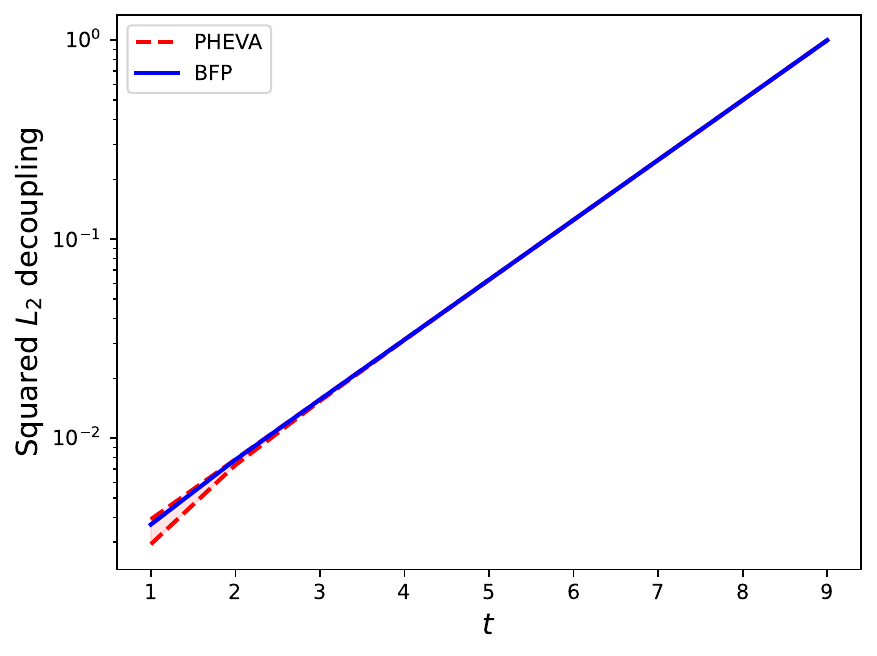}
    \end{subfigure}%
    \begin{subfigure}{.5\textwidth}
        \centering
        \includegraphics[width=\linewidth]{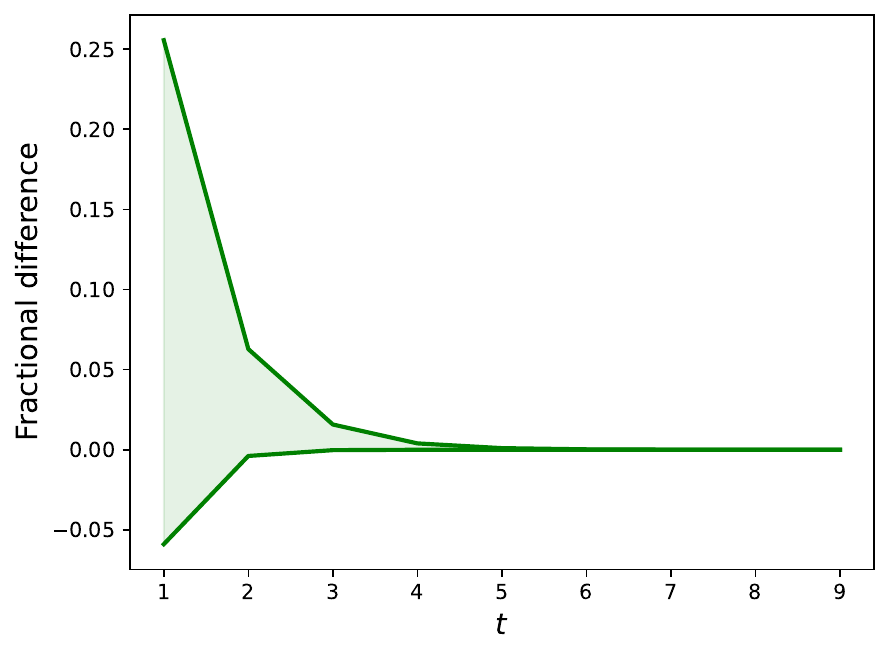}
    \end{subfigure}
    \caption{Plots comparing the squared $L_2$ decoupling from the reference system $L$ for the BFP map (\ref{eq:integral_L2_exact}) in blue and PHEVA's dynamical map (\ref{eq:PHEVA_integral_L2_exact}) in dashed red. The right panel shows the fractional difference between the two, $(\text{BFP} - \text{PHEVA})/\text{PHEVA}$. $D_{R_\text{out}}$ was set to 2 in all plots, with the full range of possible values with this choice for PHEVA's dynamical map shaded.}
    \label{fig:decoup_compare}
\end{figure}

Fig.~\ref{fig:decoup_compare} compares the exact expressions for the squared $L_2$ decoupling using the BFP map (\ref{eq:integral_L2_exact}) and PHEVA's dynamical map (\ref{eq:PHEVA_integral_L2_exact}) as a function of time during evaporation. Common parameters were set to $q=2$, $n_0 = 10$, $m_0 = 1$, and $D_{R_\text{out}} = 2$. The full range of (\ref{eq:PHEVA_integral_L2_exact}) from the maximum at $D_{rR_\text{out}} = D_{\ell R_\text{out}} = 0$ to the minimum at $D_{rR_\text{out}} = D_{\ell R_\text{out}} = 2$ is shaded in both plots. We see that the BFP map always falls within the range of possible values for PHEVA's dynamical map, performing better for certain choices of $|\hat{\Psi}\rangle$. Stronger agreement can be seen at later times.

\subsubsection*{Reconstruction on $B$}
Now that we have verified that the BFP map can be used to reconstruct $W$ anywhere on $BR_\text{out}$, we check that the reconstructions on $B$ and $R_\text{out}$ separately satisfy the requirements of entanglement wedge reconstruction. Many of the details involved in these calculations are similar to those above for the reconstruction on $BR_\text{out}$, so we will suppress them and focus on the relevant differences.

Generalizing the above theorems, our ability to use $V_\text{BFP}$ to reconstruct $W_{\ell r}$ on $B$ depends on the decoupling of $W_{\ell r}$ from both $L$ and $R_\text{out}$,
\begin{equation} \label{eq:bounded_recon_B}
    \int d\hat{U} dU \, \big\| \Psi_{LR_\text{out}} (W_{\ell r}) - \Psi_{LR_\text{out}} (\id) \big\|_1 \leq \sqrt{ |L|\cdot|R_\text{out}| \int d\hat{U} dU \, \big\| \Psi_{LR_\text{out}} (W_{\ell r}) - \Psi_{LR_\text{out}} (\id) \big\|_2^2 }.
\end{equation}
We again bound the $L_1$ decoupling using the $L_2$ decoupling, here given by
\begin{align*}
    \int d\hat{U} dU &\, \big|\big| \Psi_{LR_\text{out}} (W_{\ell r}) - \Psi_{L R_\text{out}} (\id) \big|\big|_2^2 \\
        &= \Bigg[ \frac{|B|\cdot|r|^2}{|B|^2|r|^2 - 1} - \frac{|B|\cdot|r|^2(|\ell|^2 - 1) + |\ell|\cdot|r|(|r|^2 - 1)}{(|B|^2|r|^2 - 1)(|\ell|^2|r|^2 - 1)} \Bigg] \Big[ 2 - 2\tr\big( W_{\ell r} |\hat{\Psi}\rangle\langle\hat{\Psi}| W_{\ell r}^\dagger |\hat{\Psi}\rangle\langle\hat{\Psi}| \big) \Big]. \numberthis
\end{align*}
To find an upper bound on the $L_2$ decoupling, we maximize over choice of $|\hat{\Psi}\rangle$ by setting
\begin{equation}
    \tr\big( W_{\ell r} |\hat{\Psi}\rangle\langle\hat{\Psi}| W_{\ell r}^\dagger |\hat{\Psi}\rangle\langle\hat{\Psi}| \big) = 0.
\end{equation}
Applying (\ref{eq:bounder}) and substituting into (\ref{eq:bounded_recon_B}), we find
\begin{equation} \label{eq:recon_B}
    \int dU d\hat{U} \, \big|\big| \Psi_{LR_\text{out}}(W_{\ell r}) - \Psi_{LR_\text{out}}(\id) \big|\big|_1 \leq 2 \sqrt{ \frac{|L|\cdot|R_\text{out}|}{|B|} \lp 1 + \frac{2}{|\ell|^2 |r|^2} + \frac{2}{|B|\cdot|\ell|\cdot|r|^3} \rp}.
\end{equation}
Again, the first term is exactly what was found by PHEVA in \cite{akers_black_2022} for their upper bound for reconstruction on $B$. The second and third terms are the same corrections as we found in (\ref{eq:recon_BR}) for the reconstruction on $BR_\text{out}$. Requiring that the upper bound on the $L_1$ decoupling be exponentially small in the black hole's entropy implies 
\begin{equation} \label{eq:condition_B}
    |L|\cdot|R_\text{out}| \lp 1 + \frac{2}{|\ell|^2 |r|^2} + \frac{2}{|B|\cdot|\ell|\cdot|r|^3} \rp \ll |B| \qquad \implies \qquad |L|\cdot|R_\text{out}| \ll |B|,
\end{equation}
which is true for times earlier than the Page time. Therefore interior operators $W_{\ell r}$ can be reconstructed on $B$ to exponential precision before the Page time, reminiscent of entanglement wedge reconstruction. Indeed, taking a logarithm of (\ref{eq:condition_B}) gives the condition for entanglement wedge reconstruction on $B$,
\begin{equation}
    S_2(\hat{\Psi}_{R_\text{out}}) \ll \log|B| + S_2(\hat{\Psi}_L),
\end{equation}
just as was found for PHEVA's generic holographic map in \cite{akers_black_2022}. 

\subsubsection*{Reconstruction on $R_\text{out}$}
Similarly, reconstruction of $W_{\ell r}$ on $R_\text{out}$ depends on the decoupling of both $L$ and $B$,
\begin{equation}
    \int d\hat{U} dU \, \big\| \Psi_{LB} (W_{\ell r}) - \Psi_{LB} (\id) \big\|_1 \leq \sqrt{ |L|\cdot|B| \int d\hat{U} dU \, \big\| \Psi_{LB} (W_{\ell r}) - \Psi_{LB} (\id) \big\|_2^2 },
\end{equation}
which is upper bounded by the $L_2$ norm,
\begin{align*}
    \int d\hat{U} dU \, \big|\big| \Psi_{LB} (W_{\ell r}) - \Psi_{LB} (\id) \big|\big|_2^2 &= \Bigg[ \frac{|B|^2|r|^2}{|B|^2|r|^2 - 1} - \frac{|r|^2(|\ell|^2 - 1) + |B|\cdot|\ell|\cdot|r|(|r|^2 - 1)}{(|B|^2|r|^2 - 1)(|\ell|^2|r|^2 - 1)} \Bigg] \\[0.25cm]
        &\qquad \times\Big[ 2 e^{-S_2(\hat{\Psi}_{R_\text{out}})} - 2\tr\big( W_{\ell r} \hat{\Psi}_{L\ell r} W_{\ell r}^\dagger \hat{\Psi}_{L\ell r} \big) \Big]. \numberthis
\end{align*}
This $L_2$ norm is maximized when 
\begin{equation}
    \tr\big( W_{\ell r} \hat{\Psi}_{L\ell r} W_{\ell r}^\dagger \hat{\Psi}_{L\ell r} \big) = 0,
\end{equation}
which bounds the $L_1$ decoupling of $LB$ as
\begin{equation} \label{eq:recon_Rout}
    \int dU d\hat{U} \, \big|\big| \Psi_{LB}(W_{\ell r}) - \Psi_{LB}(\id) \big|\big|_1 \leq 2 \sqrt{ \frac{|L|\cdot|B|}{e^{S_2(\hat{\Psi}_{R_\text{out}})}} \lp 1 + \frac{2}{|B|\cdot|\ell|\cdot|r|^3} + \frac{2}{|B|^2 |\ell|^2 |r|^2} \rp}.
\end{equation}
As before, the first term agrees exactly with \cite{akers_black_2022}. The second term appears as a correction in both the reconstruction on $BR_\text{out}$ from (\ref{eq:recon_BR}) and the reconstruction on $B$ from (\ref{eq:recon_B}), but the third is a new correction further suppressed by an additional factor of $1/|B|^2$. Requiring the right hand side of (\ref{eq:recon_Rout}) to be small implies
\begin{equation}
    |L|\cdot|B| \lp 1 + \frac{2}{|B|\cdot|\ell|\cdot|r|^3} + \frac{2}{|B|^2 |\ell|^2 |r|^2} \rp \ll e^{S_2(\hat{\Psi}_{R_\text{out}})} \quad \implies \quad |L|\cdot|B| \ll e^{S_2(\hat{\Psi}_{R_\text{out}})}.
\end{equation}
Again, taking a log gives the condition for entanglement wedge reconstruction on $R_\text{out}$,
\begin{equation}
    S_2(\hat{\Psi}_{R_\text{out}}) \gg \log|B| + S_2(\hat{\Psi}_L).
\end{equation}
Thus $W_{\ell r}$ can be reconstructed on $R_\text{out}$ with exponential precision at times much later than the Page time.

\section{Conclusion} \label{sec:conc}

In this work, we have further investigated properties of the backwards-forwards and BFP maps for encoding black hole interiors with non-trivial interior dynamics first proposed in \cite{dewolfe_non-isometric_2023}. Restricting to the subspace of dynamically generated states, $V_\text{BF}$ acts unitarily, $V_\text{BFP}$ acts isometrically, and both reproduce the same QES formula and Page curve as found by PHEVA in \cite{akers_black_2022}. On generic states of the effective Hilbert space, we have shown that only the BFP map satisfies all of the same criteria as PHEVA's dynamical map to be a ``good'' holographic map: it acts isometrically on average (with exponentially small fluctuations) and provides a exponentially precise state-dependent reconstruction on $BR_\text{out}$ that implies entanglement wedge reconstruction.

Furthermore, the BFP map acts as an exact isometry on dynamically generated states. Thus the set of dynamically generated states forms a unique subspace of the effective Hilbert space. This has two consequences. First, reconstruction of dynamical unitaries is exact and state-independent on this subspace. Second, dynamically generated states must be subexponentially complex according to the measure concentration results of \cite{akers_black_2022}. If we take the reference state to be the state of $R_\text{in}$ as it crosses the horizon, this is ensured by the limited lifetime of a black hole -- only a number of gates polynomial in $n_0$ can be applied before the black hole evaporates. This choice of reference state is supported by the independence of the BFP map from external interactions (as demonstrated in \cite{dewolfe_non-isometric_2023}) since any external unitary of exponential depth will drop out of the BFP map.

It is important to note that many of the calculations in this work depend on averaging the effective dynamics $\hat{U}$ over the Haar measure. We find it unlikely that the interactions present in semiclassical gravity will be strongly scrambling or complex enough to be truly modeled by Haar random unitaries. Instead, we might opt to model the effective dynamics with a unitary $k$-design that can reproduce the above integrals without being too strongly scrambling or complex. The BFP map would then involve unitary $k$-designs for the effective dynamics and pseudorandom unitaries for the fundamental dynamics (following the suggestions of \cite{kim_complementarity_2023}). We further assume that the combination of these is random enough for the measure concentration results of \cite{akers_black_2022} to apply. So long as they do, we may continue claiming that the BFP map acts isometrically on subexponentially complex generic states of the effective description, just as PHEVA's dynamical map does.

Future work is also needed to better understand how to generalize the toy models of black holes considered here to more realistic scenarios. For example, the treatment  of Hawking modes entering and exiting the model relied on a model of modes coming down from high energy to motivate the ``freezing out'' of radiation modes as they are blue-shifted towards some ultraviolet scale during the backwards evolution in the effective description. It would be interesting to place this within a more realistic model of effective field theory with a cutoff interacting within, and backreacting upon, semiclassical gravity. We also did not need to assume very much about the unitaries $\hat{U}$ used to model interior interactions in the effective description beyond that they are well-modeled by a unitary $k$-design, but one might hope that a more realistic model incorporating more of bulk locality would constrain these further; some progress could potentially be made using well understood models such as tensor networks and JT gravity. There is still much work to be done in this area.

\section*{Acknowledgements}

We are grateful to Chris Akers, Netta Engelhardt, Daniel Harlow, Isaac Kim, Tyler McMaken, John Preskill, and Graeme Smith for helpful discussions. The authors are supported by the Department of Energy under grants DE-SC0010005 and DE-SC0020360.

\appendix

\section{Integration over the Haar measure} \label{app:int}

Here we review some techniques for integrating over the Haar measure. We will only cover those formulas that are needed for the computations performed in this work; for a more general review, see for example \cite{mele_introduction_2023}.

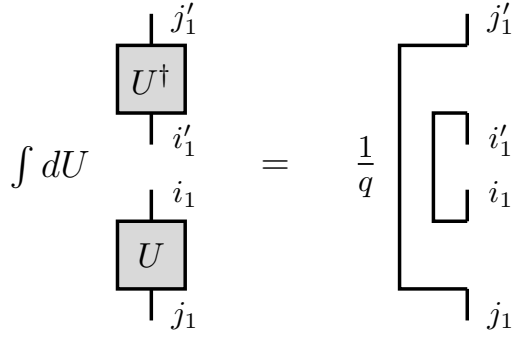
\begin{figure} 
    \centering

\scalebox{1.2}{
\begin{tikzpicture}[line width=1.1pt]

\node[font=\large] at (0,0) {$\int dU$};

\draw (1.125,1.35) -- (1.125,1.7);
\node[font=\small] at (1.5,1.65) {$j'_1$};

\draw[fill=gray!30] (0.75,0.6) rectangle (1.5,1.35);
\node at (1.125,0.975) {$U^\dagger$};

\draw (1.125,0.6) -- (1.125,0.25);
\node[font=\small] at (1.5,0.3) {$i'_1$};

\draw (1.125,-0.6) -- (1.125,-0.25);
\node[font=\small] at (1.5,-0.3) {$i_1$};

\draw[fill=gray!30] (0.75,-0.6) rectangle (1.5,-1.35);
\node at (1.125,-0.975) {$U$};

\draw (1.125,-1.35) -- (1.125,-1.7);
\node[font=\small] at (1.5,-1.65) {$j_1$};

\node[font=\large] at (2.5,0) {$=$};
\node[font=\Large] at (3.5,0) {$\frac{1}{q}$};

\draw (4.625,1.35) -- (4.625,1.7);
\node[font=\small] at (5,1.65) {$j'_1$};
\draw (4.625,0.6) -- (4.625,0.25);
\node[font=\small] at (5,0.3) {$i'_1$};
\draw (4.625,-0.6) -- (4.625,-0.25);
\node[font=\small] at (5,-0.3) {$i_1$};
\draw (4.625,-1.35) -- (4.625,-1.7);
\node[font=\small] at (5,-1.65) {$j_1$};

\draw (4.625,0.6) -- (4.25,0.6) -- (4.25,-0.6) -- (4.625,-0.6);

\draw (4.625,1.35) -- (3.875,1.35) -- (3.875,-1.35) -- (4.625,-1.35);

\end{tikzpicture}
}
    \caption{A pictorial representation of equation~(\ref{eq:dU1}) for the Haar integration over $U$ involving a single copy of $U$ and its Hermitian conjugate.}
    \label{fig:dU1}
\end{figure}

Haar measure integration of a quantum circuit allows us to understand the behavior of the circuit for ``typical'' unitaries. The Haar measure $dU$ is normalized such that
\begin{equation}
    \int dU = 1.
\end{equation}
There are generic formulas for integrals over arbitrary combinations of unitaries, but we only need two specific integrations for this work. The first involves a unitary $U$ and its Hermitian conjugate,
\begin{equation} \label{eq:dU1}
    \int dU \, U^\dagger_{j'_1 i'_1}  U_{i_1 j_1} = \frac{1}{q} \delta_{i_1 i'_1} \delta_{j_1 j'_1},
\end{equation}
where $q$ is the dimension of each index. A pictorial representation of (\ref{eq:dU1}) is shown in Fig.~\ref{fig:dU1}. The legs $i_1,i'_1,j_1,j'_1$ can be connected in any way by inserting delta functions and summing over contracted indices. For example, if we connect legs $i_1$ and $i'_1$ with $\delta_{i_1 i'_1}$, the completed loop gives a contribution of $q$ and the integral reduces to the identity $\mathbb{1}_{j'_1 j_1}$.

The second integral involves two copies of a unitary and its Hermitian conjugate,
\begin{align*}
    \int dU \, U_{j'_1 i'_1}^\dagger U_{j'_2 i'_2}^\dagger U_{i_1 j_1} U_{i_2 j_2} &= \frac{1}{q^2-1} \Big[ \delta_{i_1 i'_1} \delta_{i_2 i'_2} \delta_{j_1 j'_1} \delta_{j_2 j'_2} + \delta_{i_1 i'_2} \delta_{i_2 i'_1} \delta_{j_1 j'_2} \delta_{j_2 j'_1} \Big] \\
        &\qquad -\frac{1}{q(q^2-1)} \Big[ \delta_{i_1 i'_1} \delta_{i_2 i'_2} \delta_{j_1 j'_2} \delta_{j_2 j'_1} + \delta_{i_1 i'_2} \delta_{i_2 i'_1} \delta_{j_1 j'_1} \delta_{j_2 j'_2} \Big].\numberthis \label{eq:dU2}
\end{align*}
A pictorial representation of (\ref{eq:dU2}) is shown in Fig.~\ref{fig:dU2}. We note that connecting $i_1$, $i'_1$ via $\delta_{i_1 i'_1}$ and $i_2$, $i'_2$ via $\delta_{i_2 i'_2}$ again reduces the integral to the identity.

\begin{figure} 
    \centering

\scalebox{0.9}{
\begin{tikzpicture}[line width=1.1pt]

\node[font=\large] at (0,0) {$\int dU$};

\draw (1.125,1.35) -- (1.125,1.7);
\node[font=\small] at (1.5,1.65) {$j'_1$};

\draw[fill=gray!30] (0.75,0.6) rectangle (1.5,1.35);
\node at (1.125,0.975) {$U^\dagger$};

\draw (1.125,0.6) -- (1.125,0.25);
\node[font=\small] at (1.5,0.3) {$i'_1$};

\draw (1.125,-0.6) -- (1.125,-0.25);
\node[font=\small] at (1.5,-0.3) {$i_1$};

\draw[fill=gray!30] (0.75,-0.6) rectangle (1.5,-1.35);
\node at (1.125,-0.975) {$U$};

\draw (1.125,-1.35) -- (1.125,-1.7);
\node[font=\small] at (1.5,-1.65) {$j_1$};

\begin{scope}[shift={(1.25,0)}]
\draw (1.125,1.35) -- (1.125,1.7);
\node[font=\small] at (1.5,1.65) {$j'_2$};

\draw[fill=gray!30] (0.75,0.6) rectangle (1.5,1.35);
\node at (1.125,0.975) {$U^\dagger$};

\draw (1.125,0.6) -- (1.125,0.25);
\node[font=\small] at (1.5,0.3) {$i'_2$};

\draw (1.125,-0.6) -- (1.125,-0.25);
\node[font=\small] at (1.5,-0.3) {$i_2$};

\draw[fill=gray!30] (0.75,-0.6) rectangle (1.5,-1.35);
\node at (1.125,-0.975) {$U$};

\draw (1.125,-1.35) -- (1.125,-1.7);
\node[font=\small] at (1.5,-1.65) {$j_2$};
\end{scope}

\node[font=\large] at (3.75,0) {$=$};

\node[font=\Large] at (4.75,0) {$\frac{1}{q^2-1}$};
\draw (5.6,2) -- (5.5,2) -- (5.5,-2) -- (5.6,-2);

\begin{scope}[shift={(6.5,0)}]
\draw (0,0.6) -- (0,0.25);
\node[font=\small] at (0.375,0.3) {$i'_1$};
\draw (0,-0.6) -- (0,-0.25);
\node[font=\small] at (0.375,-0.3) {$i_1$};
\draw (0,0.6) -- (-0.375,0.6) -- (-0.375,-0.6) -- (0,-0.6);

\begin{scope}[shift={(1.25,0)}]
\draw (0,0.6) -- (0,0.25);
\node[font=\small] at (0.375,0.3) {$i'_2$};
\draw (0,-0.6) -- (0,-0.25);
\node[font=\small] at (0.375,-0.3) {$i_2$};
\draw (0,0.6) -- (-0.375,0.6) -- (-0.375,-0.6) -- (0,-0.6);

\begin{scope}[shift={(1.25,0)}]
\draw (0,1.35) -- (0,1.7);
\node[font=\small] at (0.325,1.65) {$j'_1$};
\draw (0,-1.35) -- (0,-1.7);
\node[font=\small] at (0.325,-1.65) {$j_1$};
\draw (0,1.35) -- (0,-1.35);

\begin{scope}[shift={(1.25,0)}]
\draw (0,1.35) -- (0,1.7);
\node[font=\small] at (0.325,1.65) {$j'_2$};
\draw (0,-1.35) -- (0,-1.7);
\node[font=\small] at (0.325,-1.65) {$j_2$};
\draw (0,1.35) -- (0,-1.35);
\end{scope}
\end{scope}
\end{scope}
\end{scope}

\node[font=\large] at (11.5,0) {$+$};

\begin{scope}[shift={(12.5,0)}]
\draw (0,0.6) -- (0,0.25);
\node[font=\small] at (-0.375,0.3) {$i'_1$};
\draw (0,-0.6) -- (0,-0.25);
\node[font=\small] at (-0.375,-0.3) {$i_1$};
\draw (0,0.6) -- (1.25,-0.6);

\begin{scope}[shift={(1.25,0)}]
\draw (0,0.6) -- (0,0.25);
\node[font=\small] at (0.375,0.3) {$i'_2$};
\draw (0,-0.6) -- (0,-0.25);
\node[font=\small] at (0.375,-0.3) {$i_2$};
\draw (0,0.6) -- (-1.25,-0.6);

\begin{scope}[shift={(1.25,0)}]
\draw (0,1.35) -- (0,1.7);
\node[font=\small] at (0.325,1.65) {$j'_1$};
\draw (0,-1.35) -- (0,-1.7);
\node[font=\small] at (0.325,-1.65) {$j_1$};
\draw (0,1.35) -- (1.25,-1.35);

\begin{scope}[shift={(1.25,0)}]
\draw (0,1.35) -- (0,1.7);
\node[font=\small] at (0.325,1.65) {$j'_2$};
\draw (0,-1.35) -- (0,-1.7);
\node[font=\small] at (0.325,-1.65) {$j_2$};
\draw (0,1.35) -- (-1.25,-1.35);
\end{scope}
\end{scope}
\end{scope}
\end{scope}

\draw (17.4,2) -- (17.5,2) -- (17.5,-2) -- (17.4,-2);


\node[font=\Large] at (4.5,-5) {$-\frac{1}{q(q^2-1)}$};
\draw (5.6,-3) -- (5.5,-3) -- (5.5,-7) -- (5.6,-7);

\begin{scope}[shift={(6.5,-5)}]
\draw (0,0.6) -- (0,0.25);
\node[font=\small] at (0.375,0.3) {$i'_1$};
\draw (0,-0.6) -- (0,-0.25);
\node[font=\small] at (0.375,-0.3) {$i_1$};
\draw (0,0.6) -- (-0.375,0.6) -- (-0.375,-0.6) -- (0,-0.6);

\begin{scope}[shift={(1.25,0)}]
\draw (0,0.6) -- (0,0.25);
\node[font=\small] at (0.375,0.3) {$i'_2$};
\draw (0,-0.6) -- (0,-0.25);
\node[font=\small] at (0.375,-0.3) {$i_2$};
\draw (0,0.6) -- (-0.375,0.6) -- (-0.375,-0.6) -- (0,-0.6);

\begin{scope}[shift={(1.25,0)}]
\draw (0,1.35) -- (0,1.7);
\node[font=\small] at (0.325,1.65) {$j'_1$};
\draw (0,-1.35) -- (0,-1.7);
\node[font=\small] at (0.325,-1.65) {$j_1$};
\draw (0,1.35) -- (1.25,-1.35);

\begin{scope}[shift={(1.25,0)}]
\draw (0,1.35) -- (0,1.7);
\node[font=\small] at (0.325,1.65) {$j'_2$};
\draw (0,-1.35) -- (0,-1.7);
\node[font=\small] at (0.325,-1.65) {$j_2$};
\draw (0,1.35) -- (-1.25,-1.35);
\end{scope}
\end{scope}
\end{scope}
\end{scope}

\node[font=\large] at (11.5,-5) {$+$};

\begin{scope}[shift={(12.5,-5)}]
\draw (0,0.6) -- (0,0.25);
\node[font=\small] at (-0.375,0.3) {$i'_1$};
\draw (0,-0.6) -- (0,-0.25);
\node[font=\small] at (-0.375,-0.3) {$i_1$};
\draw (0,0.6) -- (1.25,-0.6);

\begin{scope}[shift={(1.25,0)}]
\draw (0,0.6) -- (0,0.25);
\node[font=\small] at (0.375,0.3) {$i'_2$};
\draw (0,-0.6) -- (0,-0.25);
\node[font=\small] at (0.375,-0.3) {$i_2$};
\draw (0,0.6) -- (-1.25,-0.6);

\begin{scope}[shift={(1.25,0)}]
\draw (0,1.35) -- (0,1.7);
\node[font=\small] at (0.325,1.65) {$j'_1$};
\draw (0,-1.35) -- (0,-1.7);
\node[font=\small] at (0.325,-1.65) {$j_1$};
\draw (0,1.35) -- (0,-1.35);

\begin{scope}[shift={(1.25,0)}]
\draw (0,1.35) -- (0,1.7);
\node[font=\small] at (0.325,1.65) {$j'_2$};
\draw (0,-1.35) -- (0,-1.7);
\node[font=\small] at (0.325,-1.65) {$j_2$};
\draw (0,1.35) -- (0,-1.35);
\end{scope}
\end{scope}
\end{scope}
\end{scope}

\draw (17.4,-3) -- (17.5,-3) -- (17.5,-7) -- (17.4,-7);

\end{tikzpicture}
}
    \caption{A pictorial representation of equation~(\ref{eq:dU2}) for the Haar integration over $U$ involving two copies of $U$ and its Hermitian conjugate.}
    \label{fig:dU2}
\end{figure}

It is sometimes convenient to re-express (\ref{eq:dU2}) as a partition function over a spin system \cite{nahum_operator_2018,akers_black_2022}. Notice that all of the delta functions in (\ref{eq:dU2}) connect $i$ to $i'$ (we will refer to these as ``internal'' indices) and $j$ to $j'$ (referred to as ``external'' indices). There are only two ways to connect each set of $i$ and $j$ indices, leading to the four terms in (\ref{eq:dU2}). The first connects indices of the same number, such as $\delta_{i_1 i'_1} \delta_{i_2 i'_2}$. These are the vertically contracted indices in Fig.~\ref{fig:dU2}. We will denote such contractions using a spin in the $\uparrow$ state; $|\!\uparrow\rangle$ for the internal $i$ indices and $\langle\uparrow\!|$ for the external $j$ indices. The second class of delta functions connects indices of different number, such as $\delta_{i_1 i'_2} \delta_{i_2 i'_1}$, shown by swapped contractions in Fig.~\ref{fig:dU2}. These contractions will be denoted by a spin in the $\downarrow$ state; again $|\!\downarrow\rangle$ for internal $i$ indices and $\langle\downarrow\!|$ for external $j$ indices. 

We use the delta functions to define inner products between the ``spin'' basis $|\!\uparrow,\downarrow\rangle$ and the ``index'' basis $|i_1 i'_1 i_2 i'_2\rangle$ as
\begin{equation} \label{eq:index_basis}
    \langle i_1 i'_1 i_2 i'_2 |\! \uparrow \rangle = \frac{1}{q} \delta_{i_1 i'_1} \delta_{i_2 i'_2}, \quad \langle i_1 i'_1 i_2 i'_2 |\! \downarrow \rangle = \frac{1}{q} \delta_{i_1 i'_2} \delta_{i_2 i'_1}.
\end{equation}
We can then rewrite (\ref{eq:dU2}) using the spin basis as
\begin{equation} \label{eq:spin_int}
    \int dU \, U^\dagger \otimes U^\dagger \otimes U \otimes U = \frac{q^2}{q^2 - 1} \Big( |\!\uparrow\rangle\langle\uparrow\!| + |\!\downarrow\rangle\langle\downarrow\!| \Big) - \frac{q}{q^2 - 1} \Big( |\!\uparrow\rangle\langle\downarrow\!| + |\!\downarrow\rangle\langle\uparrow\!| \Big).
\end{equation}
Taking the expectation value of (\ref{eq:spin_int}) with states in the index basis reproduces (\ref{eq:dU2}). In the spirit of (\ref{eq:spin_int}), we associate two spins with the unitary $U$ we are integrating over. The first $|\!\uparrow,\downarrow\rangle$ indicates how to contract the internal indices $i$ of $U$ and its Hermitian conjugate; the second $\langle\uparrow,\downarrow\!|$ indicates how to contract the outer indices $j$. The full integration is given by a sum over all states on these two spins, analogous to a partition function. 

This technique is most effective when integrating over many different unitaries $U$. Contracting indices between different unitaries requires an inner product on the spin basis,
\begin{equation} \label{eq:spin_inner}
    \langle \uparrow | \downarrow \rangle = \frac{1}{q}
\end{equation}
which can be found by inserting an identity as a complete set of states in the index basis and using (\ref{eq:index_basis}). Multiple integrations can then be performed by taking products of the right hand side of (\ref{eq:spin_int}) for each unitary being averaged, with the ``innermost'' unitary being the leftmost product.

This permits a useful graphical interpretation for integrating over multiple unitaries in two copies of a quantum circuit and its hermitian conjugate. We may construct a spin network by replacing every unitary in the circuit with two spins: one connected to all outgoing indices, the other connected to all ingoing. Then the integration is found by summing over all configurations of these spins. The weights of each term in the sum are seen in (\ref{eq:spin_int}) and are found as follows. A unitary with $k$ indices and two matching spins gives a contribution of 
\begin{equation}
    \vcenter{
    \hbox{
    \scalebox{1.2}{
\begin{tikzpicture}[line width=1.1pt]

\draw[double] (0,0) -- (0,0.75);

\draw (0,0) -- (-1,-0.75);
\draw (0,0) -- (1,-0.75);
\draw (0,0) -- (-0.5,-0.75);
\draw (0,0) -- (0.5,-0.75);

\node[scale=1.2] at (0.05,-0.6) {$\dots$};

\draw (0,0.75) -- (-1,1.5);
\draw (0,0.75) -- (1,1.5);
\draw (0,0.75) -- (-0.5,1.5);
\draw (0,0.75) -- (0.5,1.5);

\node[scale=1.2] at (0.05,1.35) {$\dots$};

\filldraw[darkgreen] (0,0) circle (2pt);
\filldraw[darkgreen] (0,0.75) circle (2pt);

\node at (-0.5,0) {$|\textcolor{darkgreen}{\uparrow}\rangle$};
\node at (0.5,0.75) {$\langle\textcolor{darkgreen}{\uparrow}|$};

\node[scale=1.2] at (1.5,0.325) {$=$};


\begin{scope}[shift={(3,0)}]

\draw[double] (0,0) -- (0,0.75);

\draw (0,0) -- (-1,-0.75);
\draw (0,0) -- (1,-0.75);
\draw (0,0) -- (-0.5,-0.75);
\draw (0,0) -- (0.5,-0.75);

\node[scale=1.2] at (0.05,-0.6) {$\dots$};

\draw (0,0.75) -- (-1,1.5);
\draw (0,0.75) -- (1,1.5);
\draw (0,0.75) -- (-0.5,1.5);
\draw (0,0.75) -- (0.5,1.5);

\node[scale=1.2] at (0.05,1.35) {$\dots$};

\filldraw[red] (0,0) circle (2pt);
\filldraw[red] (0,0.75) circle (2pt);

\node at (-0.5,0) {$|\textcolor{red}{\downarrow}\rangle$};
\node at (0.5,0.75) {$\langle\textcolor{red}{\downarrow}|$};

\node[scale=1.2] at (1.5,0.325) {$=$};
    
\end{scope}

\end{tikzpicture}
}
    }
    }
    \frac{q^{2k}}{q^{2k} - 1}
\end{equation}
while the same unitary with two mismatching spins gives
\begin{equation}
    \vcenter{
    \hbox{
    \scalebox{1.2}{
\begin{tikzpicture}[line width=1.1pt]

\draw[double] (0,0) -- (0,0.75);

\draw (0,0) -- (-1,-0.75);
\draw (0,0) -- (1,-0.75);
\draw (0,0) -- (-0.5,-0.75);
\draw (0,0) -- (0.5,-0.75);

\node[scale=1.2] at (0.05,-0.6) {$\dots$};

\draw (0,0.75) -- (-1,1.5);
\draw (0,0.75) -- (1,1.5);
\draw (0,0.75) -- (-0.5,1.5);
\draw (0,0.75) -- (0.5,1.5);

\node[scale=1.2] at (0.05,1.35) {$\dots$};

\filldraw[darkgreen] (0,0) circle (2pt);
\filldraw[red] (0,0.75) circle (2pt);

\node at (-0.5,0) {$|\textcolor{darkgreen}{\uparrow}\rangle$};
\node at (0.5,0.75) {$\langle\textcolor{red}{\downarrow}|$};

\node[scale=1.2] at (1.5,0.325) {$=$};


\begin{scope}[shift={(3,0)}]

\draw[double] (0,0) -- (0,0.75);

\draw (0,0) -- (-1,-0.75);
\draw (0,0) -- (1,-0.75);
\draw (0,0) -- (-0.5,-0.75);
\draw (0,0) -- (0.5,-0.75);

\node[scale=1.2] at (0.05,-0.6) {$\dots$};

\draw (0,0.75) -- (-1,1.5);
\draw (0,0.75) -- (1,1.5);
\draw (0,0.75) -- (-0.5,1.5);
\draw (0,0.75) -- (0.5,1.5);

\node[scale=1.2] at (0.05,1.35) {$\dots$};

\filldraw[red] (0,0) circle (2pt);
\filldraw[darkgreen] (0,0.75) circle (2pt);

\node at (-0.5,0) {$|\textcolor{red}{\downarrow}\rangle$};
\node at (0.5,0.75) {$\langle\textcolor{darkgreen}{\uparrow}|$};

\node[scale=1.2] at (1.5,0.325) {$=$};
    
\end{scope}

\end{tikzpicture}
}
    }
    }
    \frac{q^k}{q^{2k} - 1}
\end{equation}
These simplify in a large $q$ approximation to $1$ and $1/q$, respectively. Connecting legs give a contribution according to the inner product (\ref{eq:spin_inner}): a leg connecting two matching spins gives a contribution of $1$, while a leg connecting two mismatching spins gives $1/q$. 

Dangling legs can be given a single $\uparrow$ or $\downarrow$ spin to perform certain computations. As an example relevant to Sec.~\ref{sec:page}, consider computing the average second Renyi entropy of subsystem $A$ for a density matrix $\rho_{AB}$ after processing by a Haar random unitary $U$. We can take the logarithm to perform the average:
\begin{equation} \label{eq:second_renyi}
    \int dU \, \tr \big( \tr_B U \rho_{AB} U^\dagger \big)^2 = \int dU
    \vcenter{
    \hbox{
    \scalebox{0.7}{
\begin{tikzpicture}[line width=1.1pt]

\draw (0,0) -- (0,0.75);
\draw (1,0) -- (1,0.75);

\draw[fill=gray!30] (-0.25,0.75) rectangle (1.25,1.5);
\node[font=\large] at (0.5,1.125) {$U$};

\draw (0,1.5) -- (0,2.25);
\node at (-0.25,1.875) {$A$};
\draw (1,1.5) -- (1,2.25);
\node at (1.25,1.875) {$B$};

\node[scale=1.5] at (0.5, 2.625) {$\rho$};

\draw (0,3) -- (0,3.75);
\node at (-0.25,3.375) {$A$};
\draw (1,3) -- (1,3.75);
\node at (1.25,3.375) {$B$};

\draw[fill=gray!30] (-0.25,3.75) rectangle (1.25,4.5);
\node[font=\large] at (0.5,4.125) {$U^\dagger$};

\draw (0,4.5) -- (0,5.25);
\draw (1,4.5) -- (1,5.25) -- (2,5.25) -- (2,0) -- (1,0);


\begin{scope}[shift={(3.5,0)}]

\draw (0,0) -- (0,0.75);
\draw (1,0) -- (1,0.75);

\draw[fill=gray!30] (-0.25,0.75) rectangle (1.25,1.5);
\node[font=\large] at (0.5,1.125) {$U$};

\draw (0,1.5) -- (0,2.25);
\node at (-0.25,1.875) {$A$};
\draw (1,1.5) -- (1,2.25);
\node at (1.25,1.875) {$B$};

\node[scale=1.5] at (0.5, 2.625) {$\rho$};

\draw (0,3) -- (0,3.75);
\node at (-0.25,3.375) {$A$};
\draw (1,3) -- (1,3.75);
\node at (1.25,3.375) {$B$};

\draw[fill=gray!30] (-0.25,3.75) rectangle (1.25,4.5);
\node[font=\large] at (0.5,4.125) {$U^\dagger$};

\draw (0,4.5) -- (0,5.25);
\draw (1,4.5) -- (1,5.25) -- (2,5.25) -- (2,0) -- (1,0);
    
\end{scope}

\draw (0,0) -- (3.5,5.25);
\draw (0,5.25) -- (3.5,0);
    
\end{tikzpicture}
}
    }
    }
\end{equation}
Instead of the circuit diagram on the right hand side of (\ref{eq:second_renyi}), we can use a spin state representation to integrate over the two copies of $U$ and its Hermitian conjugate:
\begin{equation}
    \int dU \, \tr \big( \tr_B U \rho_{AB} U^\dagger \big)^2 = \sum_{s_1,s_2 = \textcolor{darkgreen}{\uparrow}, \textcolor{red}{\downarrow}}
    \vcenter{
    \hbox{
    \scalebox{1}{
\begin{tikzpicture}[line width=1.1pt]

\draw[double] (0,0) -- (0,0.75);
\filldraw (0,0) circle (2pt);
\filldraw (0,0.75) circle (2pt);

\draw (0,0) -- (-0.75,-0.75);
\draw (0,0) -- (0.75,-0.75);

\draw (0,0.75) -- (-0.75,1.5);
\draw (0,0.75) -- (0.75,1.5);

\node at (-0.5,0) {$|s_2\rangle$};
\node at (0.5,0.75) {$\langle s_1|$};

\node at (-0.75,1.75) {$|\textcolor{red}{\downarrow}\rangle$};
\node at (0.75,1.75) {$|\textcolor{darkgreen}{\uparrow}\rangle$};

\draw[decorate,decoration={brace,amplitude=10pt,mirror}] (-0.85,-0.85) -- (0.85,-0.85);

\node at (0,-1.5) {$\langle\rho,\textcolor{darkgreen}{\uparrow}|$};

\end{tikzpicture}
}
    }
    }
\end{equation}
The $B$ subsystem is traced over by a vertical contraction in (\ref{eq:second_renyi}), so we fix the upper right leg in the $\uparrow$ state. The $A$ subsystem is left take the product $(\rho_A)^2$ and is then traced over, represented by a swapped contraction in (\ref{eq:second_renyi}), so we fix the $A$ spin in the $\downarrow$ state. Since the state $\rho$ connects the internal legs of $U$ and $U^\dagger$, we've inserted it into the bottom of the spin network in the $\uparrow$ state as $\langle\rho,\uparrow\!|$. Here, the contractions give
\begin{equation}
    \langle\rho,\uparrow\!|\!\uparrow\rangle = \tr\rho, \qquad \langle\rho,\downarrow\!|\!\downarrow\rangle = \tr\rho^2 = -\ln S_2(\rho).
\end{equation}
The full integration is then performed by summing over the two spins $s_1$ and $s_2$ for the unitary $U$ and using the rules described above.

\bibliographystyle{utphys}
\bibliography{Non-iso}

\end{document}